\documentclass[12pt,a4paper]{article}
\pdfoutput=1
\usepackage{jheppub}

\usepackage{amsmath,amssymb,amsfonts}
\usepackage{graphicx}
\usepackage{color}
\usepackage{tikz}
\usepackage{physics}

\usepackage{dsfont}

\usepackage{subfigure}

\usepackage{booktabs}

\def\cA{{\cal A}}
\def\cB{{\cal B}}
\def\cC{{\cal C}}
\def\cD{{\cal D}}
\def\cF{{\cal F}}

\def\cL{{\cal L}}

\def\cN{{\cal N}}
\def\cW{{\cal W}}
\def\cY{{\cal Y}}
\def\cZ{{\cal Z}}

\def\RR{{\mathds{R}}}

\DeclareMathOperator{\vol}{vol}

\DeclareMathOperator{\Li}{Li}

\makeatletter
\gdef\@fpheader{}
\makeatother

\def\Im{\mathop{\rm Im}}
\def\Re{\mathop{\rm Re}}

\title{Mapping out the internal space in AdS/BCFT with Wilson loops}

\author[a,b]{Lorenzo Coccia}
\emailAdd{l.coccia@campus.unimib.it}

\affiliation[a]{Dipartimento di Fisica, Universit\`a di Milano-Bicocca, I-20126 Milano, Italy}
\affiliation[b]{INFN, Sezione di Milano-Bicocca, I-20126 Milano, Italy}

\author[c]{and Christoph F.~Uhlemann} 

\emailAdd{uhlemann@umich.edu}

\affiliation[c]{Leinweber Center for Theoretical Physics, Department of Physics
	\\
	University of Michigan, Ann Arbor, MI 48109-1040, USA}

\preprint{LCTP-21-37}

\abstract{We study Wilson loops in string theory realizations of AdS/BCFT and wedge holography.
The field theories are based on 3d $\mathcal N=4$ {\it long quiver} gauge theories engineered by D3, D5 and NS5 branes, and on BCFTs involving 4d $\mathcal N=4$ SYM  coupled to such 3d theories. The holographic duals have geometry $AdS_4\times S^2\times S^2$ warped over a strip. 
We identify the holographic representation of antisymmetric Wilson loops associated with individual 3d gauge nodes in terms of probe D5-branes.
The expectation values obtained holographically are matched to supersymmetric localization computations. 
Our results yield an identification of regions in the internal space with individual 3d gauge nodes. 
Connecting to bottom-up braneworld models, this gives a concrete notion of which parts of the 10d solutions correspond to the end-of-the-world brane and which to the remaining bulk.
We also construct supersymmetric {\it Janus on the brane} embeddings in AdS$_5\times$S$^5$, which describe surface defects with boundaries and interfaces in $\mathcal N=4$ SYM.}

\date{\today}

\begin{document}
	
\setcounter{tocdepth}{2}
\maketitle
%\tableofcontents
\parskip 1mm

\section{Introduction}

Conformal field theories on spaces with boundaries (BCFTs) and their holographic duals  have been studied extensively in string theory, starting with \cite{Karch:2000gx,Karch:2000ct} (see also \cite{Takayanagi:2011zk,Fujita:2011fp}).
AdS/BCFT dualities have featured prominently in recent studies of the black hole information paradox, where the holographic duals of BCFTs provide fruitful models for information transfer from black holes \cite{Almheiri:2019hni,Almheiri:2019yqk,Rozali:2019day,Almheiri:2019psy,Geng:2020qvw,Chen:2020uac,Uhlemann:2021nhu}, and in models for cosmology \cite{Cooper:2018cmb,Antonini:2019qkt}.

Holographic duals for BCFTs are often modeled in a bottom-up fashion: an AdS$_{d+1}$ bulk is cut off by an end-of-the-world (ETW) brane, so that a half space remains of the conformal boundary of AdS, on which the $d$-dimensional BCFT is defined (fig.~\ref{fig:braneworld-BCFT}). The ETW brane represents $(d-1)$-dimensional boundary degrees of freedom coupled to the $d$-dimensional ambient CFT. A related construction is the wedge holography proposal of \cite{Akal:2020wfl}. It introduces a second ETW brane ending at the same point of the conformal boundary as the first one and cutting off the remaining half of the AdS$_5$ conformal boundary. This leaves a wedge of AdS$_{d+1}$ (fig.~\ref{fig:braneworld-wedge}). These setups realize holographic duals for 3d CFTs composed of two sectors represented by the two ETW branes.

The simplicity of these braneworld models makes them ideal for extracting qualitative lessons. But a microscopic understanding ultimately needs proper AdS/CFT dualities, with concrete and well-defined gravity theories and QFTs. Such dualities can be derived from brane constructions in string theory. 4d BCFTs can be realized by D3-branes ending on D5 and NS5 branes \cite{Gaiotto:2008sa,Gaiotto:2008ak}. 
The associated Type IIB supergravity solutions were constructed, based on the groundwork of \cite{DHoker:2007zhm,DHoker:2007hhe}, in \cite{Aharony:2011yc}. The full 10d solutions, though more complicated than the bottom-up models, have the advantage of being defined in a UV-complete string theory setting and having concrete QFT duals, in the form of $\mathcal N=4$ SYM coupled to the IR fixed points of 3d $\cN=4$ quiver gauge theories.
Holographic duals for pure 3d SCFTs can also be derived from these constructions, by considering D3-branes suspended between D5 and NS5 branes. The holographic duals were constructed in \cite{Assel:2011xz}, based on the general solutions of \cite{DHoker:2007zhm,DHoker:2007hhe}. This in particular allows to realize string theory versions of wedge holography \cite{Uhlemann:2021nhu}.
We  focus on the 10d Type IIB solutions in this work.

The 10d string theory duals for BCFTs can be framed in a language which incorporates the braneworld model intuition: An asymptotic AdS$_5\times $S$^5$ region -- the dual of the 4d $\cN=4$ SYM ambient CFT -- connects smoothly to a region where the geometry becomes an AdS$_4$ solution, which is the dual of the 3d boundary degrees of freedom. 
This latter part of the geometry is the 10d uplift of the ETW brane.
Instead of being cut off by an ETW brane, the geometry ends smoothly by cycles in the internal space of the AdS$_4$ solution collapsing. 
Recent studies of these solutions include \cite{Bachas:2017rch,Bachas:2018zmb,Coccia:2020cku,Raamsdonk:2020tin,Coccia:2020wtk,Reeves:2021sab,VanRaamsdonk:2021duo,Akhond:2021ffz,Uhlemann:2021nhu,DeLuca:2021ojx}.

In this work we make this intuitive picture for the 10d Type IIB  solutions associated with systems of D3, D5 and NS5 branes quantitative.
The geometry of the supergravity solutions is a warped product of AdS$_4$ and two spheres, S$_1^2$ and S$_2^2$, over a strip, $\Sigma$.
In one asymptotic region of the strip the geometry becomes AdS$_5\times$S$^5$; in the other direction the geometry closes off smoothly.
Based on the study of Wilson loop and vortex loop operators using holography and supersymmetric localization, we identify regions on $\Sigma$ which are naturally associated with 3d degrees of freedom, corresponding to the ETW brane in the braneworld models, and regions which are associated with 4d degrees of freedom, corresponding to the AdS$_5$ bulk.
More broadly speaking, we study 3d SCFTs, 4d BCFTs, and Janus interface CFTs, and use loop operators to map out the internal space in the associated supergravity solutions. 

We now introduce the results in more detail.
The 3d SCFTs we are interested in arise from D3-branes suspended between NS5 branes, augmented by additional D5 branes (fig.~\ref{fig:branes}). The 4d BCFTs arise from D3-branes ending on combinations of D5 and NS5 branes (fig.~\ref{fig:brane-BCFT}), and the Janus CFTs arise from D3-branes intersecting and partly ending on D5 and NS5 branes.
The planar limit of the 3d SCFTs corresponds to the IR fixed points of 3d $\cN=4$ quiver gauge theories with a large number of nodes. Building on similar studies in 5d \cite{Uhlemann:2019ypp,Uhlemann:2020bek}, these 3d theories were studied in \cite{Coccia:2020cku,Coccia:2020wtk}.
The 4d BCFTs are 4d $\cN=4$ SYM on a half space coupled to 3d SCFTs arising from long quiver gauge theories. The `boundary free energy' was studied in \cite{Raamsdonk:2020tin}.
Finally, the Janus interface CFTs we are interested in can be described as two 4d $\cN=4$ SYM nodes on half spaces separated by an interface which hosts a 3d long quiver SCFT.
In these theories we will study Wilson loops associated with individual 3d gauge nodes. We focus on 3d SCFTs for supersymmetric localization computations, building on our earlier work \cite{Coccia:2020wtk}. On the holographic side we study generic solutions, dual to 3d SCFTs, 4d BCFTs and Janus interface CFTs.

We identify the holographic representation of $\frac{1}{2}$-BPS Wilson loops in antisymmetric representations associated with individual 3d gauge nodes as $\frac{1}{2}$-BPS probe D5$^\prime$-branes, embedded in the $AdS_4\times S^2\times S^2\times \Sigma$ solutions in such a way that they wrap a curve in $\Sigma$.
The choice of curve encodes which gauge node the Wilson loop is associated with. How far the D5$^\prime$ extends along the curve encodes the rank of the representation.
Mirror-dual vortex loops are correspondingly represented by NS5$^\prime$ branes embedded along curves in $\Sigma$.
For 3d SCFTs we match the expectation values obtained holographically to field theory calculations using supersymmetric localization, and demonstrate perfect agreement.
The identification of curves in $\Sigma$ with individual gauge nodes connects to recent work in \cite{Akhond:2021ffz}, where certain boundary conditions on $\Sigma$ were identified with the rank function in the dual 3d gauge theory. Here the identification based on loop operators extends through $\Sigma$ and is directly connected to a brane picture. 
The D5$^\prime$-branes carry D3-brane and F1 charges, identifying, respectively, the gauge node and representation of the dual Wilson loop. The space of charges carved out by the D5$^\prime$ embeddings yields the rank function.\footnote{%
The local identification of the internal space with individual gauge nodes has a precendent in AdS$_6$/CFT$_5$, where Wilson loops are represented by D3-branes and their study identifies points in the internal space with faces in the 5-brane web construction of the 5d SCFTs \cite{Uhlemann:2020bek}.}

For the BCFT and Janus solutions there is a region in $\Sigma$ which is swept out by loop operator D5$^\prime$ and NS5$^\prime$ embeddings. 
There is also a region which does not host any such loop operator embeddings, and instead hosts surface operators associated with 4d $\cN=4$ SYM nodes (to be discussed shortly).
We propose to identify the former region as the 3d part of the geometry and the latter as the 4d part. A transition region between them hosts certain 3d loop operators but not their mirror duals. 
The first region then corresponds to the ETW brane region in braneworld models and the second to the AdS$_{d+1}$ bulk.
The results are illustrated in fig.~\ref{fig:braneworld} for BCFT duals and in fig.~\ref{fig:braneworld-Janus} for duals of Janus interface CFTs, where the 3d region is sandwiched between two 4d regions. 
For duals of 3d SCFTs the loop operator embeddings sweep out all of $\Sigma$ (e.g.\ fig.~\ref{fig:D52NS52-WLD5}), identifying all of $\Sigma$ as 3d region, as one would expect.

A special case of the general $AdS_4\times S^2\times S^2\times\Sigma$ solutions is the AdS$_5\times$S$^5$ solution of Type IIB dual to the 4d $\cN=4$ SYM theory on $\RR^{1,3}$. For this solution the 16 supersymmetries preserved by generic $AdS_4\times S^2\times S^2\times\Sigma$ solutions are enhanced to 32. The probe D5$^\prime$-brane embeddings with D3-brane and F1 charges found here, which preserve 8 supersymmetries, can be studied in this solution. They describe surface operators in $\cN=4$ SYM which themselves have a boundary (fig.~\ref{fig:AdS5S5-2}).
By combining two such operators one can realize planar surface operators with an interface on the surface. We leave more detailed studies of these operators for the future. Similar surface operator embeddings exist for BCFT and Janus solutions. They are associated with 4d degrees of freedom in these solutions, and are the D5$^\prime$ embeddings associated with 4d nodes alluded to in the previous paragraph (the blue curves leading to the 4d regions e.g.\ in fig.~\ref{fig:braneworld}).

{\bf Outline:} 
In sec.~\ref{sec:brane} we review the brane construction of 3d $T_\rho^\sigma[SU(N)]$ theories, BCFTs based on 4d $\cN=4$ SYM coupled to such 3d theories, and Janus interface CFTs, as well as the representation of Wilson loops in these theories.
In sec.~\ref{sec:loc} we compute the expectation values of antisymmetric Wilson loops in 3d long quiver SCFTs using supersymmetric localization.
In sec.~\ref{sec:hol} we discuss the holographic duals for the theories introduced in sec.~\ref{sec:brane} and identify the holographic representation of antisymmetric Wilson loops. 
We match the expectation values to field theory computations and identify, based on the Wilson loop discussion, regions in the holographic duals which are naturally associated with either 3d or 4d degrees of freedom.
{\it Janus on the brane} surface operators in AdS$_5\times$S$^5$ are discussed in sec.~\ref{sec:AdS5S5}.
In sec.~\ref{sec:braneworld} we connect the 10d supergravity solutions to the language of braneworld models and propose a concrete notion of uplift for the ETW brane.
We close with a discussion in sec.~\ref{sec:disc}.
The derivation of the D5$^\prime$ BPS conditions is given in app.~\ref{app:BPS}.

\section{4d \texorpdfstring{$\cN=4$}{N=4} SYM BCFTs \& 3d \texorpdfstring{$T_\rho^\sigma[SU(N)]$}{Trhosigma[SU(N)]}}\label{sec:brane}

The field theories we are interested in can be described as IR fixed points of 3d $\cN=4$ linear quiver gauge theories with a large number of $U(\cdot)$ gauge nodes, possibly coupled to 4d $\cN=4$ SYM on half spaces. The quiver diagrams take the following form,
\begin{align}\label{eq:3d4d-quiver}
	\widehat{SU(N_0)} - 
	U&(N_1)-U(N_2)-\ldots -U(N_{L-1})-U(N_L) - \widehat{SU(N_{L+1})}
	\nonumber\\
	&\hskip 2mm |\hskip 15mm |\hskip 27mm |\hskip 18mm |
	\\
	&[k_1] \hskip 10mm [k_2] \hskip 20mm [k_{L-1}] \hskip 11mm [k_L]
	\nonumber
\end{align}
The $SU(\cdot)$ nodes at the ends, which are distinguished by a hat, are 4d $\cN=4$ SYM nodes on a half space, while the $U(\cdot)$ nodes in the interior are 3d $\cN=4$ nodes connected by bifundamental hypermultiplets.
If $N_0$ and $N_{L+1}$ are both zero the quiver describes a 3d gauge theory, and the IR fixed point is a 3d SCFT.
If one of $N_0$ and $N_{L+1}$ is zero and the other non-zero, the quiver describes a 4d BCFT.
If $N_0$ and $N_{L+1}$ are both non-zero the quiver describes a Janus CFT, i.e.\ two $\cN=4$ SYM nodes on half spaces separated by a 3d interface. 
In the following we review the brane construction of these theories and the brane representation of Wilson and vortex loop operators.

\paragraph{Brane construction}

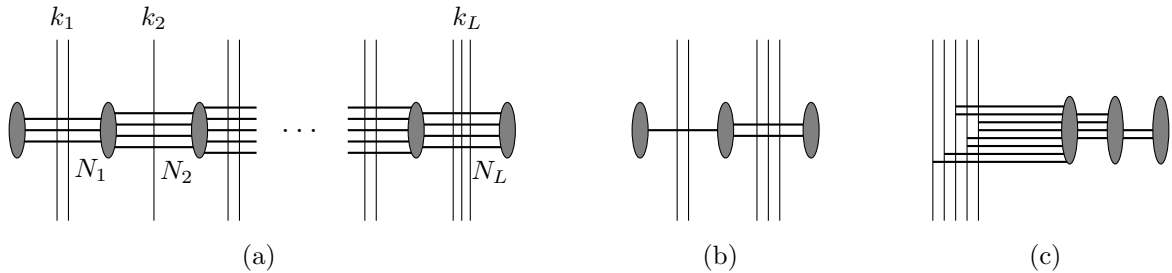
\begin{figure}
	\centering
	\subfigure[][]{\label{fig:branes}
			\begin{tikzpicture}[scale=1.5]
			\foreach \i in {-0.1,0,0.1} \draw[thick] (0,\i) -- (0.8,\i);
			\foreach \i in {-0.15,-0.05,0.05,0.15} \draw[thick] (0.8,\i) -- (1.6,\i);
			\foreach \i in {-0.2,-0.1,0,0.1,0.2} \draw[thick] (1.6,\i) -- (2.1,\i);
			
			\node at (2.5,0) {\ldots};	
			
			\foreach \i in {-0.2,-0.1,0,0.1,0.2} \draw[thick] (2.9,\i) -- (3.5,\i);
			\foreach \i in {-0.15,-0.05,0.05,0.15} \draw[thick] (3.5,\i) -- (4.3,\i);
			
			\foreach \i in {0,0.8,1.6,3.5,4.3}{ \draw[fill=gray] (\i,0) ellipse (2pt and 7pt);}
			
			\foreach \i in {-0.05,0.05} \draw (0.4+\i,-0.8) -- +(0,1.6);
			\foreach \i in {0} \draw (1.2+\i,-0.8) -- +(0,1.6);
			
			\foreach \i in {-0.075,0,0.075} \draw (3.9+\i,-0.8) -- +(0,1.6);
			
			\foreach \i in {-0.05,0.05} \draw (1.9+\i,-0.8) -- +(0,1.6);
			\foreach \i in {-0.05,0.05} \draw (3.1+\i,-0.8) -- +(0,1.6);
			
			\foreach \i in {1,2}{ \node[anchor=south] at ({-0.4+0.8*\i},0.8) {\footnotesize $k_{\i}$};}
			\node[anchor=south] at (3.95,0.8) {\footnotesize $k_{L}$};
			
			\node at (0.65,-0.35) {\footnotesize $N_1$};
			\node at (1.4,-0.37) {\footnotesize $N_2$};
			\node at (4.15,-0.37) {\footnotesize $N_{L}$};        
		\end{tikzpicture}
	}\hskip 12mm
	\subfigure[][]{\label{fig:branes-2}
		\begin{tikzpicture}[scale=1.5]
			\draw[thick] (0,0) -- (0.75,0);
			\draw[thick] (0.75,0.05) -- (1.5,0.05);
			\draw[thick] (0.75,-0.05) -- (1.5,-0.05);		
			
			\foreach \i in {0,0.75,1.5}{ \draw[fill=gray] (\i,0) ellipse (2pt and 7pt);}

			\foreach \i in {-1/2,1/2} \draw ({0.375+\i/10},-0.8) -- +(0,1.6);
			\foreach \i in {-1,0,1} \draw ({1.125+\i/10},-0.8) -- +(0,1.6);		
		\end{tikzpicture}
	}\hskip 12mm
	\subfigure[][]{\label{fig:branes-2a}
		\begin{tikzpicture}[scale=1.5]
			\draw[thick] (-0.6,0) -- (1.0,0);
			\draw[thick] (-0.6,0.07) -- (0.6,0.07);			
			
			\draw[thick] (-0.7,-0.07) -- (1.0,-0.07);						
			\draw[thick] (-0.7,-0.14) -- (0.2,-0.14);
			
			\draw[thick] (-0.8,0.14) -- (0.6,0.14);
			\draw[thick] (-0.8,0.21) -- (0.2,0.21);			
			
			\draw[thick] (-0.9,-0.21) -- (0.2,-0.21);
			\draw[thick] (-1.0,-0.28) -- (0.2,-0.28);			
			
			\foreach \i in {0.2,0.6,1.0}{ \draw[fill=gray] (\i,0) ellipse (2pt and 8.5pt);}
			
			\foreach \i in {0,...,4} \draw ({-1+\i/10},-0.8) -- +(0,1.6);
		\end{tikzpicture}
	}
	\caption{Left: Brane setups for 3d quiver gauge theories. D3-branes are shown as horizontal lines, NS5-branes as ellipses and D5-branes as vertical lines. Fig.~\ref{fig:branes-2}: $[2]-U(1)-U(2)-[3]$ theory. Moving D5-branes to one side and NS5-branes to the other, taking into account Hanany-Witten brane creation, leads to fig.~\ref{fig:branes-2a},  corresponding to $\rho=[4,2,2]$ and  $\sigma=[2,2,2,1,1]$.\label{fig:branes-3d}}
\end{figure}

The theories in (\ref{eq:3d4d-quiver}) can be engineered by configurations of D3-branes ending on and suspended between D5 and NS5 branes, following \cite{Hanany:1996ie,Gaiotto:2008sa,Gaiotto:2008ak}.
We start the discussion with 3d SCFTs, i.e.\ $N_0=N_{L+1}=0$, and then discuss the generalization to BCFTs and Janus CFTs.
The brane orientations are given by the first three lines in the following table:
\begin{equation}\label{eq:brane-tab}
	\begin{tabular}{cccc|ccccccc}
		\toprule
		& \ 0 \ & \ 1 \ & \ 2 \ & \ 3 \ & \ 4 \ & \ 5 \ & \ 6 \ & \ 7 \ & \ 8 \ & \ 9 \ \\
		\hline
		D3 & $\times$ & $\times$ & $\times$ & $\times$ &&&\\
		D5 & $\times$ & $\times$ & $\times$ & & $\times$ &  $\times$ & $\times$ & \\
		NS5 & $\times$ & $\times$ & $\times$ & &&&& $\times$ & $\times$  & $\times$ \\
		\hline
		F1 & $\times$ & & & & & & &&&$\times$ \\
		D5$^\prime$ & $\times$ & & & & $\times$ & $\times$ &$\times$ &$\times$ &$\times$ &\\
		D1 & $\times$ & & & & & & $\times$  &&& \\
		NS5$^\prime$ & $\times$ & & & &  $\times$ & $\times$ & & $\times$ & $\times$ & $\times$\\
		\bottomrule
	\end{tabular}
\end{equation}
Each 3d $U(N_t)$ gauge node is represented by $N_t$ D3-branes suspended between NS5 branes, while D5-branes intersecting the D3-branes represent fundamental matter, as illustrated in fig.~\ref{fig:branes}.

An alternative characterization of the quiver can be obtained by moving all D5-branes to one side and all NS5-branes to the other, as illustrated for an example in figs.~\ref{fig:branes-2} and \ref{fig:branes-2a}. The brane configuration is now characterized by two Young tableaux, $\rho$ and $\sigma$, which encode how the D3-branes end on the NS5-branes on one side and on the D5-branes on the other.
The theories are referred to as $T_\rho^\sigma[SU(N)]$, where $N$ denotes the total number of D3-branes suspended (we use this to refer to the IR SCFT and to the UV gauge theory).
The gauge theories were classified into good, bad and ugly in \cite{Gaiotto:2008ak}. We focus on the good theories, which have well-behaved IR fixed points without decoupled sectors. For these theories the number of flavors at each node is at least twice the number of colors. 
Expressions for the quiver data in terms of $\rho$ and $\sigma$ can be found in \cite{Nishioka:2011dq,Cremonesi:2014uva}. 
Mirror symmetry exchanges $\rho$ and $\sigma$, and corresponds to taking the S-dual brane configuration.

\begin{figure}
	\centering
	\subfigure[][]{\label{fig:brane-BCFT-0}
		\begin{tikzpicture}[scale=1.5]
			\foreach \i in {-0.05,0.05,0.15} \draw  (\i,-0.8) -- +(0,1.6);

			\foreach \i in {0.02,0.06} \draw (-0.05,\i) -- (2,\i);
			\foreach \i in {-0.02,-0.06} \draw (0.05,\i) -- (2,\i);
			
			\foreach \i in {-0.1,-0.14} \draw (0.15,\i) -- (2,\i);
		\end{tikzpicture}
	}\hskip 10mm
	\subfigure[][]{\label{fig:brane-BCFT-1}
	\begin{tikzpicture}[scale=1.5]
		\draw[white]  (0,-0.8) -- +(0,1.6);
		\foreach \i in {0.02,0.06} \draw (0,\i) -- (2,\i);
		\foreach \i in {-0.02,-0.06} \draw (0.3,\i) -- (2,\i);
		\foreach \i in {-0.1,-0.14} \draw (0.6,\i) -- (2,\i);
		\foreach \i in {0,0.3,0.6}{ \draw[fill=gray] (\i,0) ellipse (1.5pt and 6pt);}
		
	\end{tikzpicture}
	}\hskip 10mm
	\subfigure[][]{\label{fig:brane-BCFT-2}
	\begin{tikzpicture}[scale=1.5]
		\foreach \i in {-0.075,-0.025,0.025,0.075} \draw  (0.9+\i,-0.8) -- +(0,1.6);
		\foreach \i in {-0.075,-0.025,0.025,0.075} \draw (0,\i) -- (0.6,\i);
		\foreach \i in {-0.125,-0.075,-0.025,0.025,0.075,0.125} \draw (0.6,\i) -- (1.2,\i);
		\foreach \i in {-0.075,-0.025,0.025,0.075} \draw (1.2,\i) -- (1.8,\i);
		\foreach \i in {-0.025,0.025} \draw (1.8,\i) -- (3.2,\i);
		
		\foreach \i in {0,0.6,1.2,1.8}{ \draw[fill=gray] (\i,0) ellipse (2pt and 6.5pt);}
		
		\foreach \i in {-0.05,0.05} \draw  (2.2+\i,-0.8) -- +(0,1.6);
		
		\foreach \i in {0.075,0.125} \draw (2.2-0.05,\i) -- (3.2,\i);
		\foreach \i in {-0.075,-0.125} \draw (2.2+0.05,\i) -- (3.2,\i);
	\end{tikzpicture}
	}
\caption{Left: BCFT with semi-infinite D3-branes ending on D5-branes, corresponding to Nahm-pole boundary conditions. Center: S-dual configuration with semi-infinite D3-branes ending on NS5 branes, corresponding to 4d $\cN=4$ SYM coupled to a non-trivial 3d SCFT. Right: configuration with D3-branes ending on a combination of D5 and NS5 branes. \label{fig:brane-BCFT}}
\end{figure}
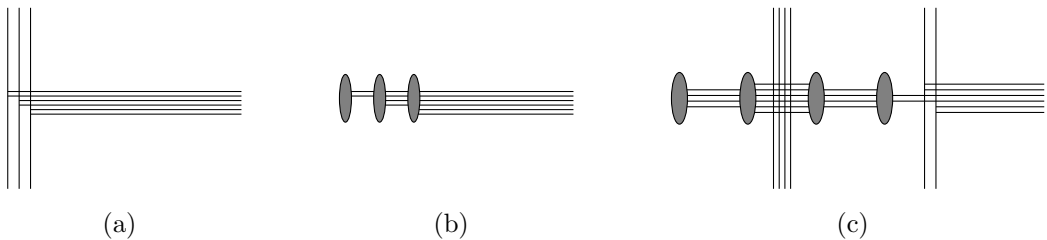

BCFTs based on 4d $\cN=4$ SYM can be realized by terminating semi-infinite D3-branes on D5 and/or NS5 branes. The simplest cases correspond to D3-branes ending on only D5 branes or on only NS5 branes (fig.~\ref{fig:brane-BCFT-0}, \ref{fig:brane-BCFT-1}). For D3-branes ending on D5-branes the BCFT can be described as a choice of boundary conditions for the $\cN=4$ SYM fields. The 4d $\cN=4$ vector multiplet can be decomposed into two sets of fields such that one yields a 3d $\cN=4$ vector multiplet on the boundary and the other a 3d hypermultiplet. 
The boundary conditions for the 3d vector multiplet are Dirichlet. The 3d hypermultiplet satisfies Nahm pole boundary conditions, in which the scalars $X^i$ behave as $X^i\sim t^i/x_3$, where $x_3$ is the direction normal to the boundary and $t^i$ are $SU(2)$ generators in a representation determined by how the D3 branes end on the D5-branes (see \cite[sec.~2.2]{Raamsdonk:2020tin} for a concise review).
The S-dual configurations are semi-infinite D3-branes ending on NS5-branes. This realizes $\cN=4$ SYM on a half space coupled to the IR fixed point of a 3d $\cN=4$ gauge theory. These two cases are illustrated in fig.~\ref{fig:brane-BCFT}. 
More general boundary conditions involve D5 and NS5 branes (fig.~\ref{fig:brane-BCFT-2}).
We will discuss examples in sec.~\ref{sec:BCFT-hol}.

The brane configurations can be further generalized by allowing semi-infinite D3-branes to emerge in both directions from the D5 and/or NS5 branes. This realizes Janus CFTs with two $\cN=4$ SYM nodes of possibly different ranks on half spaces separated by an interface. 
Depending on the brane configuration the interface can amount to imposing boundary conditions on part of the 4d $\cN=4$ SYM fields or to a 3d SCFT on the interface mediating between the two half spaces.
An example will be discussed briefly in sec.~\ref{sec:braneworld}.
A special case of these brane configurations is to simply have infinite D3-branes with no D5 or NS5-branes, corresponding  to the 4d $\cN=4$ SYM theory on $\RR^{1,3}$.

\paragraph{Wilson loops}

The brane realization of line operators in the 3d $T_\rho^\sigma[SU(N)]$ theories was discussed in \cite{Assel:2015oxa}, including half-BPS Wilson loops which preserve a $U(1)_C \times SU(2)_H$ subgroup of the $SU(2)_C \times SU(2)_H$ R-symmetry and their mirror-dual vortex loop operators (see also \cite{Dey:2021jbf,Dey:2021gbi}). The discussion generalizes to Wilson loops associated with 3d gauge nodes in 4d BCFTs or Janus CFTs. The corresponding branes and their orientations are included in the table in (\ref{eq:brane-tab}).

\begin{figure}
	\centering
	\subfigure[][]{\label{fig:HWa}
		\begin{tikzpicture}[scale=2.5]
			
			\node at (0.31,0) {\ldots};
			
			\foreach \i in {-0.08,0,0.08} \draw[thick] (0.5,\i) -- (1,\i);
			\foreach \i in {-0.18,-0.06,0.06,0.18} \draw[thick] (1,\i) -- (2,\i);
			\foreach \i in {-0.04,0.04} \draw[thick] (2,\i) -- (2.5,\i);
			
			\node at (2.7,0) {\ldots};

			\foreach \i in {1,2} \draw[line width=1.6pt] (\i,-0.6) -- +(0,1.2);
						
			\fill [red] (1.5,0.6) circle (1.5pt);
			
			\draw[thick, densely dotted] (1.48,0.558) -- (1.48, 0.18);
			\draw[thick, densely dotted] (1.52,0.558) -- (1.52, 0.055);
			
			\node at (1,-0.7) {\footnotesize NS5};       
			\node at (2,-0.7) {\footnotesize NS5};       			
			\node at (1.5,-0.4) {\footnotesize $(N_t)$ \ D3 };
			\node at (1.5,0.8) {\footnotesize D5$^\prime$};
			\node at (1.75,0.3) {\footnotesize $(k)$ F1};
		\end{tikzpicture}
	}\hskip 7mm
	\subfigure[][]{\label{fig:HWb}
		\begin{tikzpicture}[scale=2.5]
			
			\node at (0.31,0) {\ldots};
			
			\foreach \i in {-0.08,0,0.08} \draw[thick] (0.5,\i) -- (1,\i);
			\foreach \i in {-0.18,-0.06,0.06,0.18} \draw[thick] (1,\i) -- (2,\i);
			\foreach \i in {-0.04,0.04} \draw[thick] (2,\i) -- (2.5,\i);
			
			\node at (2.7,0) {\ldots};

			\foreach \i in {1,2} \draw[line width=1.6pt] (\i,-0.6) -- +(0,1.2);

			\fill [red] (1.5,0) circle (1.5pt);
			
			\node at (1,-0.7) {\footnotesize NS5};       
			\node at (2,-0.7) {\footnotesize NS5};       			
			\node at (1.5,-0.4) {\footnotesize $(N_t)$ \ D3 };
			\node at (1.75,0) {\footnotesize D5$^\prime$};
		\end{tikzpicture}
	}
	\caption{Brane realization of Wilson loops in the rank $k$ antisymmetric representation of the $t^{\rm th}$ gauge node. For simplicity we show a gauge node with no D5-branes representing fundamental matter. The Wilson loop is realized by $k$ fundamental strings stretched between the D3 branes and a D5$^\prime$ brane, as shown on the left. Via Hanany-Witten moves, the D5$^\prime$ can be brought inside the D3 stacks, annihilating all the F1s, as shown on the right.}
\end{figure}
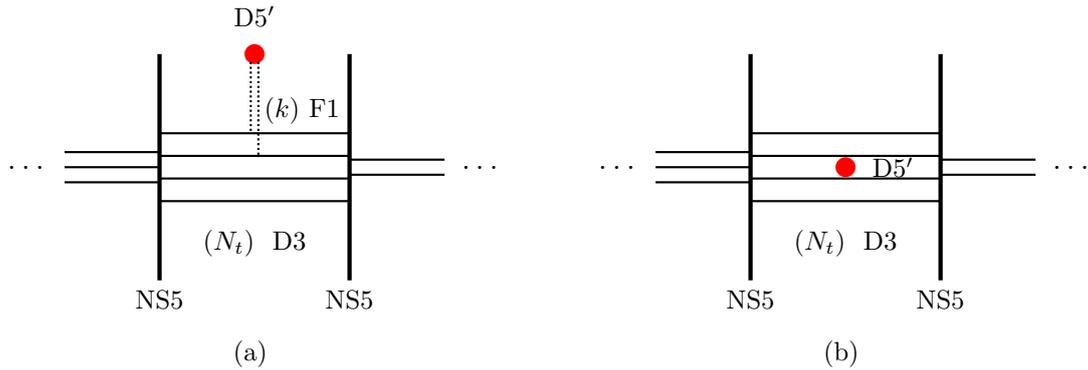

Wilson loops in the fundamental representation of the $t^{\rm th}$ gauge node are realized by a fundamental string ending on the stack of D3-branes associated with $U(N_t)$.
Wilson loops in the antisymmetric representation of rank $k$ of the $U(N_t)$ gauge group are realized by $k$ fundamental strings stretching between D5$^\prime$ branes and the D3-branes associated with $U(N_t)$, as illustrated in fig.~\ref{fig:HWa}. By the s-rule, the number of strings between the D5$^\prime$ and each D3 is at most one. 
Alternatively, one may start with a D5$^\prime$-brane within the D3-brane stack (fig.~\ref{fig:HWb}) with no strings attached. Using Hanany-Witten transitions to move it out of the D3-brane stack creates a fundamental string each time a D3-brane is crossed. If the D5$^\prime$ was initially separated from the asymptotic region by $k$ D3-branes this leads to configuration for a rank-$k$ antisymmetric Wilson loop.
In that case the rank $k$ can be interpreted as coordinate for the position of the D5$^\prime$-brane within the D3-brane stack.

The gauge node which the Wilson loop represented by a D5$^\prime$ is associated with is determined by the pair of NS5 branes which the D5$^\prime$ is in between. Labeling the NS5-branes for a quiver with $L$ nodes from $1$ to $L+1$, Wilson loops associated with the gauge node $U(N_t)$ are represented by a D5$^\prime$ between the NS5-branes labeled $t$ and $t+1$. Moving the D5$^\prime$ past NS5 branes creates new D3-branes through the Hanany-Witten effect. These D3-branes extend along the (0378) directions in the table (\ref{eq:brane-tab}).
An alternative point of view therefore is that the D5$^\prime$ associated with the $U(N_t)$ node has $t$ D3-branes oriented along (0378) ending on it. In that picture the number of D3-branes determines which gauge node the corresponding Wilson loop is associated with.
In the supergravity duals to be discussed below we will identify probe D5$^\prime$ branes with Wilson loops, and use F1 and D3 charges carried by the D5$^\prime$ to identify the gauge node and representation of the Wilson loop. This is  similar to the identification of Wilson loops in 5d long quiver SCFTs in \cite{Uhlemann:2020bek}.

Under mirror symmetry Wilson loops are exchanged with vortex loops. Vortex loops are represented in the brane construction by the S-duals of the branes representing Wilson loops. 
We will be brief in the discussion of vortex loops and focus on Wilson loops, but vortex loops will play a role in sec.~\ref{sec:braneworld}.
The orientations of D-strings, representing vortex loops dual to Wilson loops in the fundamental representation, and NS5$^\prime$-branes, representing vortex loops dual to Wilson loops in the antisymmetric representation, are included in the table in (\ref{eq:brane-tab}). For a more detailed discussion we refer to \cite{Assel:2015oxa}.

\section{Wilson loops in 3d long quivers}\label{sec:loc}

In this section we discuss Wilson loops in 3d $T_\rho^\sigma[SU(N)]$ SCFTs using supersymmetric localization.
For Wilson loops in antisymmetric representations associated with individual gauge nodes of the UV gauge theories we derive general expressions for the expectation values and discuss concrete example theories.

\subsection{Localization for long quivers}
The three dimensional SCFTs obtained from the brane setups of the previous section are the so called $T_\rho^\sigma[SU(N)]$ theories \cite{Gaiotto:2008ak}. 
This large class of 3d $\mathcal{N}=4$ SCFTs can be obtained as infrared superconformal fixed points of three-dimensional $\mathcal{N}=4$ gauge theories described by quiver diagrams of the generic form
\begin{equation}\label{quiv:gen-T_rho_sigma}
	\begin{split}	
		U&(N_1)-U(N_2)-\ldots -U(N_{L-1})-U(N_L) \\
		&\hskip 2mm |\hskip 15mm |\hskip 27mm |\hskip 18mm |
		\\
		&[k_1] \hskip 10mm [k_2] \hskip 20mm [k_{L-1}] \hskip 11mm [k_L]
	\end{split}
\end{equation}
This corresponds to \eqref{eq:3d4d-quiver} with $N_0=N_{L+1}=0$.
Using a variable $t$ to denote the position along the quiver, each node corresponds to a gauge group factor $U(N_t)$ with $k_t$ hypermultiplets in the fundamental representation attached. 
Each node is associated with an $\mathcal{N}=4$ vector multiplet and each line connecting two nodes with an $\mathcal{N}=4$ hypermultiplet in the bifundamental representation. Our focus will be on the so-called good theories, in which the number of flavors at each node is at least twice the number of colors, namely $N_{t-1}+N_{t+1}+k_t \ge 2N_t$. 

It is particularly interesting to consider $T_\rho^\sigma[SU(N)]$ theories placed on a three-sphere $S^3$, as we will do from now on. Being on a compact space, the path integral does not suffer from IR divergences and is well defined. Moreover, using supersymmetric localization, the partition function can
be reduced to a matrix model of the form \cite{Kapustin:2009kz}
\begin{equation}\label{eq:matrix_S3}
	\cZ_{S^3}  =
	\frac{1}{\left| W \right|} \int \prod_{\text{Cartan}} d\lambda\,
	\frac{\prod_{\alpha>0} \left(2 \sinh(\pi\alpha(\lambda))\right)^2}{\prod_{\substack{\text{hyper} \\ \text{in rep $R$}}}\prod_{\rho}  2 \cosh(\pi \rho(\lambda))}\equiv  \frac{1}{\left| W \right|} \int d \lambda \ e^{-\cF} \ .
\end{equation}
Here $\abs{W}$ is the order of the Weyl group of the gauge group, $\alpha>0$ denotes the positive roots of the gauge group and the denominator in the integrand takes into account the presence of hypermultiplets in a representation $R$ with weights $\rho$.

As discussed in detail in \cite{Coccia:2020wtk}, the natural planar limit of the $T_\rho^\sigma[SU(N)]$ theories amounts to taking a large length of the quiver $L\gg 1$, the rank  of the generic gauge group as $N_t=\mathcal O(L^2)$  and a large number of 
flavours. In this limit the path integral can be evaluated via a saddle point approximation, using the formalism developed in \cite{Coccia:2020cku,Coccia:2020wtk}, mimicking the 5d discussion of \cite{Uhlemann:2019ypp}. We now review the main ideas. Having large ranks, one can assume a continuous distribution for the eigenvalues of each node. This can be realized by introducing, for each $t$, a normalized eigenvalue density $\rho_t$ and substituting
\begin{equation}
	\sum_{i=1}^{N_t} \qquad \to \qquad N_t\int d\lambda \ \rho_t(\lambda)	 \ ,
\end{equation}
in the matrix model \eqref{eq:matrix_S3}. Then, since $L \gg 1$, it is convenient to replace the variable $t$ with a continuous coordinate 
\begin{equation}\label{eq:def_z}
	z=\frac{t}{L} \ , \qquad z \in [0,1] 
\end{equation}
labelling the position along the quiver. In terms of $z$, the data of the quiver $\{ N_t, k_t \}$ are replaced by the continuous functions $N(z)=N_{zL}$ and $k(z)=k_{zL}$. With these notations, the scaling we consider can be written as
\begin{align}\label{eq:lim}
	N(z)&=\mathcal O(L^2)~, & \lim_{z\rightarrow \lbrace 0,1\rbrace }L^{-2} N(z)&=0 \ ,
\end{align}
with $N(z)$ a piecewise linear function. Fundamental hypermultiplets are attached to isolated nodes $z_t$ and their total number is $\mathcal O(L)$.
With the replacement \eqref{eq:def_z}, the possible eigenvalue configurations can be conveniently written in terms of a density function $\rho(z,\lambda)$ of two continuous variables, with
\begin{equation}\label{eq:scaling_lambda}
	\rho_{zL}(\lambda) \equiv \rho(z,\lambda) \ .
\end{equation}
In order to have a non-trivial saddle point configuration, the eigenvalues have to scale with the length of the quiver as\footnote{Differently from \cite{Coccia:2020cku,Coccia:2020wtk}, we keep the chiral field R-charge $r$ fixed to its canonical value $r=1/2$.}
\begin{equation}\label{eq:scaling}
	\begin{split}
		& \lambda=xL \ , \\
		& \hat{\rho}(z,x)=L\rho(z,Lx)  \ ,
	\end{split}
\end{equation} 
with $x$ of order one and where we also introduced a rescaled density $\hat{\rho}(z,x)$. With the assumption \eqref{eq:scaling}, the exact form of the saddle point density can then be obtained by solving a two dimensional electrostatics problem, uniquely determined by the form of the quiver. Using this approach, the planar limit free energies on $S^3$ were matched with supergravity computations in \cite{Coccia:2020cku}. The scaling of the free energies was found to be $L^4$, which is quadratic in terms of the ranks of individual 3d gauge groups and the familiar $N^2$ from a 4d $\cN=4$ SYM perspective (viewing the 3d $T_\rho^\sigma[SU(N)]$ theories as IR limit of 4d $\cN=4$ SYM on an interval, fig.~\ref{fig:branes-2a}). Theories with the previously known $N^2\ln N$ scaling \citep{Assel:2012cp} can be obtained as a particular limit, which modifies the conditions in \eqref{eq:lim} such that the ranks of the 3d gauge groups are of the same order as the length of the quiver.

\subsection{Localization and Wilson loops} The formalism just discussed can also be used to compute the expectation values of other supersymmetric observables, such as supersymmetric Wilson loops. This has been carried out for 5d long quiver theories in \cite{Uhlemann:2020bek}, whose discussion we will closely follow. 

We want to compute the expectation value of Wilson loop operators associated with the $t^{\text{th}}$ gauge node, which  is given by
\begin{equation}\label{eq:Wilson_loop}
	W_R^{(t)}=\frac{1}{\text{dim} R}\text{Tr}_R \left( \mathcal{P} \exp \oint \left( i A_\mu^{(t)} \dot{x}^\mu-\abs{\dot{x}}\sigma^{(t)} \right) d \tau \right) \ ,
\end{equation} 
where $x(\tau)$ denotes the closed world-line of the Wilson loop and $\sigma^{(t)}$ and $A_\mu^{(t)}$ are, respectively, the scalar field and the gauge field of the $\mathcal{N}=2$ vector multiplet in the $\mathcal{N}=4$ vector multiplet associated with the $t^{\text{th}}$ node.\footnote{We recall that an $\mathcal{N}=4$ vector multiplet consists of an $\mathcal{N}=2$ vector and an $\mathcal{N}=2$ adjoint chiral multiplet. Similarly, an $\mathcal{N}=4$ hypermultiplet decomposes into a pair of chiral and anti-chiral multiplets.} In \eqref{eq:Wilson_loop}, we also used $\mathcal{P}$ to indicate the path-ordering operator and $R$ to denote a representation of the gauge group $U(N_t)$. For a 3d $\mathcal{N}=4$ SCFT, one can place the Wilson loop \eqref{eq:Wilson_loop} on a great circle on $S^3$, preserving half of the supersymmetries and $U(1)_C\times SU(2)_H$ of the original R-symmetry group. More details can be found in \cite{Assel:2015oxa, Drukker:2019bev}.

From supersymmetric localization techniques \cite{Kapustin:2009kz}, the expectation value of BPS Wilson loop operators at the $t^{\text{th}}$ node can be obtained  by inserting, in the integral of the matrix model \eqref{eq:matrix_S3}, an appropriate factor 
\begin{equation}
	\langle W_{R}^{(t)} \rangle= \frac{1}{\cZ_{S^3}}\frac{1}{\left| W \right|} 
	\int d \lambda \  e^{-\cF}
	\left( \frac{1}{\text{dim} \ R}\sum_{w \in R} e^{2 \pi w \cdot \lambda^{(t)}} \right) \ ,
\end{equation}
where $w$ runs over the weights of the representation $R$ of the gauge group. With the assumption \eqref{eq:scaling_lambda} on the scaling of the eigenvalues, one can argue that the insertion of the Wilson loop operator in the matrix model does not affect the saddle point configurations of the $S^3$ free energy. The Wilson loop expectation value then becomes
\begin{equation}\label{eq:W_gen}
	\langle W_{R}^{(t)} \rangle=\frac{1}{\text{dim} \ R}\sum_{w \in R} e^{2 \pi w \cdot \lambda^{(t)}} \bigg \lvert_{\text{saddle}} .
\end{equation}
The saddle point configurations for a number of quivers can be directly read from \cite{Coccia:2020wtk}.

A first application of this formula can be obtained for the fundamental representation of the gauge group $U(N_t)$.
In the continuous limit, with the rescaling \eqref{eq:scaling},
\begin{equation}\label{eq:W_fund}
	\langle W_{f}^{(t)} \rangle=\frac{1}{N_t}\sum_{i=1}^{N_t} e^{2 \pi \lambda_i^{(t)}} \bigg \lvert_{\text{saddle}} \qquad \xrightarrow[\text{limit}]{\text{continuous
	}} \qquad	\langle W_f (z) \rangle=\int dx \ \hat{\rho}(z,x)e^{2 \pi x L} \ .
\end{equation}
In this case, the expectation value of the Wilson loops has a scaling determined by the largest eigenvalue. 
For the theories to be discussed below, at the majority of nodes the eigenvalue densities have exponential tails and are not bounded. This signals a logarithmically enhanced scaling of the fundamental Wilson loop expectation values. This was discussed in sec.~IIA of \cite{Uhlemann:2020bek}, to which we refer for more details.

In the rest of this section we focus on Wilson loops in antisymmetric representations of rank $k$, denoted by $\wedge^k$, with $k$ large. For such operators, eq.~\eqref{eq:W_gen} becomes
\begin{equation}\label{eq:Wilson_anti}
	\langle W_{\wedge^k}^{(t)} \rangle=
	\begin{pmatrix}
		N_t \\
		k
	\end{pmatrix}^{-1}
	\sum_{j_1 < j_2 < \dots < j_k} e^{2 \pi \sum_{\ell=1}^k\lambda^{(t)}_{j_\ell}}\bigg \lvert_{\text{saddle}} \ .
\end{equation}
Note that this expression is invariant under the change $k \to N_t - k$. For large representations, with $k= O(N_t)$, the leading order contribution to the expectation value comes from the $k$ distinct largest eigenvalues, which we need to identify in the continuous formalism. Let us recall that we introduced a density $\rho_t$ to describe the distribution of the eigenvalues for each node $t$. So, given an integer $\ell \le N_t$, we can define a quantity $b_{t,\ell}$ such that \cite{Uhlemann:2020bek}
\begin{equation}\label{eq:b_discr}
	N_t\int_{b_{t,\ell}}^\infty d\lambda \ \rho_t(\lambda)=\ell \ .
\end{equation} 
This expression provides the cutoff $b_{t,\ell}$ to isolate the $\ell$ largest eigenvalues. In other words, we can think of $b_{t,1}$ as giving the value of the largest eigenvalue, $b_{t,2}$ the value of the second-largest and so on. This means that
\begin{equation}\label{eq:lnW_discr}
	\ln \langle W_{\wedge^k}^{(t)} \rangle=2 \pi \sum_{\ell=1}^k b_{t,\ell} \ .
\end{equation} 
It is interesting to observe that all the information of the original theory is now encoded in the saddle point configuration and in the scaling of the eigenvalues, which determine $b_{t,\ell}$ via \eqref{eq:b_discr}. As noted in Section VI of \cite{Coccia:2020wtk}, some 3d theories share the same saddle point configurations with the 5d theories studied in \cite{Uhlemann:2019ypp}. Hence, the Wilson loop expectation value \eqref{eq:lnW_discr} for such theories can only differ from the 5d analogs by an overall factor, depending on the different scaling of $\lambda$.

For a large representation, we can replace the sum in \eqref{eq:lnW_discr} with an integral. In order to do that, let us make some redefinitions. First, introducing $y=\ell/N_t$ and performing the changes of variable $\lambda=L x$ and $z=t/L$, equation \eqref{eq:b_discr} becomes
\begin{equation}\label{eq:b_cont}
	\int_{b(z,y)}^\infty dx \ \hat{\rho}(z,x)=y \ ,
\end{equation} 
where we also replaced $b_{t, \ell}$ with $b(z,y)$, which is a function of two continuous variables. Then we introduce another continuous parameter to encode the rank of the representation,
\begin{equation}
	\mathds{k}\equiv \frac{k}{N_t} \ ,
\end{equation}
so that the expectation value \eqref{eq:lnW_discr} becomes
\begin{equation}\label{eq:anti_sym_1}
	\ln \langle W_\wedge (z,\mathds{k})\rangle=2 \pi L N(z) \int_0^{\mathds{k}}dy \ b(z,y)~.
\end{equation}
Following the notation of \cite{Uhlemann:2020bek}, we redefined $W_\wedge(z,\mathds{k}) \equiv W_{\wedge^{\mathds{k}N(z)}}^{(z,L)} $ on the left hand side. Equation \eqref{eq:anti_sym_1} can be rewritten as
\begin{equation}\label{eq:anti_sym_2}
	\ln \langle W_\wedge (z,\mathds{k})\rangle=2 \pi L N(z) \int_{b(z,0)}^{b(z,\mathds{k})} y'(b) db \ b ~,
\end{equation}
which turns out to be more convenient from a computational perspective. This expression is, up to an overall factor, equivalent to equation (2.18) in \cite{Uhlemann:2020bek} for 5d Wilson loops.

\subsection{General balanced quivers}\label{sec:hol-balanced}
As a concrete realization of the previous discussion, we consider theories described by balanced quivers, i.e.\ satisfying for each node $t$ the condition $N_{t-1}+N_{t+1}+k_t = 2N_t$. The three-sphere free energy for such theories, with the scaling in \eqref{eq:lim}, has been studied in \cite{Coccia:2020wtk} and the saddle point densities are
\begin{align}\label{eq:varrho-s}
	\hat{\rho}_s(z,x) =-\frac{L}{2\pi N(z)}\sum_{t=2}^{L-1} k_t
	\ln \left(\frac{\cosh (2 \pi  x)-\cos\left(\pi(z-z_t)\right)}{\cosh (2 \pi  x)-\cos\left(\pi(z+z_t)\right)}\right) ~,
\end{align}
where $z_{t}$ denotes the positions of the flavours. The function $b(z,y)$ defined in \eqref{eq:b_cont} is determined by the equation
\begin{equation}\label{eq:y_bal_th}
	y=\frac{L}{2\pi^2 N(z)}\sum_{t=2}^{L-1} k_t \Re \left[\Li_2 \left(w \tau_t\right)-\Li_2\left( w/\tau_t\right) \right]
\end{equation}
with
\begin{equation}\label{eq:w-tau-def}
	w=e^{-2 \pi b(z,y)-i \pi z} \ , \qquad \tau_t=e^{i \pi z_t} \ .
\end{equation}
The expression for the expectation value of the Wilson loop from \eqref{eq:anti_sym_2} is
\begin{equation}\label{eq:W_bal_th}
	\begin{split}
		&\ln \langle W_\wedge\rangle=\frac{L^2}{2\pi^2}\sum_{t=2}^{L-1} k_t \Re \left[\Li_3\left(w \tau_t\right)-\Li_3\left(w/\tau_t \right) -\ln\abs{w}\left( \Li_2\left(w \tau_t \right)-\Li_2\left(w/\tau_t\right)\right)\right]\Bigg \lvert^{b(z,\mathds{k})}_{b(z,0)} \ .
	\end{split}
\end{equation}
This can be read as follows: One first fixes $b\in\RR$. Then $\mathds{k}$ is given by the expression in (\ref{eq:y_bal_th}) by setting $y=\mathds{k}$ and the Wilson loop expectation value is given by (\ref{eq:W_bal_th}). More precisely, from \eqref{eq:b_cont} we have $b(z,0)=\infty$ (which can also be seen from (\ref{eq:y_bal_th})). As a result, the contribution from the lower bound in (\ref{eq:W_bal_th}) drops out.
We arrive at
\begin{equation}\label{eq:W_field_bal}
	\begin{split}
		\mathds{k} & = \frac{L}{2\pi^2N(z)}\sum_{t=2}^{L-1} k_t \Re \left[\Li_2 \left(w \tau_t\right)-\Li_2\left( w/\tau_t\right) \right]\,, \\
		\ln \langle W_\wedge\rangle&=\frac{L^2}{2\pi^2}\sum_{t=2}^{L-1} k_t \Re \left[\Li_3\left(w \tau_t\right)-\Li_3\left(w/\tau_t \right) -\ln\abs{w}\left( \Li_2\left(w \tau_t \right)-\Li_2\left(w/\tau_t\right)\right)\right] \ .
	\end{split}
\end{equation}

\paragraph{Example} Theories described by balanced quivers can be realized by partitions $\rho$, $\sigma$ of the schematic form
\begin{equation}
	\sigma=[M_1^{N_1},\dots ,M_n^{N_n}] \ , \qquad \rho=[\hat{M}^{\hat{N}}] \ , \qquad 
\end{equation}
For the planar limit all the quantities involved are large and of the same order, as discussed in more detail in \cite{Coccia:2020wtk}.
Here we explicitly discuss the simple choice
\begin{equation}\label{eq:part_delta_0}
	\sigma=\rho=\left[N^{2N} \right] \ .
\end{equation}
From the brane perspective this corresponds to D3-branes ending on one side on a single group of D5-branes, with an equal number of D3-branes ending on each D5-brane, and on the other side on a single group of NS5-branes, with an equal number of D3-branes ending on each NS5-brane. The corresponding quiver has $2N$ flavors at the central node of the quiver and the rank of the gauge groups is encoded in the function
\begin{equation}\label{eq:delta0-rank}
	N(z)= 2 N^2 \left(\frac{1}{2}-\abs{z-\frac{1}{2}}\right) \ .
\end{equation}
The saddle point density is
\begin{equation}
	\hat{\rho}_s(z,x) = 
	-\frac{2 N^2}{\pi N(z)} 
	\ln \left(\frac{\cosh (2 \pi  x)-\sin\left(\pi z \right)}{\cosh (2 \pi  x)+\sin\left(\pi z \right)}\right) \ .
\end{equation} 
We can focus on the first half $z \le \frac{1}{2}$ of the quiver since the results in the second half can be obtained by symmetry, so that $N(z)=2 N^2z$.
Plugging this expression into \eqref{eq:W_field_bal} one finds
\begin{align}\label{eq:Wilson-delta0-summary}
	\mathds{k} &= \frac{1}{\pi^2 z}\Re\left[\Li_2\left( i w\right)-\Li_2 \left(-i w\right) \right], \hskip 15mm w=e^{-2 \pi b-i \pi z}~,
	\nonumber\\
	\ln \langle W_\wedge\rangle&=
	\frac{4 N^3}{\pi^2} \Re\left[\Li_3\left(i w \right)-\Li_3\left(-iw\right)-\ln|w|\left( \Li_2\left(i w \right)-\Li_2\left(-i w\right)\right) \right] \ .
\end{align}

\subsubsection{\texorpdfstring{$T[SU(N)]$}{T[SU(N)]}}

A limiting case of the theories discussed above amounts to considering the ranks of the gauge groups of the same order as the length of the quivers and accumulating all flavors at one end of the quiver. This leads to the theories with $N^2\ln N$ scaling of the free energy discussed in \cite{Assel:2012cp}.
A more complete discussion of the limit can be found in \cite{Coccia:2020wtk}. The simplest example of such theories is $T[SU(N)]$, described by the partitions
\begin{equation}
	\rho=\sigma=[1^N] \ .
\end{equation}
We could obtain the expectation value of Wilson loops as a limiting case of \eqref{eq:W_field_bal}. However, we can take a short-cut. Indeed, $T[SU(N)]$ has the same saddle point configuration as the $5d$ $T_N$ theory considered in \cite{Uhlemann:2020bek}, namely
\begin{equation}
	\hat{\rho}_s(z,x)=\frac{\sin(\pi z)}{z}\frac{1}{\cosh(2\pi x)+\cos(\pi z)} \ .
\end{equation}
Hence, as argued before, we can import the 5d results of \cite{Uhlemann:2020bek}, adjusting the overall factors.  The function $b(z,y)$ defined in \eqref{eq:b_cont} is given by 
\begin{equation}
	b(z,y)=\frac{1}{2\pi}\ln \left[\sin(\pi(1-y)z)\csc(\pi y z) \right]
\end{equation}
so that, using \eqref{eq:anti_sym_1} or \eqref{eq:anti_sym_2}, we obtain
\begin{equation}
	\ln \langle W_\wedge (z,\mathds{k})\rangle=\frac{1}{\pi}N^2 D\left(e^{2 \pi b(\mathds{k},z)+i \pi(1-z)} \right) \ ,
\end{equation}
where $D$ is the Bloch-Wigner function, $D(u)=\Im(\Li_2(u)+\ln|u|\ln(1-u))$.

\subsection{An unbalanced quiver}
We conclude the field theory analysis with a discussion of the unbalanced quiver described by the partitions
\begin{equation}\label{eq:part_delta}
	\rho=\sigma=[(2N-\Delta N)^{N},(\Delta N)^{N}] \ .
\end{equation}
This theory is a generalization of the balanced theory \eqref{eq:part_delta_0}, which is recovered by choosing $\Delta=1$. The quiver  associated with the choice \eqref{eq:part_delta} has length $L=2N-1 \sim 2N$ with an unbalanced node at $t=N$ and fundamental flavors
\begin{equation} \label{eq:zandk}
	\begin{split}
		&\mathsf{k}_1=N \qquad \text{at} \quad \mathsf{z}_1=\frac{\Delta}{2} \ , \\
		&\mathsf{k}_2=N \qquad \text{at} \quad \mathsf{z}_2=1-\mathsf{z}_1 \ .
	\end{split}
\end{equation}  
The explicit expression of the quiver is given in (\ref{eq:D52NS52-quiver}) below.
The saddle point eigenvalue density can be obtained from equation (96) in \cite{Coccia:2020wtk}
\begin{equation} \label{eq:varrho_unb_unb}
	\hat{\rho}_s=-\frac{N^2}{\pi N(z)} \sum_{ a\in \{1, 2 \}} \ln \abs{\frac{i \sqrt{-v}-i \sqrt{-v_a}}{i \sqrt{-v}+i\sqrt{-\bar v_a})}}^2 \ ,
\end{equation} 
with
\begin{equation}
	v(u)=\frac{ue^{4 \pi x_1}+1}{u+e^{4 \pi x_1}} \ , \qquad \qquad u(x)=e^{4 \pi x+2\pi i z} \ , \qquad v_a=\frac{e^{2\pi i \mathsf{z}_a}  e^{4 \pi x_1}+1}{e^{2\pi i \mathsf{z}_a} +e^{4 \pi x_1}} \ , 
\end{equation}
and $x_1$ determined by
\begin{equation}\label{eq:fix_x1}
	\frac{i}{2} \sum_{a\in\{ 1,2 \} } \ln\left(\frac{\sqrt{-v_a}+e^{2 \pi  x_1}}{1+e^{2 \pi  x_1}\sqrt{-v_a} } \right)=\pi \mathsf{z}_1 \ .
\end{equation}
Before writing the solution of the previous equation, it is convenient to introduce a variable $\delta$ such that 
\begin{equation}\label{eq:sub_delta}
	\mathsf{z}_1=\frac{\Delta}{2}= \frac{1}{4}+\frac{1}{\pi}\arctan e^{-2 \delta} \ .
\end{equation}
This parameter will appear naturally in the supergravity description of Section \ref{sec:D52NS52}. With this substitution, the solution of equation \eqref{eq:fix_x1} is
\begin{equation}
	x_1=-\frac{1}{2\pi}\ln \left(\tanh\left(\frac{\delta}{2}\right) \right)  \ .
\end{equation}
This solution holds in the interval $\mathsf{z}_1 \in [1/4,3/4]$, which actually represents the interval of the possible values of $\mathsf{z}_1$, consistently with equation \eqref{eq:sub_delta}. In the limit $\delta \to 0$, in which the quiver becomes balanced,  the value of $x_1$ goes to infinity, as expected for a balanced quiver \cite{Coccia:2020wtk}. For generic $\delta$, instead, the density \eqref{eq:varrho_unb_unb} has compact support at the unbalanced node $z=1/2$, with $x \in [-x_1,x_1]$. 
With an eigenvalue density of compact support, the leading behavior of Wilson loops at $z=1/2$ in the fundamental representation is given, from equation \eqref{eq:W_fund}, by the largest eigenvalue
\begin{equation}\label{eq:WF-unbalanced}
	\ln \left \langle W_{f}\left(\frac{1}{2} \right) \right \rangle= -2N \ln \tanh\left(\frac{\delta}{2}\right) \ , \qquad \delta \neq 0 \ .
\end{equation}
For the other nodes with unbounded support, we expect the logarithmic enhancement discussed below eq.~(\ref{eq:W_fund}).

The expectation values of large-rank antisymmetric Wilson loops can be obtained from the general expressions (\ref{eq:b_cont}), (\ref{eq:anti_sym_2}) with the eigenvalue density (\ref{eq:varrho_unb_unb}). This will be used to numerically compare the expectation values to supergravity results in sec.~\ref{sec:D52NS52}.

\section{Wilson loops in AdS/(B)CFT}\label{sec:hol}

We review in sec.~\ref{sec:sugra-sol} the $AdS_4\times S^2\times S^2\times\Sigma$ supergravity solutions associated with the brane setups of sec.~\ref{sec:brane}, and discuss BPS string and D5$^\prime$-brane embeddings in sec.~\ref{sec:BPS-F1D5}, before moving on to concrete examples and the discussion of Wilson loops. 
AdS$_5\times$S$^5$ is discussed as a warm-up in sec.~\ref{sec:AdS5S5}, where we also identify {\it Janus on the brane} D5-brane embeddings.
For duals of balanced 3d SCFTs we compare the Wilson loop expectation values obtained holographically with field theory calculations in sec.~\ref{sec:balanced-hol}.
BCFTs corresponding to D3-branes terminating on a symmetric combination of D5 and NS5 branes are discussed in sec.~\ref{sec:BCFT-hol}.
In sec.~\ref{sec:D52NS52} we discuss 3d SCFTs with an unbalanced central node, which provide 10d realizations of wedge holography.
The solutions of sec.~\ref{sec:BCFT-hol} and sec.~\ref{sec:D52NS52} were used in \cite{Uhlemann:2021nhu} to study information transfer from black holes and the entropy of Hawking radiation; we will make connections to that discussion in sec.~\ref{sec:D52NS52} and sec.~\ref{sec:braneworld}.

\subsection{Supergravity solutions}\label{sec:sugra-sol}

The geometry of the general type IIB supergravity solutions constructed in  \cite{DHoker:2007zhm,DHoker:2007hhe} is a warped product of $AdS_4\times S^2\times S^2$ over a Riemann surface $\Sigma$.
They preserve 16 supersymmetries.
The Einstein frame metric and dilaton are given by
\begin{align}\label{eq:ds2-IIB}
	ds^2&=f_4^2 ds^2_{AdS_4}+f_1^2 ds^2_{S_1^2}+f_2^2 ds^2_{S_2^2}+4\rho^2 |dz|^2~, & e^{4\phi}&=\frac{N_2}{N_1}~,
\end{align}
where we follow the dilaton convention of \cite{DHoker:2007zhm,DHoker:2007hhe} in  using $\tau=\chi+i e^{-2\phi}$.
The 3-form and 5-form field strengths are
\begin{align}\label{eq:fieldstrengths-IIB}
	H_{(3)}&=\vol_{S_1^2}\wedge\, db_1~, \qquad\qquad
	F_{(3)}=\vol_{S_2^2}\wedge\, db_2~,
	\nonumber\\
	F_{(5)}&=-4  \vol_{AdS_4}\wedge\, dj_1+4 f_1^2f_2^2f_4^{-4}\vol_{S_1^2}\wedge \vol_{S_2^2}\wedge \star_2 dj_1~,
\end{align}
where $\star_2$ denotes Poincar\'e duality on $\Sigma$ and $b_1$, $b_2$, $j_1$ are functions on $\Sigma$.

\smallskip

The solutions are parametrized by a pair of harmonic functions $h_1$, $h_2$ on $\Sigma$, which may be written in terms of holomorphic functions $\cA_{1}$, $\cA_2$ as
\begin{align}\label{eq:h12-cA12}
	h_1&=-i (\cA_1-\bar \cA_1)~, & h_2&=\cA_2+\bar \cA_2~,
	\nonumber\\
	h_1^D&=\cA_1+\bar \cA_1~, & h_2^D&=i(\cA_2-\bar \cA_2)~.
\end{align}
Composite quantities are defined as
\begin{align}
	W&=\partial\bar\partial (h_1 h_2)~, & N_i &=2h_1 h_2 |\partial h_i|^2 -h_i^2 W~.
\end{align}
The metric functions are
\begin{align}
	f_4^8&=16\frac{N_1N_2}{W^2}~, & f_1^8&=16h_1^8\frac{N_2 W^2}{N_1^3}~, & f_2^8&=16 h_2^8 \frac{N_1 W^2}{N_2^3}~,
	&
	\rho^8&=\frac{N_1N_2W^2}{h_1^4h_2^4}~.
\end{align}
The remaining quantities appearing in the field strengths (\ref{eq:fieldstrengths-IIB}) are functions $b_1$, $b_2$ and $j_1$, given by
\begin{align}\label{eq:b1b2}
	b_1&=\frac{2h_1^2h_2}{N_1}\cY+2h_2^D~,
	&
	b_2&=\frac{2h_1 h_2^2}{N_2}\cY-2h_1^D~,
	&
	j_1&=3\cC+3\bar\cC-3\cD+\frac{h_1h_2}{W}\cY~,
\end{align}
where
\begin{align}
	\cY&=i(\partial h_1\bar\partial h_2-\bar\partial h_1\partial h_2)~,
	&
	\cD&=\bar\cA_1\cA_2+\cA_1\bar\cA_2~,
\end{align}
and $\cC$ is defined by
\begin{align}\label{eq:cC-def}
	\partial\cC&=\cA_1\partial\cA_2-\cA_2\partial\cA_1~.
\end{align}

\paragraph{Harmonic functions:}
Depending on the choice of $h_{1/2}$ and $\Sigma$, solutions with different holographic interpretations can be constructed. We focus on duals for 3d SCFTs, 4d BCFTs and 4d Janus interface CFTs of  the type discussed in sec.~\ref{sec:brane}. All solutions describe  D3-branes suspended between, ending on, or intersecting combinations of D5 and NS5 branes. 
For these solutions the form of the harmonic functions $h_1$, $h_2$ on the strip
\begin{align}
	\Sigma&=\left\lbrace z\in \mathds{C}\, \vert \, 0\leq\Im(z)\leq \frac{\pi}{2}\right\rbrace
\end{align}
is
\begin{align}\label{eq:h1h2-gen}
	h_1&=-\frac{i \pi \alpha^\prime}{4} (K e^z-Le^{-z})-\frac{\alpha^\prime}{4} \sum_{a=1}^AN_{\rm D5}^{(a)}\ln\tanh\left(\frac{i\pi}{4}-\frac{z-\delta_a}{2}\right)+\rm{c.c.}
	\nonumber\\
	h_2&=\frac{\pi \alpha^\prime}{4} (K e^z+Le^{-z})-\frac{\alpha^\prime}{4}\sum_{b=1}^B N_{\rm NS5}^{(b)}\ln\tanh\left(\frac{z-\delta_b}{2}\right)+\rm{c.c.}
\end{align}
These solutions describe $A$ groups of D5-branes with $N_{\rm D5}^{(a)}$ D5-branes in the $a^{\rm th}$ group and $B$ groups of NS5-branes with $N_{\rm NS5}^{(b)}$ NS5-branes in the $b^{\rm th}$ group.
D3-branes are suspended between the 5-branes, as illustrated in figs.~\ref{fig:branes-3d}, \ref{fig:brane-BCFT} for 3d SCFTs and BCFTs, respectively.
The numbers of semi-infinite D3-branes emerging on the left and right are controlled by $L$ and $K$, respectively.
For $L=K=0$ the solutions are dual to 3d SCFTs. If one of $K$ and $L$ is zero and the other non-zero the solutions are dual to 4d BCFTs, and if $K$ and $L$ are both non-zero the solutions are dual to 4d Janus interface CFTs.
For the general identification of the brane configurations and their linking numbers we refer to \cite{Aharony:2011yc,Assel:2011xz} and recent discussions in \cite{Coccia:2020wtk,Raamsdonk:2020tin}. We will discuss concrete examples and the associated brane configurations below.
We note that S-duality amounts to exchanging $h_1$ and $h_2$.

\subsection{BPS strings and 5-branes}\label{sec:BPS-F1D5}
The BPS conditions for fundamental strings and D5$^\prime$ branes relevant for the discussion of Wilson loops are derived and partly solved in appendix \ref{app:BPS}; here we summarize the results.

\subsubsection{Strings}
Wilson loops in the fundamental representation are represented by fundamental strings wrapping AdS$_2$ in AdS$_4$ and otherwise localized at a point in the internal space formed out of $\Sigma$ and the two spheres (such strings were studied in \cite{Estes:2012nx}). The induced metric is $g=f_4^2 ds^2_{AdS_2}$. The action (with dilaton convention $\tau=\chi+ie^{-2\phi}$) becomes
\begin{align}\label{eq:SF1-0}
	S_{\rm F1}&=-T\, V_{AdS_2}f_4^2e^\phi~,
\end{align}
where $T=(2\pi\alpha')^{-1}$ and $V_{AdS_2}$ is the renormalized volume of AdS$_2$.
The BPS condition for the string to preserve half of the 16 supersymmetries (cf.\ (\ref{eq:position_F1})) reads
\begin{align}\label{eq:F1-BPS}
	h_1&=0~, & \partial h_2&=0~.
\end{align}
$h_1=0$ implies that $f_1^2=0$, i.e.\ the sphere $S_1^2$ collapses and the associated isometries are preserved.
Since the string is localized on $S_2^2$, which does not collapse, the $SU(2)$ symmetry associated with $S_2^2$ is broken to  $U(1)$.

For the supergravity solutions (\ref{eq:h1h2-gen}), $h_1$  vanishes only on the boundary component $\Im(z)=0$, where $\partial h_2$ has poles.
On that boundary component $h_2$ satisfies the Neumann boundary condition $(\partial-\bar\partial) h_2=0$.
Supersymmetric string embeddings can be found at points where in addition $(\partial+\bar\partial)h_2=0$.

The BPS string embeddings can be related to the brane configuration in  (\ref{eq:brane-tab}) as follows. In the solutions (\ref{eq:h1h2-gen}), the sphere $S_1^2$ is associated with the (456) directions in (\ref{eq:brane-tab}), while the sphere $S_2^2$ is associated with the (789) directions:
The sphere $S_1^2$ in the geometry (\ref{eq:ds2-IIB}) collapses on the boundary component $\Im(z)=0$, where $h_1$ vanishes and $\partial h_2$ has poles corresponding to NS5 branes. The NS5-branes thus wrap $S_2^2$ but not $S_1^2$.
Likewise, on the boundary component $\Im(z)=\frac{\pi}{2}$, where the D5 sources are, the sphere $S_2^2$ collapses; the D5-branes wrap $S_1^2$ but not $S_2^2$.
The BPS string embeddings satisfying (\ref{eq:F1-BPS}) are in line with this identification and the orientation of F1 strings in (\ref{eq:brane-tab}); they preserve the symmetries of $S_1^2$, which is wrapped by the D5-branes, and break the symmetries of $S_2^2$, which is wrapped by the NS5-branes.

The circular Wilson loop expectation value is given by the on-shell action (\ref{eq:SF1-0}) with AdS$_2$ in global coordinates, such that $V_{AdS_2}=-2\pi$.
Noting that $f_4^8e^{4\phi}=16N_2^2W^{-2}$ and using the BPS conditions, the on-shell action simplifies to 
\begin{align}\label{eq:S-BPS-F1}
	S_{\rm F1}&=4\pi T\, |h_2|~.
\end{align}
The BPS conditions and action for D1-branes representing vortex loops can be obtained by exchanging $h_2$ and $h_1$.

\subsubsection{5-branes}

Wilson loops in antisymmetric representations can be realized by D5$^\prime$-branes, as discussed in sec.~\ref{sec:brane}.
The appropriate embedding ansatz for the D5$^\prime$ branes can be identified as follows.
As discussed for the BPS strings, the $S_1^2$ in the supergravity solutions corresponds to the (456) directions in (\ref{eq:brane-tab}), while the $S_2^2$ corresponds to the (789) directions. The D5$^\prime$-branes in table (\ref{eq:brane-tab}) should therefore wrap the entire $S_1^2$ and a one-dimensional part of $S_2^2$. To preserve the appropriate R-symmetry they have to wrap an $S^1$ in $S_2^2$.
They should also wrap an AdS$_2$ in AdS$_4$. This leaves a curve in $\Sigma$.

For a D5$^\prime$ wrapping AdS$_2$ in AdS$_4$, the $S_1^2$, a circle in $S_2^2$, and a curve in $\Sigma$, the entire embedding can be parametrized by a complex function $z(\xi)$ specifying a curve in $\Sigma$ and a real function $\theta(\xi)$ specifying the $S^1$ in $S_2^2$ with metric
\begin{align}
	 ds^2_{S_2^2}&=d\theta^2+\sin^2\!\theta\,d\phi^2~.
\end{align}
The induced metric on the D5$^\prime$  becomes
\begin{align}
	g&=(f_2^2\theta'^2+4\rho^2|z'|^2)d\xi^2+f_1^2ds^2_{S_1^2}+f_4^2ds^2_{AdS_2}+f_2^2\sin^2\!\theta\,ds^2_{S^1}~.
\end{align}
We expect BPS configurations to carry D3 and F1 charges, and therefore include worldvolume electric fields on $AdS_2$ and magnetic fields on $S_1^2$,
\begin{align}
	F&=F_{\rm el}\vol_{AdS_2}+F_1\vol_{S_1^2}~.
\end{align}
We expect the D3-brane charge induced by $F_1$ to control the embedding along $\Sigma$, following the discussion in sec.~\ref{sec:brane}.
The general D5-brane action with the appropriate Wess-Zumino terms reads
\begin{align}
	S_{\rm D5}&=T_{\rm D5}\int d^6\xi e^{-2\phi}\sqrt{\det(\tilde g+\cF)}
	-Q_{\rm D5}\int \left(C_{(4)}\wedge \cF +\frac{1}{2} C_{(2)}\wedge \cF\wedge \cF\right),
\end{align}
where $\cF=F-B_2$ and $\tilde g$ is the string frame induced metric $\tilde g=e^{\phi}g$ (with $\tau=\chi+ie^{-2\phi}$).
The tension is given by $T_{\rm D5}^{-1}=(2\pi)^5{\alpha^\prime}^3$.
We write the RR potentials corresponding to the field strengths in (\ref{eq:fieldstrengths-IIB}) as
\begin{align}
	C_{(2)}&=t(\theta) d\phi \wedge d b_2~,
	&
	C_{(4)}&=4f_1^2f_2^2\vol_{S_1^2}\wedge t(\theta) d\phi\wedge \star_2 (f_4^{-4}dj_1)+\ldots~,
\end{align}
where
\begin{align}\label{eq:t-def}
	t'(\theta)&=\sin\theta~.
\end{align}
The omitted terms in $C_{(4)}$ are not relevant for the D5$^\prime$ embedding. 
With the explicit expressions for the background fields and the induced metric the action becomes
\begin{align}\label{eq:D5-action}
	S_{\rm D5}=&\,T_{\rm D5}V_{{\rm AdS}_2}V_{S^2} V_{S^1}\!\int \! d\xi\sin\theta e^{-\phi} f_2 \sqrt{(f_4^4e^{2\phi}-F_{\rm el}^2)(f_1^4e^{2\phi}+(F_1-b_1)^2)\left(f_2^2{\theta^\prime}^2+4\rho^2|z'|^2\right)}
	\nonumber\\
&	-Q_{\rm D5}\int \frac{4f_1^2f_2^2}{f_4^{4}}\vol_{S_1^2}\wedge\, t(\theta) d\phi \wedge (\star_2 dj_1) \wedge F_{\rm el} \vol_{AdS_2}
	\nonumber\\
&-Q_{\rm D5}\int \frac{1}{2}\,t(\theta) d\phi \wedge db_2 \wedge F_{\rm el}\vol_{AdS_2}\wedge (F_1-b_1)\vol_{S_1^2}~,
\end{align}
where $db_2=\partial_z b_2 dz+\partial_{\bar z}b_2d\bar z$ and
$\star_2 dj_1=-i dz\partial_z j_1+id\bar z \partial_{\bar z}j_1$.

\paragraph{BPS configurations}
The BPS conditions for the D5$^\prime$ branes are derived and solved in app.~\ref{app:BPS}.
They imply that, along the embedding,
\begin{align}\label{eq:D5-BPS}
	h_2^D&={\rm const}~, & F_{\rm el}& = 2 \lambda h_2 \cos\theta~,
	&
	 F_1&=2h_2^D~,
\end{align}
with $\lambda=\pm 1$.
If these conditions are satisfied the embedding preserves half the supersymmetries.
The first equation fixes a curve in $\Sigma$ along which the D5$^\prime$-brane extends. The second equation fixes $\theta$ in terms of the electric field and the location on $\Sigma$. The embedding closes off smoothly with the $S^1$ in $S_2^2$ collapsing at the point where $|2h_2|=|F_{\rm el}|$. 
We verified in a number of examples that the BPS configurations satisfy the equations of motion as well.
The BPS conditions for NS5$^\prime$-branes can be obtained by exchanging $h_1$ and $h_2$.

When evaluating the on-shell action (\ref{eq:D5-action}), the integration constant in $t(\theta)$ defined in (\ref{eq:t-def}) has to be chosen judiciously.
It has to be such that the two-form potential is well defined with no sources in the hemisphere in which the $S^1$ wrapped by the D5$^\prime$-brane collapses. This leads to
\begin{align}\label{eq:C2-gauge}
	t(\theta)&=-\cos\theta \pm 1~.
\end{align}
The signs of $F_{\rm el}$, $\lambda$ and $h_2$ in the BPS conditions determine the sign of the constant.

The D3-brane and F1 charges carried by the D5$^\prime$ branes can be expressed concisely in terms of the harmonic functions. 
The D3-brane charge is determined by $F_1$ and given by 
\begin{align}\label{eq:ND3}
	N_{\rm D3}&=\frac{1}{2\pi}\int_{S_1^2}(2\pi \alpha')^{-1}F_1=\frac{2}{\pi\alpha'}h_2^D~,
\end{align}
where factors $2\pi\alpha'$ which had been absorbed into the definition of the worldvolume gauge fields have been taken into account.
For the F1 charge we find
\begin{align}\label{eq:NF1-int}
	N_{\rm F1}&=2\pi \alpha'\int \frac{\delta S_{\rm D5}}{\delta F_{\rm el}}=16\pi \alpha^\prime \lambda T_{\rm D5}V_{S^2}V_{S^1}\int d\xi\, h_1(\partial_z h_2)z'~.
\end{align}
Upon expanding $h_1$, $h_2$ into holomorphic and anti-holomorphic components, and using that $h_2^D$ is constant along the embedding, the integral can be expressed as
\begin{align}\label{eq:NF1}
	N_{\rm F1}&=\frac{4\lambda}{\pi^2{\alpha'}^2}\left[\Im\left(\cA_1\cA_2+\cC\right)\right]_{z_0}^{z_1}~,
\end{align}
where $z_0$ and $z_1$ are the start and end points of the curve along which the D5$^\prime$ extends in $\Sigma$ and $\cC$ was defined in (\ref{eq:cC-def}).

The Wilson loop expectation value is given by the Legendre-transformed on-shell action, 
which is discussed in app.~\ref{app:D5-action-charge}.
We find
\begin{align}\label{eq:I-int}
	S_{\rm D5}-F_{\rm el}\frac{\delta S_{\rm D5}}{\delta F_{\rm el}}
	&=\frac{8}{\pi^2{\alpha'}^3}I~, & I&=\int d\xi\, h_1 h_2 (\partial_z h_2) z'~.
\end{align}
As shown in app.~\ref{app:D5-action-charge} the integral can be expressed in terms of a holomorphic function $\cW$ as
\begin{align}\label{eq:I-def}
	I&=\left[\Im\left(2\cA_1\cA_2^2-2\cW+ih_2^D(\cA_1\cA_2+\cC)\right)\right]_{z_0}^{z_1}~,
\end{align}
where $\cW$ is defined up to a constant by
\begin{align}
	\partial\cW&=\cA_2^2\partial\cA_1~.
\end{align}
Upon identifying the Legendre-transformed on-shell action as expectation value of the antisymmetric Wilson loop we find
\begin{align}\label{eq:W-exp-D5}
	\ln\langle W_\wedge\rangle&=
	\frac{8}{\pi^2{\alpha'}^3}
	\left[\Im\left(2\cA_1\cA_2^2-2\cW+ih_2^D(\cA_1\cA_2+\cC)\right)\right]_{z_0}^{z_1}~.
\end{align}

\subsection{\texorpdfstring{AdS$_5\times$S$^5$}{AdS5xS5} and {\it Janus on the brane}}\label{sec:AdS5S5}

As a warm-up example we start with AdS$_5\times S^5$, which corresponds to a particularly simple choice of harmonic functions $h_{1/2}$. We rephrase familiar results on Wilson loops in $\cN=4$ SYM in the language of the $AdS_4\times S^2\times S^2\times\Sigma$ solutions, and identify a new class of D5$^\prime$ embeddings which describe surface defects with boundaries or interfaces on them. To our knowledge these embeddings have not been discussed before. They describe supersymmetric versions of the {\it Janus on the brane} embeddings discussed in \cite{Gutperle:2020gez}.

The AdS$_5\times S^5$ solution of Type IIB corresponds to $K=L$ and $N_{\rm D5}^{(a)}=N_{\rm NS5}^{(a)}=0$ in (\ref{eq:h1h2-gen}).
The holomorphic functions $\cA_{1/2}$ can then be taken as
\begin{align}
	\cA_1&=\frac{\pi\alpha^\prime}{2}K\sinh z~, & \cA_2&=\frac{\pi \alpha^\prime}{2}K\cosh z~,
\end{align}
and the function $\cC$ defined up to a constant in (\ref{eq:cC-def}) is given by
\begin{align}
	\cC&=-\frac{\pi^2{\alpha^\prime}^2}{4}K^2z+\cC_0~.
\end{align}
The metric (\ref{eq:ds2-IIB}) takes a simpler form in real coordinates $z=x+i y$, which leads to
\begin{align}
	ds^2&=2\pi\alpha^\prime K\left[\cosh^2\!x\,ds^2_{AdS_4}+\sin^2\!y\,ds^2_{S_1^2}+\cos^2\!y\,ds^2_{S_2^2}+dx^2+dy^2\right]~.
\end{align}
This is AdS$_5\times$S$^5$ with AdS$_5$ in AdS$_4$ slicing.
The curvature radius is $R^4=4\pi^2{\alpha'}^2K^2$ and the number of D3-branes $N_{\rm D3}=K^2/\pi$.

For AdS$_5\times $S$^5$ the 16 supersymmetries preserved by the general solutions (\ref{eq:h1h2-gen}) are enhanced to 32. The BPS conditions for string and 5-brane embeddings derived in app.~\ref{app:BPS} imply that 8 of the 16 supersymmetries present for all $AdS_4\times S^2\times S^2\times\Sigma$ solutions are preserved.
This does not capture all embeddings which preserve half of the enhanced 32 supersymmetries in $AdS_5\times S^5$. We therefore do not expect to recover all the half-BPS Wilson loop embeddings for $\cN=4$ SYM (studied in \cite{Rey:1998ik,Maldacena:1998im,Yamaguchi:2006tq}) from the BPS conditions derived here.
Indeed, the BPS conditions (\ref{eq:F1-BPS}) for a fundamental string imply $x=0$ and $y=\frac{\pi}{2}$, while in AdS$_5\times$S$^5$ an F1 wrapping AdS$_2$ is half-BPS at any point of the $S^5$. So only the condition $x=0$ should be required for half-BPS strings and we recover a subset of the $\cN=4$ SYM Wilson loops, as expected.

\begin{figure}
	\centering
	\subfigure[][]{\label{fig:AdS5S5-1}
		\begin{tikzpicture}
			\node at (0,0) {\includegraphics[width=0.4\linewidth]{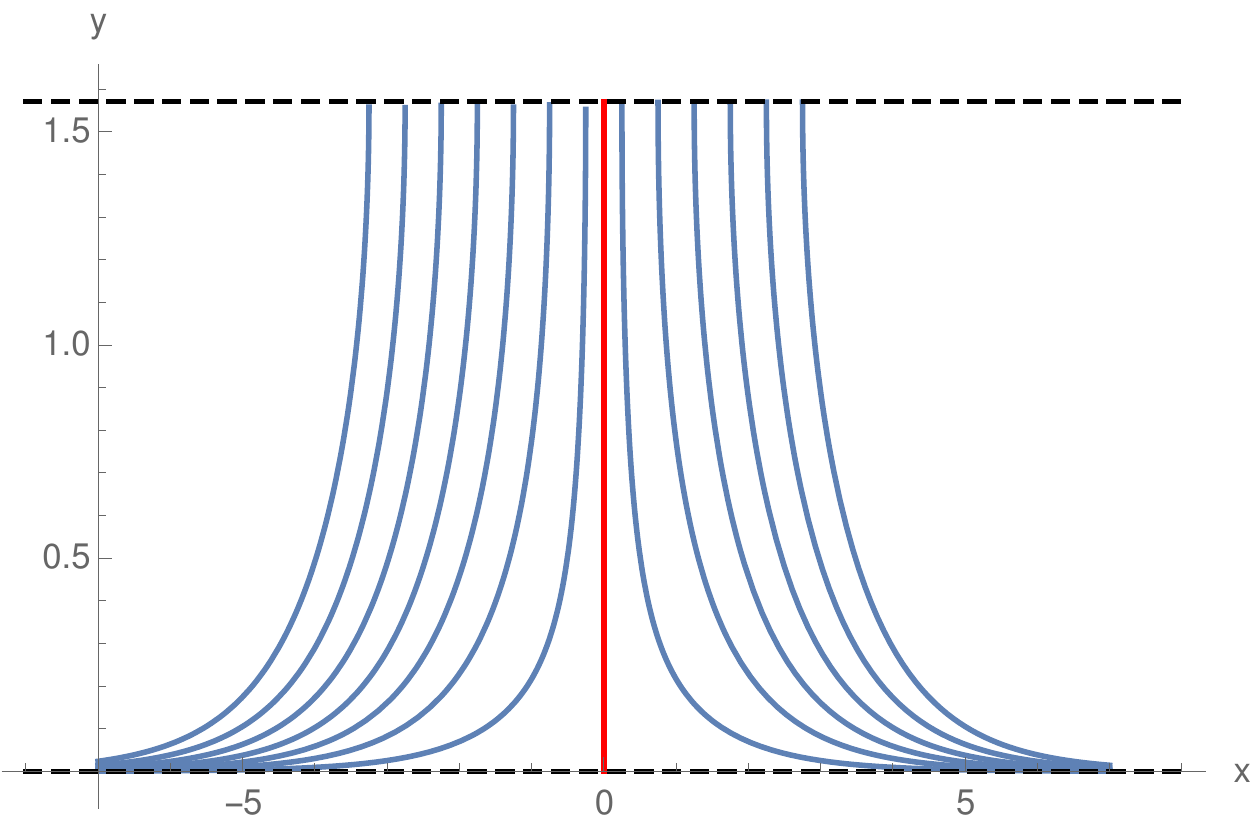}};
			\node at (2,1) {{\boldmath{$\Sigma$}}};
		\end{tikzpicture}
	}\hskip 10mm
	\subfigure[][]{\label{fig:AdS5S5-2}
		\begin{tikzpicture}[scale=1.1]\label{fig:wedge-2}
			
			\draw[very thick] (-2.5,0) -- (2.5,0);	
			\foreach \i in {0,...,18}{
				\draw (0,0) -- ({2.5*cos(10*\i)},-{2.5*sin(10*\i)});
			}
			\draw[thick,red] (0,0) -- (0,-2.5);
			\node at (1.25,0.2) {\scriptsize $x=+\infty$};
			\node at (-1.25,0.2) {\scriptsize $x=-\infty$};
			\node at (0,-2.7) {\scriptsize $x=0$};
			
			\node at (3.2,-1.25) {\small AdS$_5$};
			\node at (3.2,0) {\scriptsize $\partial$AdS$_5$};
			
			\draw[white,fill=gray,opacity=0.5] (0,0) -- (2.5,0) arc (0:-50:2.5);
			\node at (2.35,-2) {\scriptsize $x=\sinh^{-1}c$};
			\node at (1.5,-0.7) {\small D5$^\prime$};
		\end{tikzpicture}
	}
\caption{Left: D5$^\prime$ embeddings in AdS$_5\times$S$^5$ on the strip. Half-BPS Wilson loop D5$^\prime$-branes are embedded along the red curve $x=0$, corresponding to $c=0$ in (\ref{eq:BPS-AdS5S5}). The blue curves are {\it Janus on the brane} embeddings. Right: Schematic illustration of the embeddings in AdS$_5$. Lines starting at the conformal boundary are AdS$_4$ slices, of which the D5$^\prime$ wrap an AdS$_2$. The angular coordinate is $x$, with $x=\pm\infty$ corresponding to the conformal boundary and $x=0$ to the vertical slice. The shaded region is an example of a {\it Janus on the brane} D5$^\prime$ embedding.\label{fig:AdS5S5}}
\end{figure}
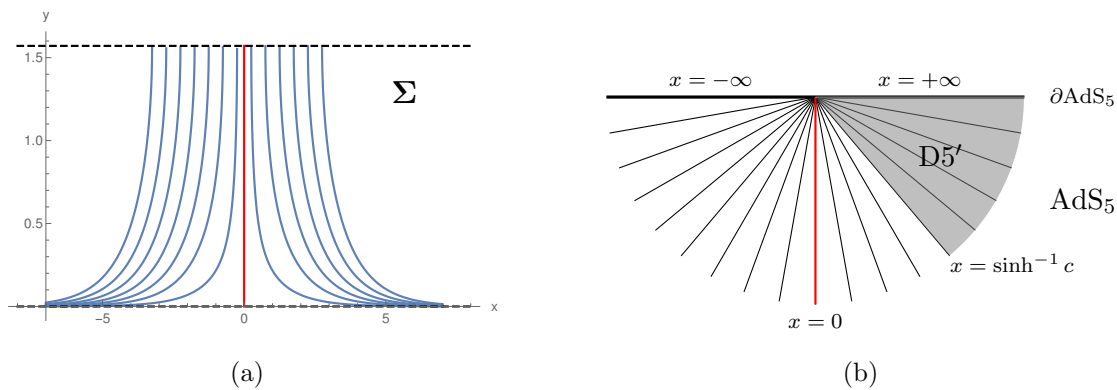

The probe D5$^\prime$ branes are more interesting, as we will find a new class of surface defects for $\cN=4$ SYM.
The BPS conditions for D5$^\prime$ embeddings in (\ref{eq:D5-BPS}) become
\begin{align}\label{eq:BPS-AdS5S5}
	h_2^D&=\pi\alpha^\prime K\sinh x\sin y=\pi\alpha^\prime Kc~,
	\nonumber\\
	F_{\rm el}&=2\pi\alpha^{\prime}K\lambda \cos\theta\cosh x\cos y~,
\end{align}
with a constant $c$.
We start with embeddings without D3-brane charge, $c=0$. A non-degenerate embedding needs $x=0$, otherwise the $S_1^2$ wrapped by the D5$^\prime$ collapses due to the first condition. The second condition can then be solved for $\theta$ as function of $y$,
\begin{align}
	\cos\theta&=\frac{F_{\rm el}}{2\pi \alpha^\prime K\lambda \cos y}~.
\end{align}
This restricts $F_{\rm el}$ to $|F_{\rm el}|\leq 2\pi \alpha' K$. On the lower boundary of $\Sigma$, where the $S_1^2$ collapses, $\cos y=1$. 
For $F_{\rm el}=0$ the D5$^\prime$ wraps an equatorial $S^1$ in $S_2^2$ and reaches the upper boundary of $\Sigma$, where the $S_2^2$ collapses.
More generally the D5$^\prime$ reaches up to 
\begin{align}
	y_\star&=\cos^{-1}\left(\frac{|F_{\rm el}|}{2\pi\alpha^\prime K}\right)~,
\end{align}
where the $S^1$ wrapped in $S_2^2$ collapses.
The $c=0$ D5$^\prime$ branes carry no D3 charge, while the F1 charge is given via (\ref{eq:NF1}) by
\begin{align}
	N_{\rm F1}&=\lambda K^2\left(\sin y_\star\cos y_\star-y_\star\right)~,
\end{align}
with $\lambda=\pm 1$ depending on the sign of $F_{\rm el}$.
The $c=0$ embeddings correspond to the $AdS_2\times S^4$ D5-branes representing antisymmetric Wilson loops in  4d $\cN=4$ SYM, discussed in \cite{Yamaguchi:2006tq}.
The embeddings are shown in red in fig.~\ref{fig:AdS5S5-1}.
The construction here implies that they preserve $8$ of the $16$ supersymmetries present for all solutions (\ref{eq:h1h2-gen}). They preserve half of the additional supersymmetries present for AdS$_5\times$S$^5$ as well.
For $y_\star =\frac{\pi}{2}$ the F1 charge is half the number of D3-branes and the rank of the representation is half the rank of the gauge group.
This is the `maximal' antisymmetric Wilson loop.	

Now to the embeddings with D3-brane charge, $c\neq 0$. From (\ref{eq:BPS-AdS5S5}) and (\ref{eq:ND3}), $N_{\rm D3}=2cK$. The first equation in (\ref{eq:BPS-AdS5S5}) fixes $y$ in terms of $x$, leading to the blue curves shown in fig.~\ref{fig:AdS5S5-1}. The D5$^\prime$ captures a range in $x$ which is constrained by $|c|<|\sinh x|<\infty$, with $x$ positive/negative for positive/negative $c$. 
For $|x|\rightarrow\infty$ we have $y\rightarrow 0$.
The second equation in (\ref{eq:BPS-AdS5S5}) fixes $\theta$ in terms of $x$,
\begin{align}
	\cos\theta&=\frac{F_{\rm el}}{2\pi\alpha^\prime K \lambda}\frac{1}{\sqrt{\cosh^2\! x-c^2\coth^2\! x}}~.
\end{align}
For $|x|\rightarrow\infty$ we have $\theta\rightarrow \frac{\pi}{2}$ and the D5$^\prime$ wraps an equatorial $S^1$ in $S_2^2$.
For $F_{\rm el}=0$ the D5$^\prime$ wraps the equatorial $S^1$ in $S_2^2$ for all $x$, and the embedding reaches all the way to $\sinh x=c$, where $y=\frac{\pi}{2}$.
For $F_{\rm el}\neq 0$, the $S^1$ starts to slip towards a pole on $S_2^2$ as $|x|$ is decreased, and the D5$^\prime$ caps off with the $S^1$ collapsing before reaching $\sinh x=c$. 
The limits $x\rightarrow \pm \infty$ lead to the boundary of AdS$_5$ and the D5$^\prime$ extend to one of these regions. The embeddings thus describe a ``half surface defect" of the form $\RR^+\times \RR $ in $\cN=4$ SYM, as shown in fig.~\ref{fig:AdS5S5-2}. 
The embeddings describe bound states of D5-branes, D3-branes and fundamental strings; they preserve 8 supersymmetries by construction and we do not expect a further enhancement from the additional supersymmetries of AdS$_5\times$S$^5$.

We briefly compare these D5$^\prime$ surface operators to the half-BPS surface operators in $\cN=4$ SYM studied in \cite{Drukker:2008wr}.
The latter can be realized holographically by D3-branes wrapping AdS$_3$ in AdS$_5$ and an $S^1$ in $S^5$.
They describe planar surface operators extending along an entire $\RR^2$ in $\RR^4$, which preserve 16 supersymmetries.
Non-supersymmetric {\it Janus} versions of these surface operators, which have a 1d interface on the 2d surface, were studied in \cite{Gutperle:2020gez}.
The embeddings we find here, on the other hand, describe surface defect operators which themselves have a boundary, and preserve 8 supersymmetries. Janus interfaces on surfaces can be realized by combining two such D5$^\prime$ embeddings, one with $c>0$ and one with $c<0$, such that two half surface defects are joined along a 1d interface. 
We focus here on line operators and leave a more detailed investigation of the surface operators for the future. We will encounter similar surface-type D5$^\prime$ embeddings for the Janus and BCFT solutions below, where they are naturally associated with the 4d gauge nodes.

\subsection{General balanced 3d \texorpdfstring{$T_\rho^\sigma[(SU(N)]$}{Trhosigma[SU(N)]}}\label{sec:balanced-hol}

\begin{figure}
	\centering
	\begin{tikzpicture}[scale=1.5]
		\draw[thick] (2,-0.8) -- +(0,1.6);
		\foreach \i in {-1.5,-0.5,0.5,1.5} \draw[thick] ({2.5+\i/10},-0.8) -- +(0,1.6);
		\draw[thick] (3,-0.8) -- +(0,1.6);
		
		\foreach \i in {0,1,2} \draw[thick] (0,0.3-0.025-0.05*\i) -- +(2,0);
		\foreach \i in {3,4} \draw[thick] (0.4,0.3-0.025-0.05*\i) -- +(1.95,0);
		\foreach \i in {5} \draw[thick] (0.4,0.3-0.025-0.05*\i) -- +(2.05,0);		
		\foreach \i in {6} \draw[thick] (0.8,0.3-0.025-0.05*\i) -- +(1.65,0);				
		\foreach \i in {7,8} \draw[thick] (0.8,0.3-0.025-0.05*\i) -- +(1.75,0);				
		\foreach \i in {9,10} \draw[thick] (1.2,0.3-0.025-0.05*\i) -- +(1.45,0);				
		\foreach \i in {11} \draw[thick] (1.2,0.3-0.025-0.05*\i) -- +(1.8,0);				
		\foreach \i in {0,0.4,0.8,1.2}{ \draw[fill=gray] (\i,0) ellipse (2pt and 9pt);}
	\end{tikzpicture}
\caption{Brane configuration for a balanced quiver. It involves one group of 4 NS5-branes with 3 D3-branes ending on each, and three groups of D5-branes: a group of one D5-brane on which 3 D3-branes end, one group of 4 D5-branes with 2 D3-branes ending on each, and a group of one D5-brane with one D3-brane ending on it.
This corresponds to $\rho=[3^4]$, $\sigma=[3^1,2^4,1^1]$.
	\label{fig:brane-balanced}}
\end{figure}
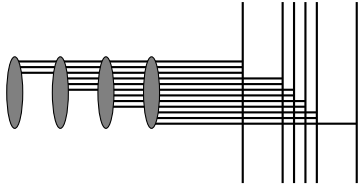

We now consider general balanced 3d quivers. The supergravity duals have 5-brane sources but no asymptotic AdS$_5\times$S$^5$ regions. We will discuss D5$^\prime$ embeddings, compute the Wilson loop expectation values holographically, and compare the results to the field theory analysis of sec.~\ref{sec:loc}.
The brane configurations for general balanced quivers are illustrated in fig.~\ref{fig:brane-balanced}. They involve one group of NS5 branes, with the same number of D3-branes ending on each NS5-brane of the group, and an arbitrary number of D5-brane groups, where within each group the numbers of D3-branes ending on each D5-brane are identical.

For the harmonic functions $h_1$, $h_2$ in (\ref{eq:h1h2-gen}), this amounts to $K=L=0$ and only one group of NS5 branes, $B=1$. The NS5 source can then be placed at $z=0$ without loss of generality, leading to
\begin{align}\label{eq:h1h2-balanced}
	h_1&=-\frac{\alpha'}{4}\sum_{a=1}^AN_{\rm D5}^{(a)} \ln\tanh\left(\frac{i\pi}{4}-\frac{z-\delta_a}{2}\right) + \rm{c.c.}
	\nonumber\\
	h_2&=-\frac{\alpha'}{4}N_{\rm NS5}\ln\tanh\left(\frac{z}{2}\right)+\rm{c.c.}
\end{align}
These supergravity solutions describe brane configurations with $A$ groups of D5-branes, with $N_{\rm D5}^{(a)}$ D5-branes in each group and an equal number of D3-branes ending on each D5-brane within a given group. The total number of D3-branes ending on each D5-brane group is set by $\delta_a$,
\begin{align}
	N_{\rm D3}^{(a)}&=N_{\rm NS5}N_{\rm D5}^{(a)}\frac{2}{\pi}\arctan e^{-\delta_a}~.
\end{align}
The identification of the supergravity parameters with field theory data was spelled out explicitly in sec.~V.A of \cite{Coccia:2020wtk}. 
The dual is a quiver of the form (\ref{quiv:gen-T_rho_sigma}) with $N_{\rm NS5}-1$ nodes and $N_{\rm D5}^{(a)}$ flavors at gauge nodes ${\sf t}_a$ with 
\begin{align}\label{eq:ta-balanced}
	{\sf t}_a&=\frac{2}{\pi}N_{\rm NS5}\arctan e^{\delta_a}~.
\end{align}
Since all nodes are balanced and $N_0=N_{L+1}=0$, the entire quiver can be reconstructed from this information.

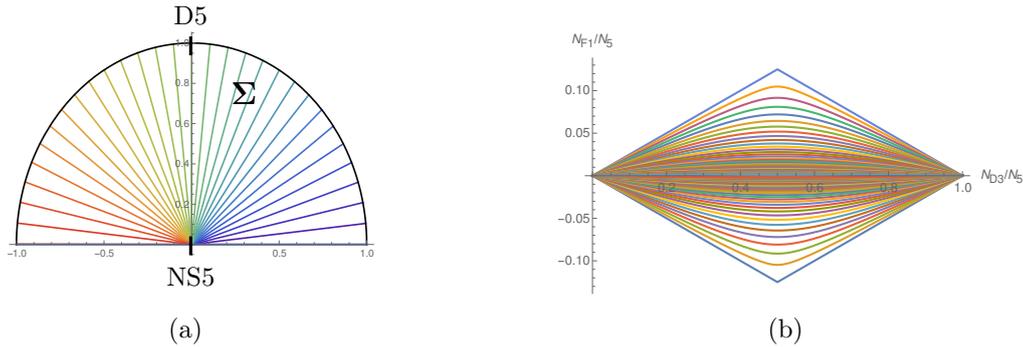
\begin{figure}
	\centering
	\subfigure[][]{\label{fig:WL-D5-NS5-1}
		\begin{tikzpicture}
			\node at (0,0) {\includegraphics[width=0.3\linewidth]{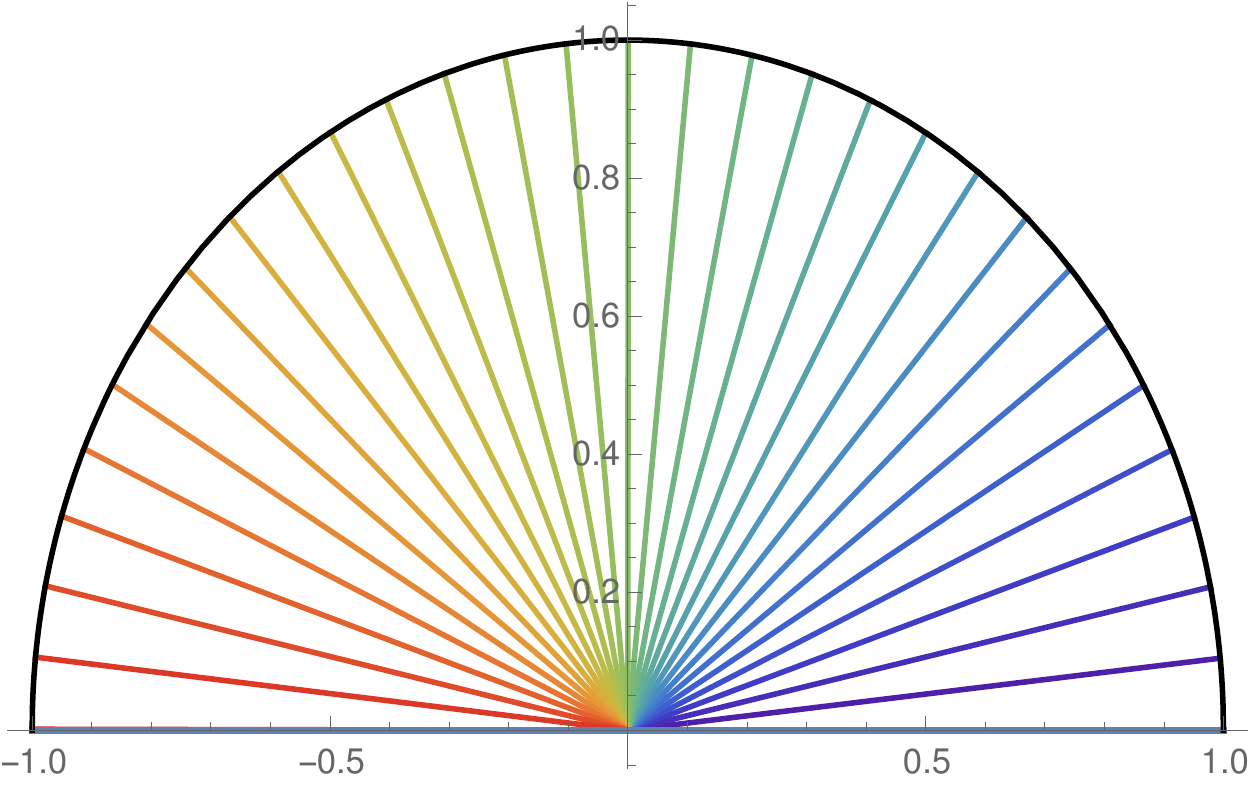}};
			\node at (0.7,0.7) {$\mathbf{\Sigma}$};
			\draw[very thick] (0,-1.2) -- (0,-1.45) node [anchor=north]{\footnotesize NS5};
			\draw[very thick] (0,1.2) -- (0,1.45) node [anchor=south]{\footnotesize D5};
		\end{tikzpicture}
	}\hskip 20mm
	\subfigure[][]{	\label{fig:WL-D5-NS5-2}
		\includegraphics[width=0.38\linewidth]{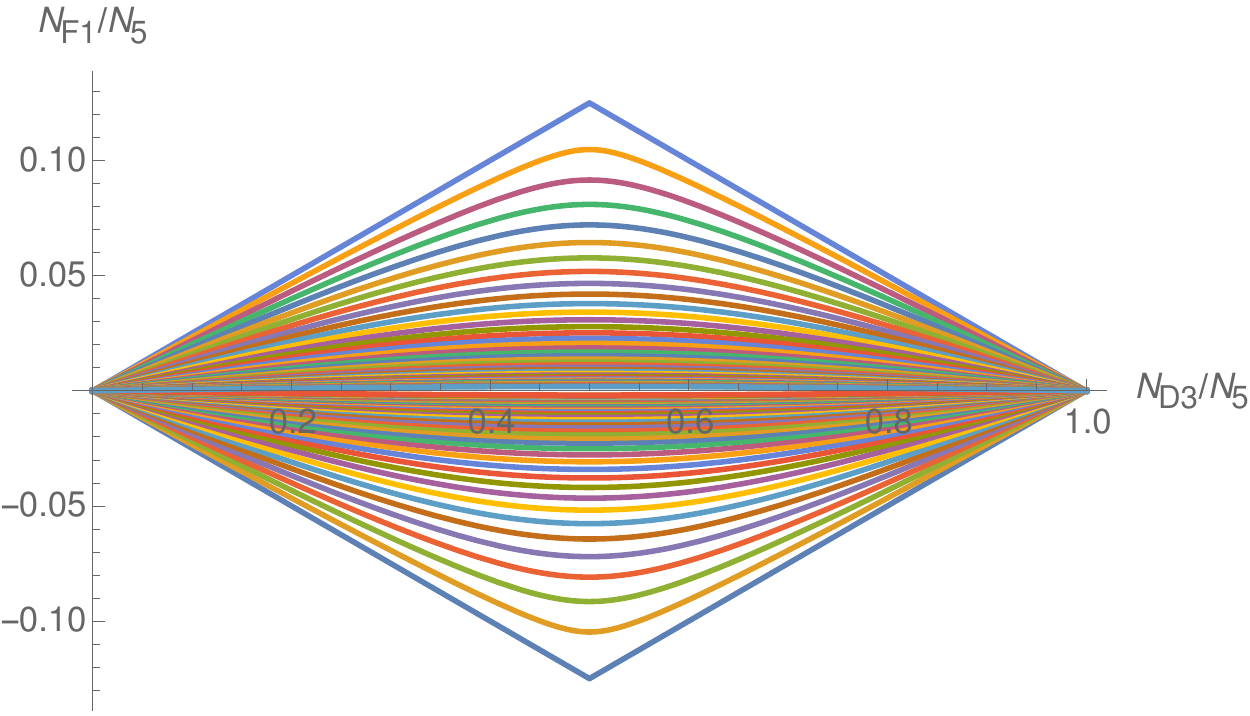}
	}
	\caption{Left: $\Sigma$ as half disc in the $u$ coordinate defined in (\ref{eq:coord-change}), with D5$^\prime$ Wilson loop embeddings. For concreteness, NS5/D5 poles are shown at $u=0$/$u=i$, corresponding to the D5/NS5 theory. Right: Space of $({\rm D3},{\rm F1})$ charges carved out by the D5$^\prime$ embeddings for the D5/NS5 theory. The curves vary $d$ for fixed $F_{\rm el}$. From top to center the curves correspond to $F_{\rm el}$ increasing from $0$ to $\infty$. The bottom half of the curves corresponds, from bottom to center, to $F_{\rm el}$ decreasing from $0$ to $-\infty$. The shape reproduces the shape of the quiver (\ref{eq:delta0-quiver}).
		\label{fig:WL-D5-NS5}}
\end{figure}

To simplify the discussion of D5$^\prime$ embeddings we change coordinates as follows,
\begin{align}\label{eq:coord-change}
	z&=\ln w~,
	&
	\frac{w-1}{w+1}&=-u~.
\end{align}
The first transformation maps the strip to the upper right quadrant in the complex plane.
The second maps the upper right quadrant to the upper half disc. The boundary component $\Im(z)=0$, where $\partial h_2$ has poles, becomes the diameter of the half disc; the boundary component $\Im(z)=\frac{\pi}{2}$, where $\partial h_1$ has poles, becomes the circumference.
The harmonic functions become
\begin{align}\label{eq:h1h2-balanced-u}
	h_1&=-\frac{\alpha'}{4}\sum_{a}N_{\rm D5}^{(a)} \ln\left|\frac{u-\sigma_a}{1-\sigma_a u}\right|^2~,
	&
	h_2&=-\frac{\alpha'}{4}N_{\rm NS5}\ln|u|^2 \ ,
\end{align}
where $\sigma_a$ are phases determined from $\delta_a$ by
\begin{align}\label{eq:sigma-balanced}
	\sigma_a&=\frac{ie^{\delta_a}-1}{ie^{\delta_a}+1}~.
\end{align}
For the special case with one D5-brane pole at $\delta_1=0$ the solution is shown in fig.~\ref{fig:WL-D5-NS5}.
In that case $\sigma_1=i$.
The holomorphic functions $\cA_{1/2}$ corresponding to (\ref{eq:h1h2-balanced-u}) can be chosen as
\begin{align}
	\cA_1&=-\frac{i\alpha'}{4}\sum_{a}N_{\rm D5}^{(a)}\left[\ln(1-u/\sigma_a)-\ln(1-\sigma_au)\right]~,
	\nonumber\\
	\cA_2&=-\frac{\alpha'}{4}N_{\rm NS5}\ln u~.
\end{align}
For the function $\cC$ defined in (\ref{eq:cC-def}) we find
\begin{align}\label{eq:cC-balanced}
	\cC&=\frac{i{\alpha'}^2}{8}N_{\rm NS5}\sum_a N_{\rm D5}^{(a)}\left[ \Li_2(u\sigma_a)-\Li_2(u/\sigma_a)\right]-\cA_1\cA_2+\cC_0~.
\end{align}

\subsubsection{D5$^\prime$ embeddings}

The half-BPS D5$^\prime$-brane embeddings are along curves with constant $h_2^D$. In the $u$ coordinate the requirement simply becomes that $\arg(u)$ should be constant. 
The embeddings are straight lines in the $u$-coordinate starting at the origin, and we introduce a parameter $d$ specifying the angle as
\begin{align}
&&	u&=re^{i\pi d}~,& d\in(0,1)~.
\end{align}
The embeddings start at $r=0$ and end at the point along the line where
\begin{align}
	|F_{\rm el}|&=2|h_2| = \alpha'N_{\rm NS5} |\ln|u||=-\alpha^\prime N_{\rm NS5}\ln r~.
\end{align}
More explicitly, they end at $u_\star$ given by
\begin{align}\label{eq:delta0-u1}
	u_\star&=\exp\left\{-\frac{|F_{\rm el}|}{\alpha' N_{\rm NS5}}+i\pi d\right\}~.
\end{align}
These D5$^\prime$ embeddings all correspond to line operators; there are no surface defect embeddings of the form discussed in sec.~\ref{sec:AdS5S5} since the AdS$_5\times$S$^5$ regions are closed off.

The D3 and F1 charges carried by the D5$^\prime$-branes can be determined from (\ref{eq:ND3}) and (\ref{eq:NF1}).
For the D3-brane charge we find
\begin{align}\label{eq:D5NS5-D3charge}
	N_{\rm D3}&=\frac{2}{\pi \alpha^\prime}h_2^D=N_{\rm NS5} d~.
\end{align}
The remaining ingredient is the F1 charge. From (\ref{eq:NF1}) with $\cC$ in (\ref{eq:cC-balanced}) we find
\begin{align}\label{eq:D5NS5-F1charge}
	N_{\rm F1}&=\frac{\lambda N_{\rm NS5}}{2\pi^2}\sum_a N_{\rm D5}^{(a)}\Re\left[ \Li_2(u_\star\sigma_a)-\Li_2(u_\star/\sigma_a)\right]~,
\end{align}
with $\lambda = \pm1$ depending on the sign of $F_{\rm el}$.

There are no solutions to the BPS conditions for F1 strings, (\ref{eq:F1-BPS}), at regular points of $\partial\Sigma$. Instead, the fundamental strings representing Wilson loops in the fundamental representation are all located at the NS5 pole. This can be seen as follows: for small rank of the antisymmetric representations, the D5$^\prime$ embeddings degenerate to points at the NS5 pole. This is the singular point where F1 strings can be embedded. The F1 action is logarithmically divergent at the NS5 pole, which signals logarithmically enhanced scaling of the fundamental Wilson loop expectation value, matching the field theory expectations discussed below eq.~(\ref{eq:W_fund}). We focus here on the large-rank antisymmetric Wilson loops.

\subsubsection{Wilson loop expectation values}

To obtain the expectation values for the Wilson loops represented by the D5$^\prime$ embeddings from (\ref{eq:W-exp-D5}) we have to solve for $\cW$, defined by
\begin{align}
	\partial_u\cW&=\cA_2^2\partial_u \cA_1=
	-\frac{i{\alpha'}^3}{64}\sum_a N_{\rm D5}^{(a)}N_{\rm NS5}^2\,\frac{\ln^2 u}{u-\sigma_a} -(\sigma_a\rightarrow 1/\sigma_a)~.
\end{align}
This leads to
\begin{align}
	\cW &=
	\frac{i{\alpha'}^3}{32}N_{\rm NS5}^2\sum_a N_{\rm D5}^{(a)}\left[-\Li_3(\sigma_a u)+\ln u\Li_2(\sigma_a u)-(\sigma_a\rightarrow 1/\sigma_a)\right]+\cA_1\cA_2^2
	~.
\end{align}
Noting that $ih_2^D=\frac{\alpha'}{4}N_{\rm NS5}(\ln u-\ln \bar u)$, the Wilson loop expectation value as defined in (\ref{eq:W-exp-D5}) becomes
\begin{align}\label{eq:D5NS5-Wexp}
	\ln\langle W_\wedge\rangle &=\frac{N_{\rm NS5}^2}{2\pi^2}\sum_a N_{\rm D5}^{(a)}
	\Re\left[\Li_3(\sigma_a u_\star)-\Li_3(u_\star/\sigma_a)-\ln |u_\star|\,(\Li_2(\sigma_a u_\star)-\Li_2(u_\star/\sigma_a ))\right].
\end{align}
The expression for $u_\star$ was given in (\ref{eq:delta0-u1}), and from (\ref{eq:ta-balanced}), (\ref{eq:sigma-balanced}) we have
\begin{align}
\sigma_a&= \exp\left\{-\frac{i\pi{\sf t}_a}{N_{\rm NS5}}\right\}~.
\end{align}

The remaining task is to identify which Wilson loop a D5$^\prime$ with parameters $d$ (specifying the embedding on $\Sigma$) and $F_{\rm el}$ (the electric field on AdS$_2$) corresponds to in the dual field theory.
From the discussion in sec.~\ref{sec:brane} we identify the D3-brane charge (\ref{eq:D5NS5-D3charge}) with the quiver coordinate labeling the gauge nodes. Indeed, the quiver has $N_5-1$ nodes, which is the range covered by the D3-brane charge in (\ref{eq:D5NS5-D3charge}) for $d\in(0,1)$.
The rank of the representation of the Wilson loop is determined by the F1 charge, (\ref{eq:D5NS5-F1charge}).

\paragraph{Comparison to field theory:}
We now compare in more detail to the field theory computation of sec.~\ref{sec:hol-balanced}.
In the field theory computations $z=t/L\in(0,1)$ was used as coordinate on the quiver, and $k=N(z)\mathds{k}$ was used to denote the rank of the representation. 
From the discussion in sec.~\ref{sec:brane} the D3-brane charge carried by the D5$^\prime$ should be related to the gauge node label $t$. This leads to $t=N_{\rm D3}$ or $z=d$.
The labels $z_t$ appearing e.g.\ in (\ref{eq:w-tau-def}) for the gauge nodes with fundamental hypermultiplets attached correspond on the supergravity side to ${\sf t}_a$ in (\ref{eq:ta-balanced}). So we have
\begin{align}
	z&=d~, &	z_t&=\frac{{\sf t}_a}{N_{\rm NS5}}~.
\end{align}
The coordinate $z$ along the quiver is proportional to the dual harmonic function $h_2^D$. This is in accordance with the identification in Appendix B of \cite{Akhond:2021ffz}, where the variable $\eta$ plays the role of our $z$.

In both calculations (holographic and field theory), the Wilson loop expectation value is given implicitly: for a given gauge node (fixed by $d$ or $z$) the representation $k$ and the expectation value are both given in terms of one real parameter, which is $F_{\rm el}$ in the holographic results (\ref{eq:D5NS5-F1charge}), (\ref{eq:D5NS5-Wexp}) and $b$ in the field theory results (\ref{eq:W_field_bal}).
If the real parameters $b$ and $F_{\rm el}$ can be identified in a way which makes the representation and the expectation value match simultaneously between field theory and supergravity, we will have demonstrated perfect agreement.
Indeed, the identification is simply
\begin{align}
	\frac{F_{\rm el}}{\alpha' N_{\rm NS5}}=2\pi b~.
\end{align}
For the quantities appearing in (\ref{eq:D5NS5-F1charge}), (\ref{eq:D5NS5-Wexp}) and (\ref{eq:W_field_bal}) this leads to the identification
\begin{align}\label{eq:balanced-match}
	u_\star\sigma_a^{\pm 1}&=\exp\left\{-\frac{F_{\rm el}}{\alpha' N_{\rm NS5}}+i\pi \left(d\mp\frac{{\sf t}_a}{N_{\rm NS5}}\right)\right\}
	=
	e^{-2\pi b+i \pi (z\mp z_t)}=\overline{w}\, \tau_t^{\mp 1}~.
\end{align}
There is a slight subtlety in the identification of the F1 charge $N_{\rm F1}$ with the representation $k$. For $2\pi b>0$, corresponding to $\lambda=1$, the  identification (\ref{eq:balanced-match}) leads to a perfect match between (\ref{eq:D5NS5-F1charge}), (\ref{eq:D5NS5-Wexp}) and (\ref{eq:W_field_bal}), with $N_{\rm F1}=k=\mathds{k} N(z)$.
Since the Wilson loop expectation values are symmetric in $\mathds{k}\rightarrow 1-\mathds{k}$, which amounts to $b\rightarrow -b$, this should extend to a match for general $b$ and $F_{\rm el}$. The only subtlety is that the general identification of $N_{\rm F1}$ and $k$ has to take into account that the F1 charge (\ref{eq:NF1}) flips sign for $F_{\rm el}<0$ ($\lambda=-1$).
This leads to the natural identification $k= N_{\rm F1} \mod N(z)$,
which indeed results in a complete match between the holographic and field theory calculations.

\subsubsection{D5/NS5 and D5$^2$/NS5 theories}

To make the formulas concrete, and also compare in detail the space of $(N_{\rm F1},N_{\rm D3})$ charges carved out by the D5$^\prime$ embeddings to field theory, we will discuss two simple examples.

\begin{figure}
	\centering
	\subfigure[][]{
		\begin{tikzpicture}
			\node at (0,0) {\includegraphics[width=0.3\linewidth]{WL-NS5-D5-K0.pdf}};
			\node at (0.7,0.7) {$\mathbf{\Sigma}$};
			\draw[very thick] (0,-1.2) -- (0,-1.45) node [anchor=north]{\footnotesize NS5};
			\draw[very thick] (0+1.68,-1.35+1.68) -- (0+1.85,-1.35+1.85) node [anchor=south west]{\footnotesize D5};
			\draw[very thick] (0-1.68,-1.35+1.68) -- (0-1.85,-1.35+1.85) node [anchor=south east]{\footnotesize D5};			
		\end{tikzpicture}
	}\hskip 20mm
	\subfigure[][]{
		\includegraphics[width=0.32\linewidth]{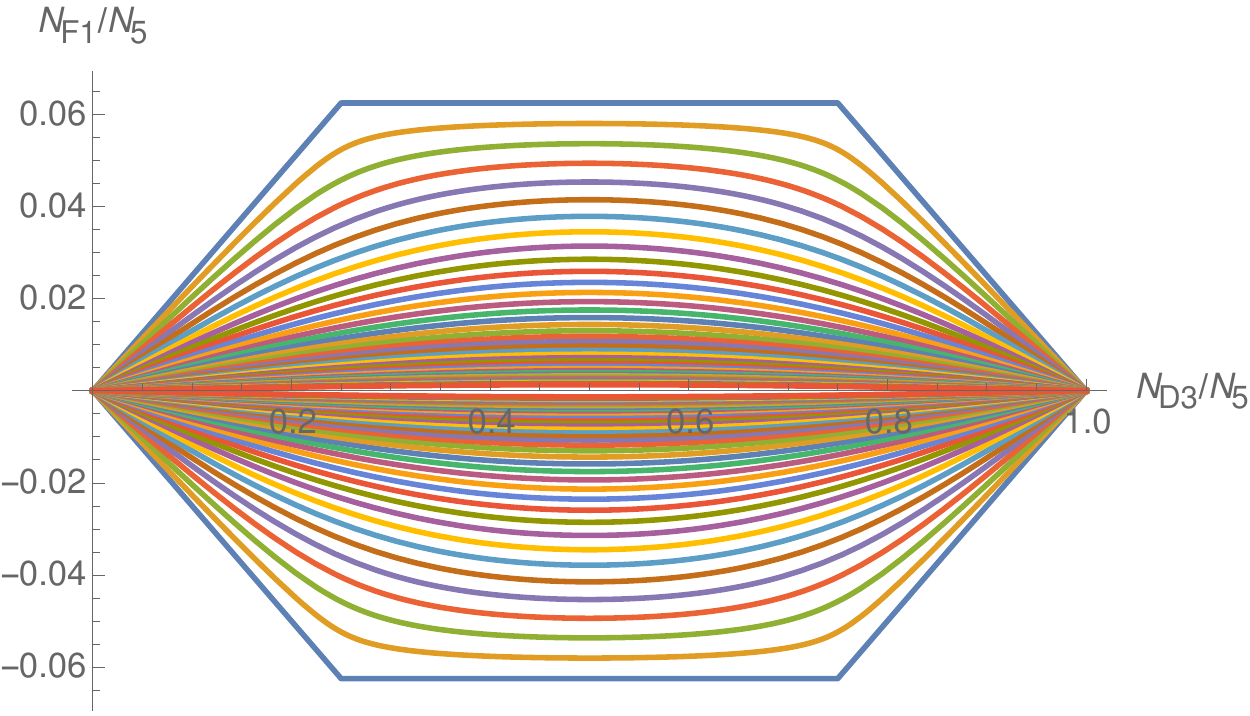}
	}
	\caption{Left: Wilson loop embeddings for a balanced theory with two D5 poles, $N_{\rm D5}^{(1)}=N_{\rm D5}^{(2)}=\frac{1}{2}N_{\rm NS5}$ and $\sigma_2=i\sigma_1=e^{3i\pi/4}$. Right: Space of (D3,F1) charges carved out by the embeddings.
		\label{fig:WL-balanced}}
\end{figure}

The first is the `D5/NS5 theory', which corresponds to having only one group of D5-branes at $\delta_1=0$, with a number of D5-branes which equals the number of NS5 branes. 
The brane configuration is symmetric under S-duality.
The setup corresponds to $A=1$, $\delta_1=0$ and $N_{\rm D5}^{(1)}=N_{\rm NS5}\equiv N_5$ in (\ref{eq:h1h2-balanced}).
This leads to $\sigma_1=i$.
The quiver gauge theory has $N_5-1$ nodes with $N_5$ flavors at the central node. 
The rank function was given in (\ref{eq:delta0-rank}).
With $N\equiv N_5/2$ the quiver is given by
\begin{align}\label{eq:delta0-quiver}
	U(N)-U(2N)-U(3N)-\ldots - &U(N^2)-U((N-1)N)-\ldots -U(2N) -U(N)
	\nonumber\\
	&\ \ \ \ \vert\\
&	\ \ [2N] \nonumber
\end{align}
The D5$^\prime$ embeddings and the space of charges carved out by the embeddings are shown in fig.~\ref{fig:WL-D5-NS5}.
The charges carved out in fig.~\ref{fig:WL-D5-NS5-2} precisely match the shape of the quiver diagram. 
In fig.~\ref{fig:WL-D5-NS5-2}, $F_{\rm el}=0$ is the maximal Wilson loop, with the rank of the representation half that of the gauge group. The rank of the representation can be decreased/increased, corresponding to positive/negative $F_{\rm el}$, up to a maximum of half the rank. 
The F1 charge and expectation value for this theory are given by
\begin{align}\label{eq:D5NS5-logW}
	N_{\rm F1}
	&=\frac{\lambda N_5^2}{2\pi^2}\Re\left(\Li_2(iu_\star)-\Li_2(-iu_\star)\right)~,
\nonumber\\
	\ln\langle W_\wedge\rangle
	&=\frac{N_5^3}{2\pi^2}\Re\Big[
	\Li_3(iu_1)-\Li_3(-iu_\star)-\left(\Li_2(iu_\star)-\Li_2(-iu_\star)\right)\ln |u_1|
	\Big].
\end{align}
Comparing to field theory, $d$ corresponds to $z$, as before, leading to the identification $u_\star=e^{-2\pi b+i\pi z}= \bar w$.
The holographic results then precisely match the field theory results (\ref{eq:Wilson-delta0-summary}) with $k = N_{\rm F1} \mod N(z)$.

As a further example we briefly discuss the case with one group of $N_5$ NS5 branes and two groups of D5 branes, with $N_5/2$ branes in each group and $\delta_1=-\delta_2 =\sinh^{-1}(1)$ in (\ref{eq:h1h2-balanced}). We refer to this theory as D5$^2$/NS5 theory.
The $\sigma_a$ parameters are given by $\sigma_1=e^{i\pi/4}$ and $\sigma_2=e^{3i\pi/4}$. 
The dual quiver has $N_5-1$ nodes and from (\ref{eq:ta-balanced}) we see that there are $N_5/2$ flavors at the ${\sf t}_1=\frac{1}{4}N_5$ node and at the ${\sf t}_2=\frac{3}{4}N_5$ node.
With $N\equiv N_5/2$ the complete quiver reads
\begin{align}
	U(N) - U(2N) -\ldots -U&(\tfrac{1}{2}N^2)-\ldots -U(\tfrac{1}{2}N^2) -\ldots -U(2N)-U(N)
	\nonumber\\
	&\ \  \, \vert \hskip 28mm \vert
	\\ \nonumber
	&\,[N]\hskip 24mm [N]
\end{align}
Along the first ellipsis the rank increases in steps of $N$, along the second ellipsis the rank is constant, and along the third the rank decreases in steps of $N$.
The charges carved out by the D5$^\prime$ branes are shown in fig.~\ref{fig:WL-balanced}. They again reproduce the quiver diagram.

\subsection{D5/NS5 \texorpdfstring{$\mathcal N=4$}{N=4} SYM BCFT}\label{sec:BCFT-hol}

\begin{figure}
	\centering
		\begin{tikzpicture}[y={(0cm,1cm)}, x={(0.707cm,0.707cm)}, z={(1cm,0cm)}, scale=1.5]
		\draw[gray,fill=gray!100,rotate around={-45:(0,0,2)}] (0,0,2) ellipse (1.8pt and 3.5pt);
		\draw[gray,fill=gray!100] (0,0,0) circle (1.5pt);
		
		\foreach \i in {-0.05,0,0.05}{ \draw[thick] (0,-1,\i) -- (0,1,\i);}

		\foreach \i in {-0.075,-0.025,0.025,0.075}{ \draw (-1.1,\i,2) -- (1.1,\i,2);}
		
		\foreach \i in {-0.045,-0.015,0.015,0.045}{ \draw (0,1.4*\i,0) -- (0,1.4*\i,2+\i);}
		\foreach \i in  {-0.075,-0.045,-0.015,0.015,0.045,0.075}{ \draw (0,1.4*\i,2+\i) -- (0,1.4*\i,4);}
		
		\node at (-0.18,-0.18,3.4) {\scriptsize $2N_5 K$};
		\node at (1.0,0.3,2) {\scriptsize $N_5$ D5};
		\node at (0,-1.25) {\footnotesize $N_5$ NS5};
		\node at (0.18,0.18,0.9) {{\scriptsize $N_5 K+\tfrac{N_5^2}{2}$}};
	\end{tikzpicture}
	\caption{Brane configuration for semi-infinite D3-branes ending on a combination of $N_5$ D5-branes and $N_5$ NS5-branes. These are the non-gravitating bath solutions of \cite{Uhlemann:2021nhu}.\label{fig:brane-D5NS5-D3}}
\end{figure}
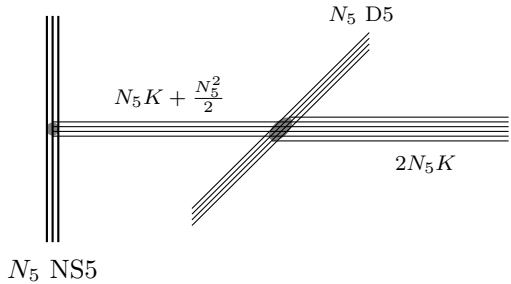

As a third example we turn to a BCFT setup with a 3d SCFT coupled to 4d $\cN=4$ SYM on a half space.
The brane setup comprises one group of $N_5$ D5-branes, one group of $N_5$ NS5-branes, and $2N_5K$ semi-infinite D3-branes ending on the 5-branes, as shown in fig.~\ref{fig:brane-D5NS5-D3}.  This setup was used in \cite{Uhlemann:2021nhu} to realize a black hole coupled to a non-gravitating bath and study the entropy of Hawking radiation, revealing a rich phase structure of entropy curves which was recently also explored from the braneworld model perspective in \cite{Geng:2021mic}.

The type of BCFT described by the setup in fig.~\ref{fig:brane-D5NS5-D3} depends on the value of $K$.
For each D5-brane the number of D3-branes to the right minus the number of D3-branes to the left is $K-N_5/2$.
For $K>N_5/2$, there are more D3-branes to the right of the D5-branes than to the left. 
Of the $2N_5K$ semi-infinite D3-branes, $N_5K-N_5^2/2$ end on the $N_5$ D5-branes.
This imposes Nahm pole boundary conditions on the corresponding parts of the adjoint scalars of the 4d $SU(N_5K)$ $\cN=4$ SYM node.
The remaining $N_5K+N_5^2/2$ D3-branes end on the NS5-branes. The corresponding part of the $\cN=4$ SYM adjoint fields is coupled to a 3d SCFT.
The field theory is 
\begin{align}\label{eq:D5NS5K-quiver-1}
	U(R)-U(2R)-\ldots - U((N_5-1)R) - \widehat{SU(2N_5K)}
\end{align}
where $R=K+N_5/2$.
For  $K<N_5/2$, on the other hand, the D5-branes have more D3-branes ending on them from the left than from the right. Using Hanany-Witten transitions the D5-branes can then be moved past NS5 branes until they have no more D3-branes attached. They end up in the 3d SCFT part of the brane construction and describe fundamental fields. We get a 3d SCFT with $N_5$ flavors at the $R^{\rm th}$ 3d node from the left, 
\begin{align}\label{eq:D5NS5K-quiver-2}
	U(R)-U(2R)-\ldots - &U(R^2) - U(R^2-S)-\ldots - U(2N_5K+S) - \widehat{SU(2N_5K)}
	\nonumber\\
	&\ \ \ \vert\\
	\nonumber & \ [N_5]	
\end{align}
where $S=N_5/2-K$.
Along the first/second ellipsis the rank increases in steps of $R$/decreases in steps of $S$.
Instead of D5-brane boundary conditions for part of the $\cN=4$ SYM fields, all 4d $\cN=4$ SYM fields are now coupled at the boundary to a 3d SCFT, which has larger-rank gauge groups and additional flavors compared to the previous case. The transition value $K=N_5/2$ will play a role in the supergravity duals as well.

The harmonic functions for the supergravity dual of the brane configuration in fig.~\ref{fig:brane-D5NS5-D3} are
\begin{align}\label{eq:h1h2-D5NS5-BCFT-z}
	h_1&=-\frac{i\pi\alpha'}{4}K e^z-\frac{\alpha'}{4}N_{5}\ln\tanh\left(\frac{i\pi}{4}-\frac{z}{2}\right)+\rm{c.c.}
\nonumber\\	
	h_2&=\frac{\pi\alpha'}{4}K e^z-\frac{\alpha'}{4}N_{5} \ln\tanh\left(\frac{z}{2}\right)+\rm{c.c.}
\end{align}
The brane configuration and supergravity solution are invariant under S-duality.
We will find it convenient to again use the coordinate transformation  (\ref{eq:coord-change}), which leads to
\begin{align}
	h_1&=-\frac{i\pi\alpha'}{4}K\left(\frac{1-u}{1+u}-\rm{c.c.}\right)-\frac{\alpha'}{4}N_{5} \ln\left|\frac{1-iu}{1+i u}\right|^2~,
	\nonumber\\
	h_2&=\frac{\pi\alpha'}{4}K\left(\frac{1-u}{1+u}+\rm{c.c.}\right)-\frac{\alpha'}{4}N_{5}\ln|u|^2~.
\end{align}
As holomorphic functions we choose
\begin{align}\label{eq:A12-BCFT}
	\cA_1&=\frac{\pi\alpha'}{4}K\frac{1-u}{1+u}+\frac{i\alpha'}{4}N_{5}\left[\ln(1+iu)-\ln(1-iu)\right]~,
	\nonumber\\
	\cA_2&=\frac{\pi\alpha'}{4}K\frac{1-u}{1+u}-\frac{\alpha'}{4}N_{5}\ln u~.
\end{align}
For the function $\cC$ defined in (\ref{eq:cC-def}) we then find
\begin{align}\label{eq:cC-BCFT}
	\cC=\,&-\frac{i{\alpha'}^2}{8}N_{5}^2\left[\Li_2(iu)-\Li_2(-iu)\right]-\frac{\pi{\alpha'}^2}{8}N_5 K \left(\ln(1+u^2)+\ln u-4\ln(1+u)\right)
	\nonumber\\ &
	-\left(\frac{\pi\alpha^\prime}{4}K\frac{1-u}{1+u}\right)^2-\cA_1\left(\cA_2-\frac{\pi\alpha'}{2}K\frac{1-u}{1+u}\right)\,.
\end{align}

\subsubsection{Fundamental Wilson loops}
We start with the BPS conditions for fundamental strings in (\ref{eq:F1-BPS}).
The requirement $h_1=0$ restricts admissible embeddings to the real axis. The condition $\partial h_2=0$ leads to
\begin{align}\label{eq:D5NS5K-F1}
	\frac{1}{u}+\frac{2\pi K}{N_5(u+1)^2}&=0~.
\end{align}
For $K\neq 0$ this equation has exactly one real solution in the interval $(-1,1)$. The 3d gauge nodes are all balanced; the fundamental Wilson loops associated with the 3d gauge nodes have enhanced scaling as discussed below (\ref{eq:W_fund}), and the corresponding strings are located at the NS5 pole, similar to the discussion in sec.~\ref{sec:balanced-hol}.
The fundamental string obtained from solving (\ref{eq:D5NS5K-F1}) represents the Wilson loop associated with the 4d $\cN=4$ SYM node (within the half space on which the $\cN=4$ SYM degrees of freedom propagate, the corresponding Wilson loop is still located on the boundary).

\subsubsection{\texorpdfstring{D5$^\prime$}{D5'} embeddings}

\begin{figure}
	\centering
	\subfigure[][]{\label{fig:WL-D5-NS5-K-a}
		\begin{tikzpicture}
			\node at (0,0) {\includegraphics[width=0.3\linewidth]{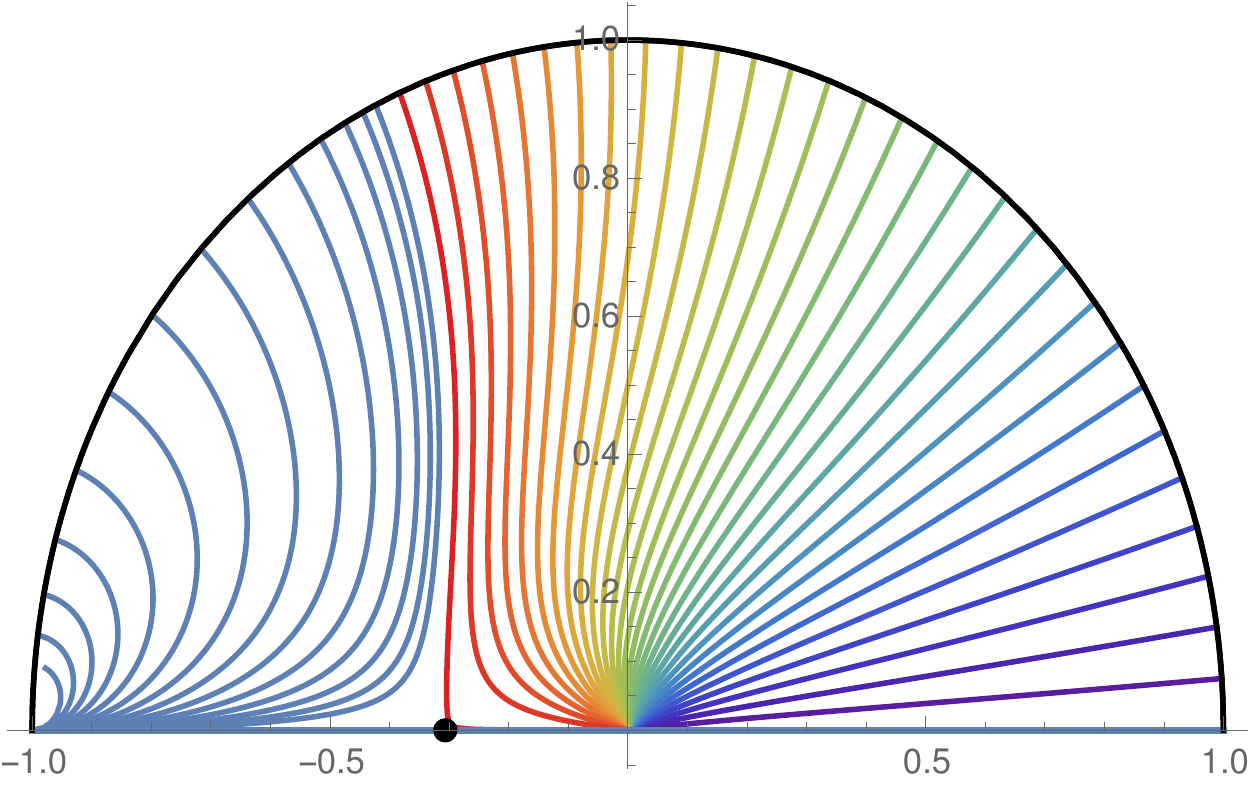}};
			\node at (0.7,0.7) {$\mathbf{\Sigma}$};
			\draw[very thick] (0,-1.2) -- (0,-1.45) node [anchor=north]{\ \ \ \footnotesize NS5};
			\draw[very thick] (0,1.2) -- (0,1.45) node [anchor=south]{\footnotesize D5};
			\draw[very thick] (-2.4,-1.4) -- (-2.2,-1.2);
			\node at (-2.35,-1.6) {\footnotesize D3};
			\node at (-0.75,-1.57) {\footnotesize F1};
		\end{tikzpicture}
	}\hspace*{-3.5mm}
	\subfigure[][]{
	\begin{tikzpicture}
		\node at (0,0) {\includegraphics[width=0.3\linewidth]{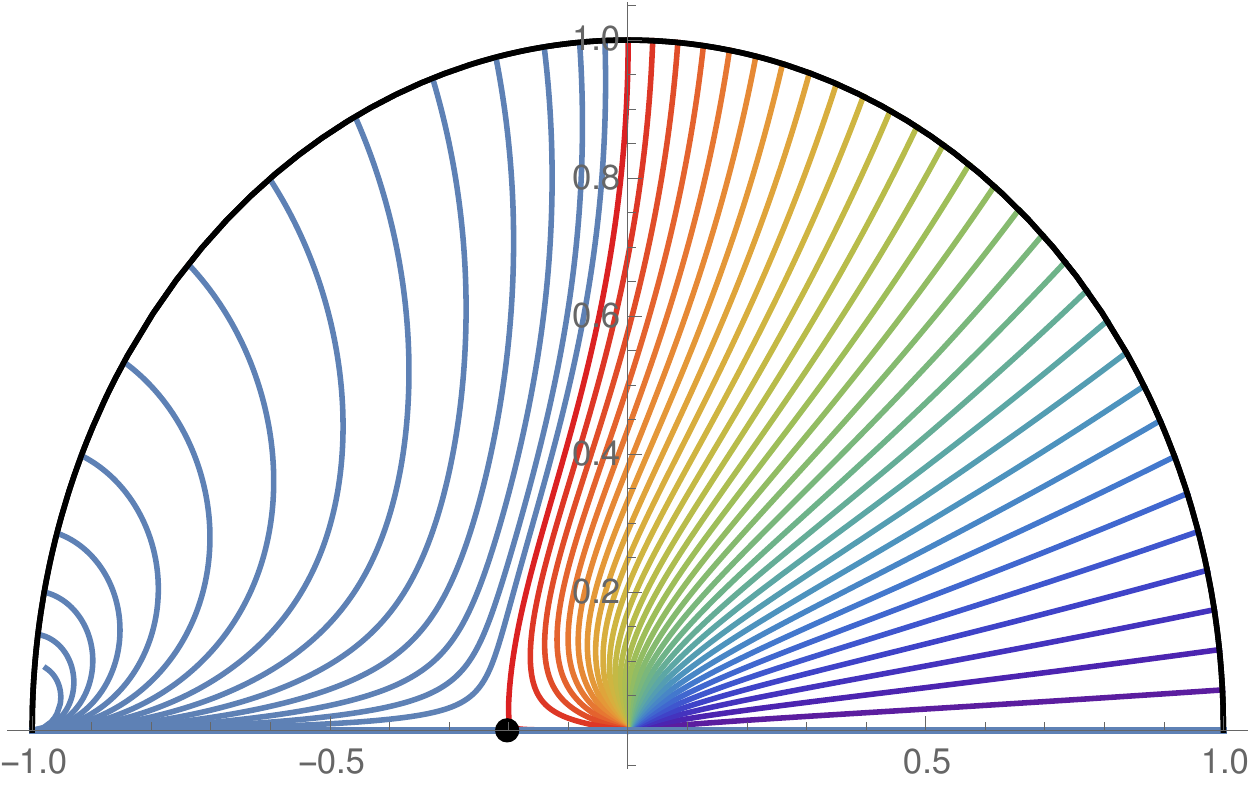}};
		\node at (0.7,0.7) {$\mathbf{\Sigma}$};
		\draw[very thick] (0,-1.2) -- (0,-1.45) node [anchor=north]{\ \ \ \footnotesize NS5};
		\draw[very thick] (0,1.2) -- (0,1.45) node [anchor=south]{\footnotesize D5};
		\draw[very thick] (-2.4,-1.4) -- (-2.2,-1.2);
		\node at (-2.35,-1.6) {\footnotesize D3};
		\node at (-0.5,-1.57) {\footnotesize F1};
	\end{tikzpicture}
	}\hspace*{-3.5mm}
	\subfigure[][]{
			\begin{tikzpicture}
		\node at (0,0) {\includegraphics[width=0.3\linewidth]{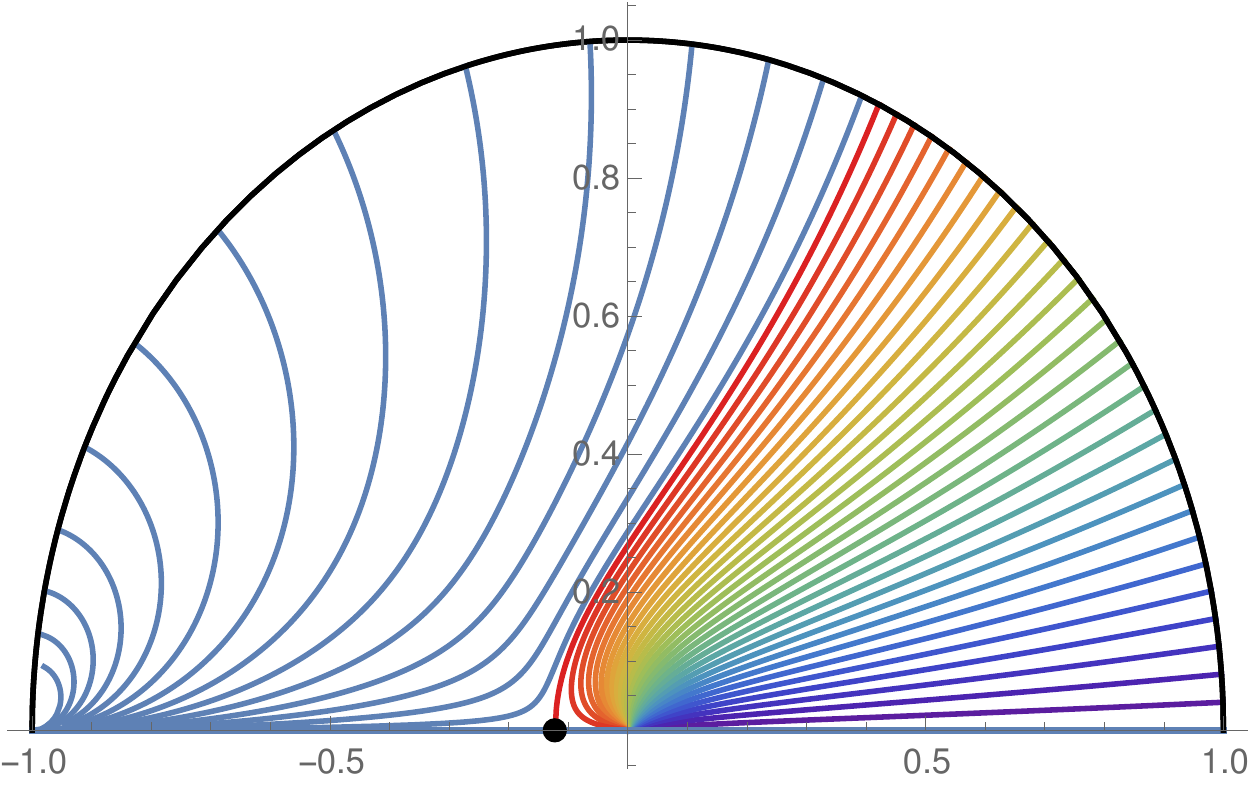}};
		\node at (0.7,0.7) {$\mathbf{\Sigma}$};
		\draw[very thick] (0,-1.2) -- (0,-1.45) node [anchor=north]{\ \ \ \footnotesize NS5};
		\draw[very thick] (0,1.2) -- (0,1.45) node [anchor=south]{\footnotesize D5};
		\draw[very thick] (-2.4,-1.4) -- (-2.2,-1.2);
		\node at (-0.37,-1.57) {\footnotesize F1};
		\node at (-2.35,-1.6) {\footnotesize D3};
	\end{tikzpicture}
	}
	\caption{Wilson loop embeddings from left to right for $K/N_5\in \lbrace \frac{1}{4},\frac{1}{2},1 \rbrace$. The D5/NS5 poles are at $u=0$/$u=i$. The semi-infinite D3-branes emerge at $u=-1$. The curves ending at the NS5 pole correspond to $\phi_0$ in the range $(0,\pi)$ and represent Wilson loop embeddings. The curves reaching to the AdS$_5\times$S$^5$ region emerging at $u=-1$ correspond to surface operator embeddings associated with the $\cN=4$ SYM node; they have $\phi_0>\pi$.	\label{fig:WL-D5-NS5-K}}
\end{figure}
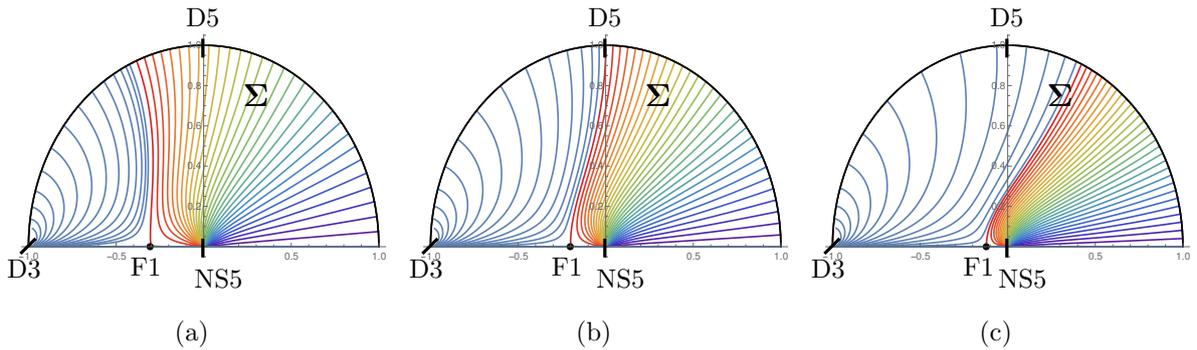

We now turn to antisymmetric Wilson loops represented by D5$^\prime$-branes.
The curves along which the D5$^\prime$-branes are embedded in $\Sigma$ are determined from the constraint that $h_2^D$ be constant, (\ref{eq:D5-BPS}). We use a constant $\phi_0$ to parametrize the embedding,
\begin{align}\label{eq:BCFT-curve}
	h_2^D&=\frac{1}{2}\alpha^\prime N_5 \phi_0~.
\end{align}
Using real coordinates this condition may be written as
\begin{align}\label{eq:r-phi-bcft}
	u&= re^{i\phi}~, & 
	\frac{\pi K}{2iN_5}\left(\frac{1-re^{i\phi}}{1+re^{i\phi}}-\frac{1-re^{-i\phi}}{1+re^{-i\phi}}\right)&=\phi-\phi_0~.
\end{align}
One may solve this condition in closed form as a quadratic equation for $r(\phi)$. However, we will only discuss limiting cases analytically.
For $r\rightarrow 0$, assuming that the embeddings reach the origin, we find $\phi\rightarrow \phi_0$. 
We therefore have $0<\phi_0<\pi$ for embeddings within $\Sigma$ which start at the origin.
These embeddings will be identified with Wilson loops; additional embeddings will be discussed below.
For $r\rightarrow 1$, (\ref{eq:r-phi-bcft}) becomes
\begin{align}
	\phi(r=1)&=\phi_0-\frac{\pi K}{N_5}\tan\left(\frac{\phi(r=1)}{2}\right)~.
\end{align}
This condition implies that for the curve with the maximal value of $\phi$ at $r\rightarrow 0$, $\phi_0=\pi$, the value of $\phi$ at $r=1$ is less than $\pi$ for non-zero $K$. There is thus a region of the boundary at $r=1$ which has no D5$^\prime$-brane embeddings that start at the origin ending on it.

Sample embeddings are shown in fig.~\ref{fig:WL-D5-NS5-K}. The plots show that the curves starting at $r=0$ discussed above do not reach to $u=-1$ ($\phi=\pi$ at $r=1$).
The end point $u_1$ of the D5$^\prime$ embeddings along the curves is determined in terms of the electric field by 
\begin{align}\label{eq:BCFT-u1}
	|F_{\rm el}|&=2|h_2(u_1)|~.
\end{align}
The figure shows additional embeddings which do not start at the origin and instead reach to the D3-brane source at $u=-1$; these correspond to $\phi_0>\pi$. These additional embeddings are analogous to the surface defect embeddings discussed for AdS$_5\times$S$^5$ in sec.~\ref{sec:AdS5S5}, and are associated with the 4d $\cN=4$ SYM node.
There is one embedding which starts at a regular point on the real line -- namely, at the location of the F1 string obtained from (\ref{eq:D5NS5K-F1}). 
This curve describes the Wilson loop at the ``last gauge node", which is the 4d $\cN=4$ SYM node.
The Wilson loop curve associated with the 4d node separates the region around $u=-1$ which corresponds to 4d $\cN=4$ SYM from the region around the 5-brane sources which corresponds to the 3d SCFT.
We may call the region with no Wilson loop embeddings in it a {\it Wilson loop shadow}.\footnote{The features resemble those of {\it entanglement shadows}, discussed e.g.\ in \cite{Balasubramanian:2017hgy}.}

The transition value $N_5=2K$, at which the form of the field theories changes from (\ref{eq:D5NS5K-quiver-1}) to (\ref{eq:D5NS5K-quiver-2}), plays a role here as well: for $2K>N_5$ the end points of the Wilson loop D5$^\prime$ embeddings on the half circle boundary component at $r=1$ do not reach past the D5-brane pole at $u=i$, while for $2K<N_5$ they do (as seen in fig.~\ref{fig:WL-D5-NS5-K}).

For $F_{\rm el}=0$ the D5$^\prime$ embeddings extend along the entire curves shown in fig.~\ref{fig:WL-D5-NS5-K}. For $F_{\rm el}\neq 0$ they still start either at the NS5-brane pole at $u=0$, at the location of the F1 string, or at the D3-brane sources at $u=-1$. But they cap off smoothly with the $S^1$ in $S_2^2$ collapsing before reaching the other end point of the curves.

\begin{figure}
	\centering
	\subfigure[][]{
		\includegraphics[width=0.3\linewidth]{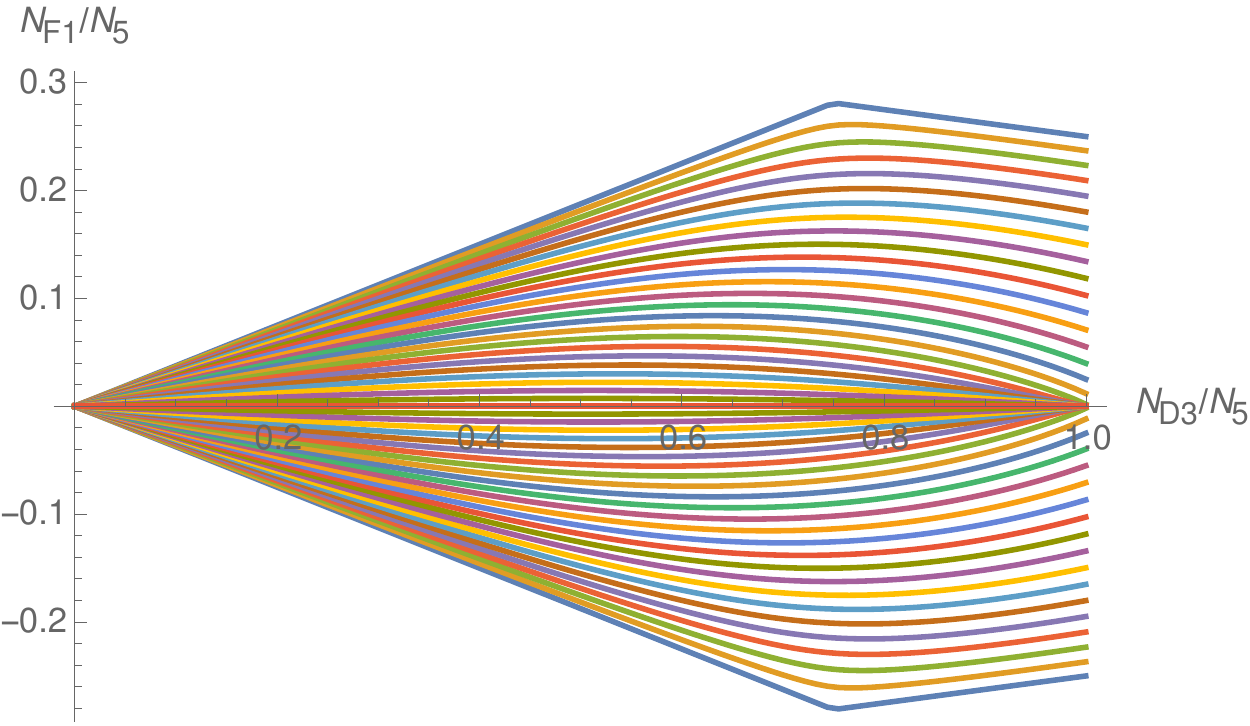}
	}\hspace*{3mm}
	\subfigure[][]{
		\includegraphics[width=0.3\linewidth]{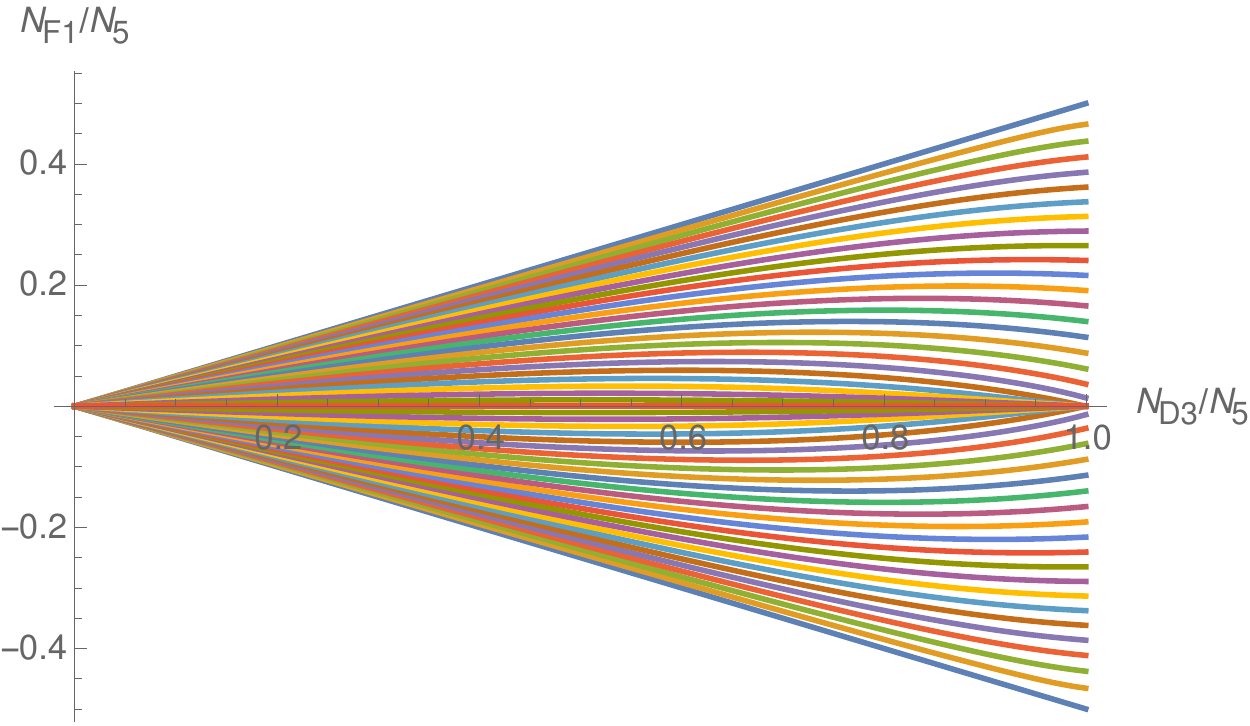}
	}\hspace*{3mm}
	\subfigure[][]{
		\includegraphics[width=0.3\linewidth]{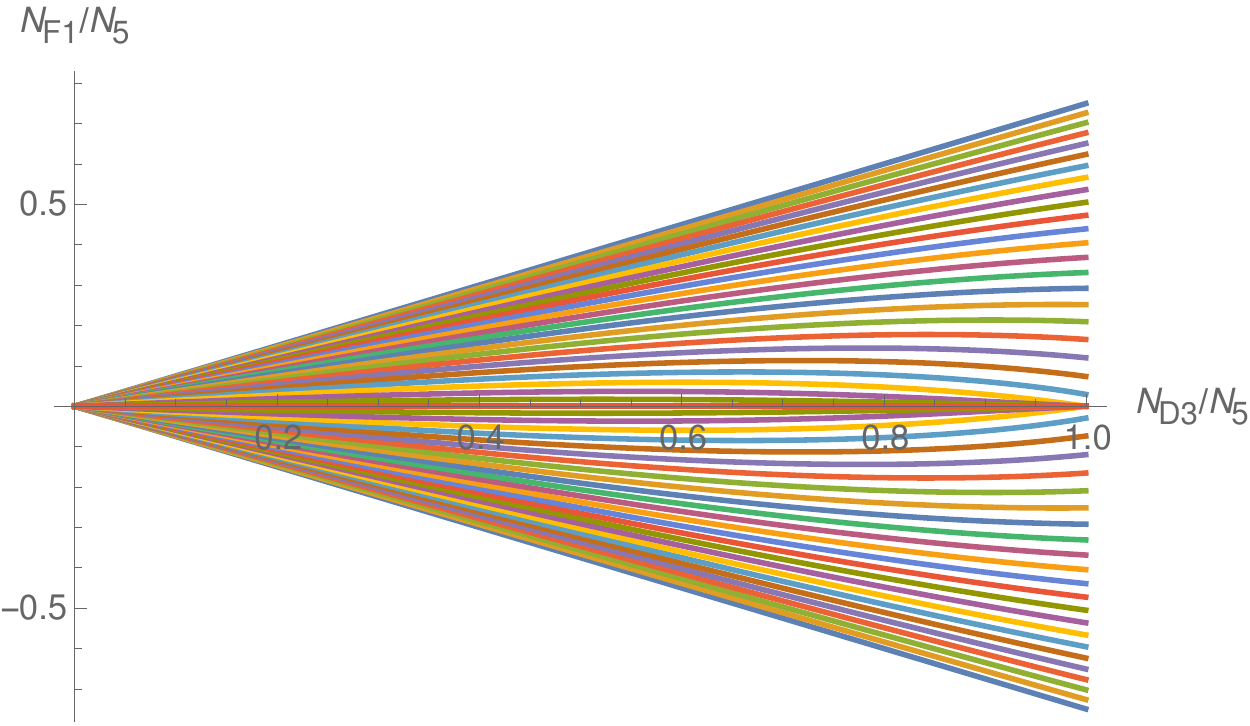}
	}
	\caption{$(N_{\rm D3},N_{\rm F1})$ charges for $K/N_5\in \lbrace \frac{1}{4},\frac{1}{2},1 \rbrace$ from left to right. For $K/N_5<\frac{1}{2}$ the D5-branes describe flavors in the (balanced) 3d quiver; for $K/N_5>\frac{1}{2}$ they provide boundary conditions for part of the 4d $\cN=4$ SYM fields.\label{fig:WL-D5-NS5-K-charges}}
\end{figure}

\subsubsection{Wilson loop expectation values}

We now collect the expressions for the Wilson loop expectation values. 
The D3-brane and F1 charges determined from (\ref{eq:ND3}), (\ref{eq:NF1}) are given by
\begin{align}\label{eq:BCFT-NF1-ND3}
	N_{\rm D3}&=\frac{\phi_0}{\pi}N_5~,
	\nonumber\\
	N_{\rm F1}&=\frac{4\lambda}{\pi^2{\alpha'}^2}\left[\Im\left(\cA_1\cA_2+\cC\right)\right]_{u_0}^{u_1}~,
\end{align}
with $\cC$ as given in (\ref{eq:cC-BCFT}) and the end point $u_1$ determined by (\ref{eq:BCFT-u1}).
The range of the D3-brane charges for Wilson loop embeddings is $N_{\rm D3}\in (0,N_5)$. This reflects the lengths of the quivers in (\ref{eq:D5NS5K-quiver-1}), (\ref{eq:D5NS5K-quiver-2}). The space of $(N_{\rm D3}, N_{\rm F1})$ charges is shown for example solutions in fig.~\ref{fig:WL-D5-NS5-K-charges}.
They precisely carve out the shape of the quivers (\ref{eq:D5NS5K-quiver-1}) and (\ref{eq:D5NS5K-quiver-2}).

To obtain the on-shell action we have to solve $\partial\cW=\cA_2^2\partial\cA_1$ for $\cW$. We find
\begin{align}
	\cW=\,\frac{{\alpha^\prime}^3}{32}\Bigg[&
	iN_5^3\left(\cL_3(u,i)-\cL_3(u,-i)\right)
	\nonumber\\ &
	-
	\pi K N_5^2\left(\cL_2(u,i)+\cL_2(u,-i)-4\cL_2(u,-1)
	+\frac{u\ln^2\! u}{1+u}\right)
	\nonumber\\ &	+\pi^2 K^2N_5\left(\tan^{-1}u+\frac{2u\ln u}{(1+u)^2}+\frac{4}{1+u}\right)
	+\frac{1}{6}\pi^3K^3\left(\frac{1-u}{1+u}\right)^3\Bigg]\, ,
\end{align}
where
\begin{align}
	\cL_2(u,\sigma)&=\Li_2(u\sigma)+\ln u \ln(1-u\sigma)~,
	\nonumber\\
	\cL_3(u,\sigma)&=\Li_3(u\sigma)-\ln u \Li_2(u\sigma)-\frac{1}{2}\ln^2\!u\,\ln(1-u\sigma)~.
\end{align}
With that expression the Wilson loop expectation value is given by
\begin{align}
\ln\langle W_\wedge\rangle&=
\frac{8}{\pi^2{\alpha'}^3}
\left[\Im\left(2\cA_1\cA_2^2-2\cW+ih_2^D(\cA_1\cA_2+\cC)\right)\right]_{u_0}^{u_1}~.
\end{align}
The points $u_0$ and $u_1$ are the start and end points of the curves (\ref{eq:BCFT-curve}), with $u_1$ determined by (\ref{eq:BCFT-u1}).
Together with the D3 and F1 charges in (\ref{eq:BCFT-NF1-ND3}), which identify the gauge node and rank of the representation as in the previous example, this gives the complete set of $\tfrac{1}{2}$-BPS Wilson loop expectation values.

\subsection{3d \texorpdfstring{D5$^2$/NS5$^2$}{D5**2/NS5**2} theories}\label{sec:D52NS52}

\begin{figure}
	\centering
	\begin{tikzpicture}[y={(0cm,1cm)}, x={(0.707cm,0.707cm)}, z={(1cm,0cm)}, scale=1.1]
	\draw[gray,fill=gray!100] (0,0,-0.5) circle (1.8pt);
	\draw[gray,fill=gray!100] (0,0,1) ellipse (1.8pt and 3pt);
	\draw[gray,fill=gray!100,rotate around={-45:(0,0,2.5)}] (0,0,2.5) ellipse (1.8pt and 3.5pt);
	\draw[gray,fill=gray!100] (0,0,4) circle (1.8pt);				
	
	\foreach \i in {-0.05,0,0.05}{ \draw[thick] (0,-1,-0.5+\i) -- (0,1,-0.5+\i);}
	\foreach \i in {-0.05,0,0.05}{ \draw[thick] (0,-1,1+\i) -- (0,1,1+\i);}

	\foreach \i in {-0.075,-0.025,0.025,0.075}{ \draw (-1.1,\i,2.5) -- (1.1,\i,2.5);}
	\foreach \i in {-0.075,-0.025,0.025,0.075}{ \draw (-1.1,\i,4) -- (1.1,\i,4);}
	
	\foreach \i in {-0.03,0,0.03}{ \draw (0,1.4*\i,-0.5) -- (0,1.4*\i,1);}
	\foreach \i in {-0.075,-0.045,-0.015,0.015,0.045,0.075}{ \draw (0,1.4*\i,1) -- (0,1.4*\i,2.5+\i);}
	\foreach \i in {-0.03,0,0.03}{ \draw (0,1.4*\i,2.5) -- (0,1.4*\i,4);}
	
	\node at (0,-1.25,-0.5) {\footnotesize $N$ NS5};
	\node at (0,-1.25,1) {\footnotesize $N$ NS5};
	\node at (1.0,0.35,2.5) {\scriptsize  $N$ D5};
	\node at (1.0,0.35,4) {\scriptsize  $N$ D5};
	\node at (0.2,0.2,1.75) {{\scriptsize $2N^2$}};
	\node at (0,0.22,0.25) {{\scriptsize $N^2\Delta$}};
	\node at (0,0.22,3.5) {{\scriptsize $N^2\Delta$}};
\end{tikzpicture}
\caption{D3-branes suspended between two groups of D5-branes and two groups of NS5-branes. The associated supergravity solutions is given by (\ref{eq:h1h2-3d-grav}).\label{fig:D52NS52-brane}}
\end{figure}
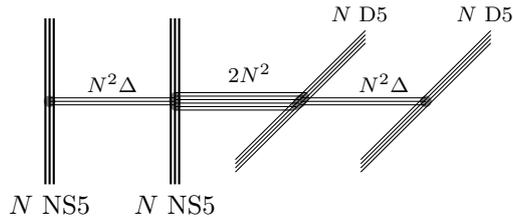

As a last example we consider a class of 3d SCFTs with an unbalanced central node. The supergravity duals were used to study information transfer from a black hole to a gravitating bath in \cite{Uhlemann:2021nhu}.
The solutions also serve as a string theory realization of the wedge holography proposal of \cite{Akal:2020wfl}, as discussed in \cite{Uhlemann:2021nhu}. We will return to this discussion below.
The brane construction involves two groups of NS5 branes, with $N$ branes in each, and two groups of D5-branes with $N$ branes in each.
The brane configuration is shown in fig.~\ref{fig:D52NS52-brane}. A total of $2N^2$ D3-branes is suspended between the 5-branes, with a parameter
\begin{align}
\Delta&=\frac{1}{2}+\frac{2}{\pi}\arctan e^{-2\delta}
\end{align}
taking values in $(\frac{1}{2},1)$ controlling how the D3-branes terminate on the 5-branes.
The harmonic functions corresponding to this brane configuration are 
\begin{align}\label{eq:h1h2-3d-grav}
	h_1&=-\frac{\alpha^\prime}{4}N\left[\ln\tanh\left(\frac{i\pi}{4}-\frac{z-\delta}{2}\right)
	+\ln\tanh\left(\frac{i\pi}{4}-\frac{z+\delta}{2}\right)\right]+\rm{c.c.}
	\nonumber\\
	h_2&=-\frac{\alpha^\prime}{4}N\left[\ln\tanh\left(\frac{z-\delta}{2}\right)+
	\ln	\tanh\left(\frac{z+\delta}{2}\right)\right]+\rm{c.c.}
\end{align}

The 3d gauge theory engineered by this brane configuration has $2N-1$ nodes, with $N$ flavors at the node ${\sf s}=\Delta N$ and $N$ flavors at the node ${\sf t}=2N-{\sf s}=(2-\Delta) N$. The quiver is given by
\begin{align}\label{eq:D52NS52-quiver}
	U({\sf s})-\ldots -U&({\sf s}^2)- \ldots -U(X+{\sf s})- U(X) - U(X+{\sf s}) -\ldots - U({\sf s}^2) -\ldots -U({\sf s})
	\nonumber\\
	&\ \,\vert\hskip 92mm \vert
	\\
	&[N]\nonumber\hskip 87mm [N]
\end{align}
where $X={\sf s}^2-(N-{\sf s})^2$ and $U(X)$ is the central node.
Along the first ellipsis the rank increases in steps of ${\sf s}$, along the second ellipsis it decreases in steps of $N-{\sf s}$.
The quiver is symmetric under reflection across the central node.
Along the third ellipsis the rank increases in steps of $N-{\sf s}$ and along the fourth it decreases in steps of ${\sf s}$.

With an unbalanced central node we expect a fundamental Wilson loop with regular scaling at that node, following the comments below (\ref{eq:W_fund}). There is indeed one embedding for fundamental strings at a regular point on $\partial\Sigma$, namely at $z=0$. 
The on-shell action obtained from (\ref{eq:S-BPS-F1}) evaluates to
\begin{align}\label{eq:NS52D52-WF}
	\ln\langle W_F\rangle&=S_{\rm F1}=-2N \ln\tanh\left(\frac{\delta}{2}\right)~.
\end{align}
This matches the expectation value of the fundamental Wilson loop associated with the central node of the quiver (\ref{eq:D52NS52-quiver}) obtained from supersymmetric localization, (\ref{eq:WF-unbalanced}).
The F1 strings representing Wilson loops associated with all other gauge nodes are at the NS5 poles, and have logarithmically enhanced scaling, in line with the discussion below (\ref{eq:W_fund}).

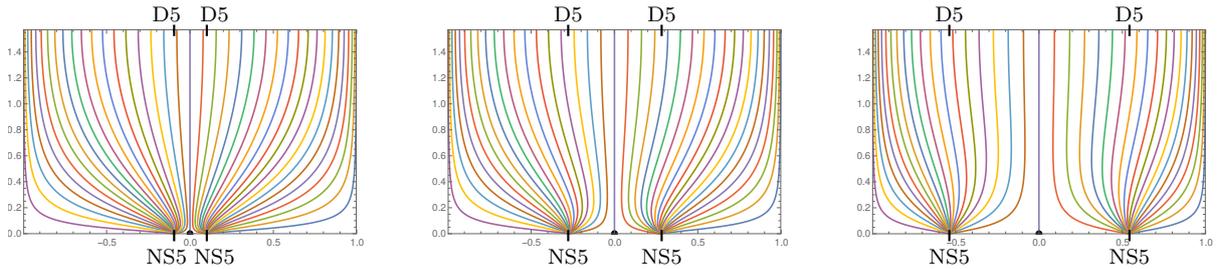
\begin{figure}
	\centering
	\begin{tikzpicture}
		\node at (0,0) {\includegraphics[width=0.29\linewidth]{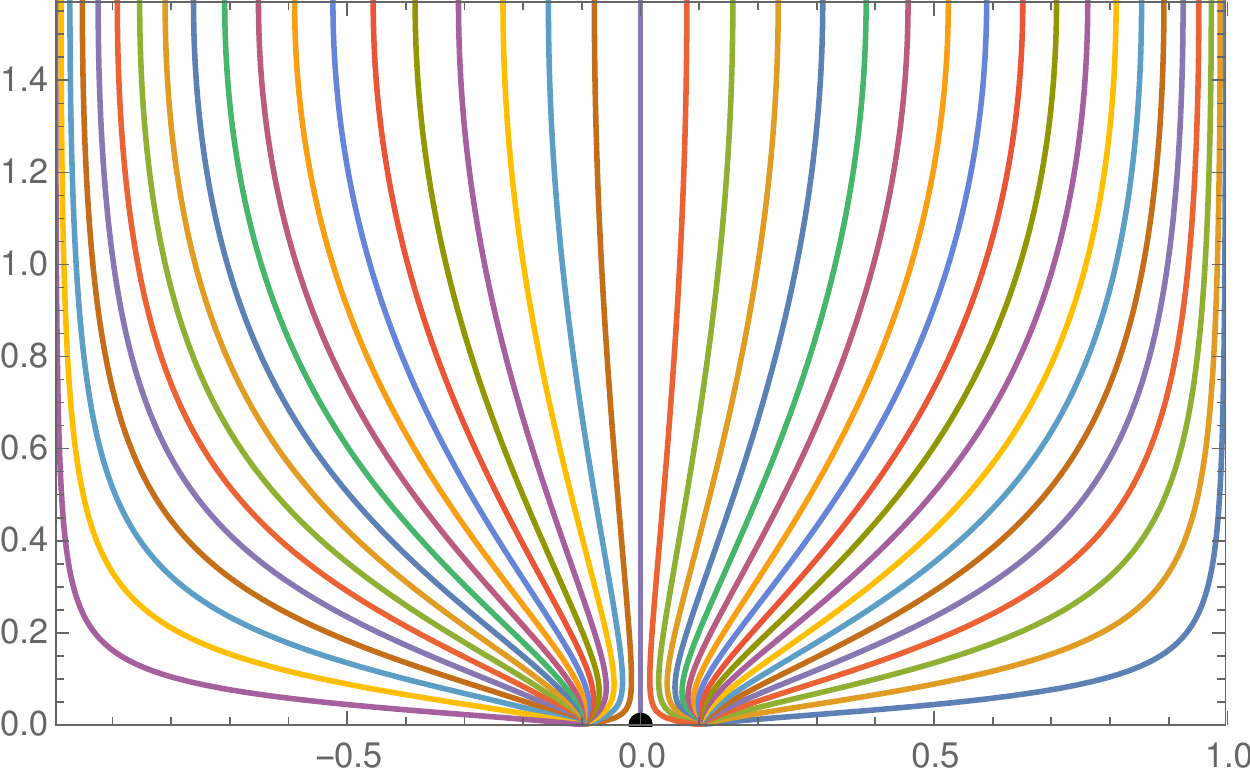}};
		\draw [thick] (-0.15,-1.2) -- +(0,-0.15) node [anchor=north, yshift=0.5mm,xshift=-1mm] {\scriptsize NS5};
		\draw [thick] (0.28,-1.2) -- +(0,-0.15) node [anchor=north, yshift=0.5mm,xshift=1mm] {\scriptsize NS5};
		
		\draw [thick] (-0.15,1.52) -- +(0,-0.15) node [anchor=south, yshift=0.5mm,xshift=-1mm] {\scriptsize D5};
		\draw [thick] (0.28,1.52) -- +(0,-0.15) node [anchor=south, yshift=0.5mm,xshift=1mm] {\scriptsize D5};
	\end{tikzpicture}
	\hfill
	\begin{tikzpicture}
		\node at (0,0) {\includegraphics[width=0.29\linewidth]{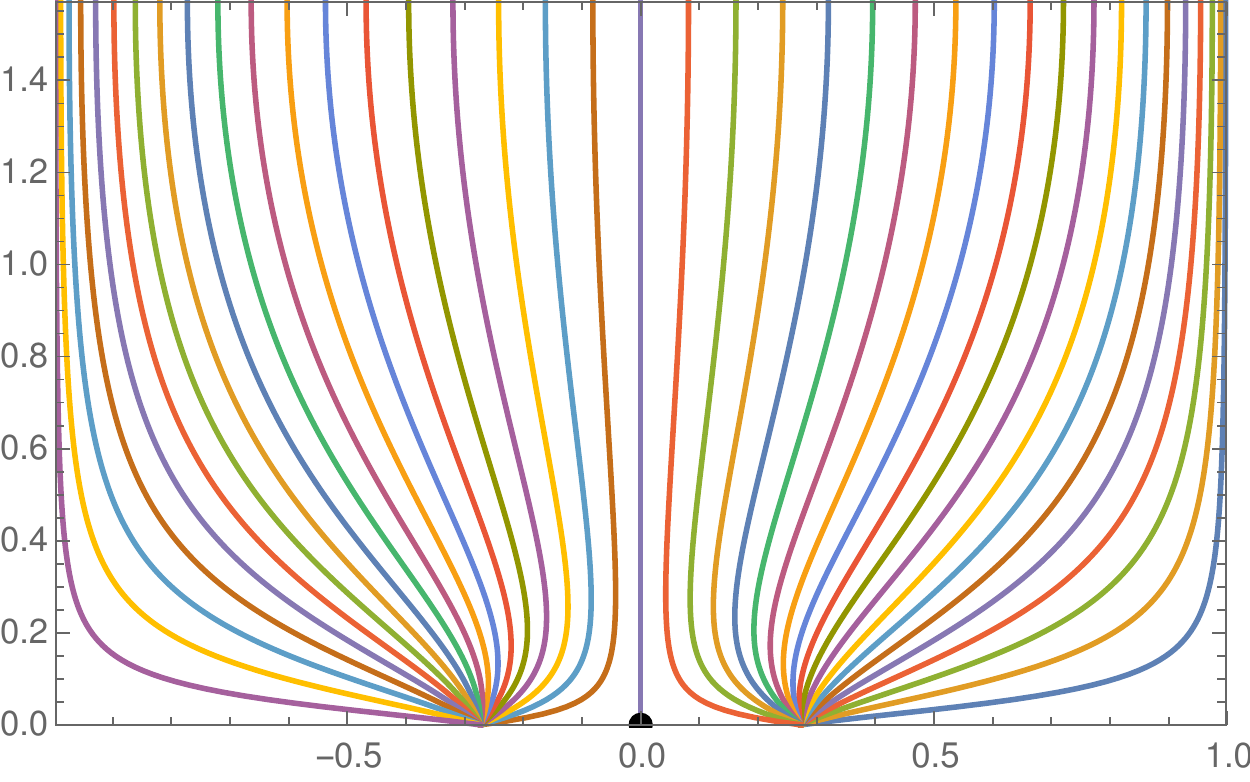}};
		\draw [thick] (-0.15-0.4,-1.2) -- +(0,-0.15) node [anchor=north, yshift=0.5mm] {\scriptsize NS5};
		\draw [thick] (0.28+0.4,-1.2) -- +(0,-0.15) node [anchor=north, yshift=0.5mm] {\scriptsize NS5};

		\draw [thick] (-0.15-0.4,1.52) -- +(0,-0.15) node [anchor=south, yshift=0.5mm] {\scriptsize D5};
		\draw [thick] (0.28+0.4,1.52) -- +(0,-0.15) node [anchor=south, yshift=0.5mm] {\scriptsize D5};

	\end{tikzpicture}
	\hfill
	\begin{tikzpicture}
		\node at (0,0) {\includegraphics[width=0.29\linewidth]{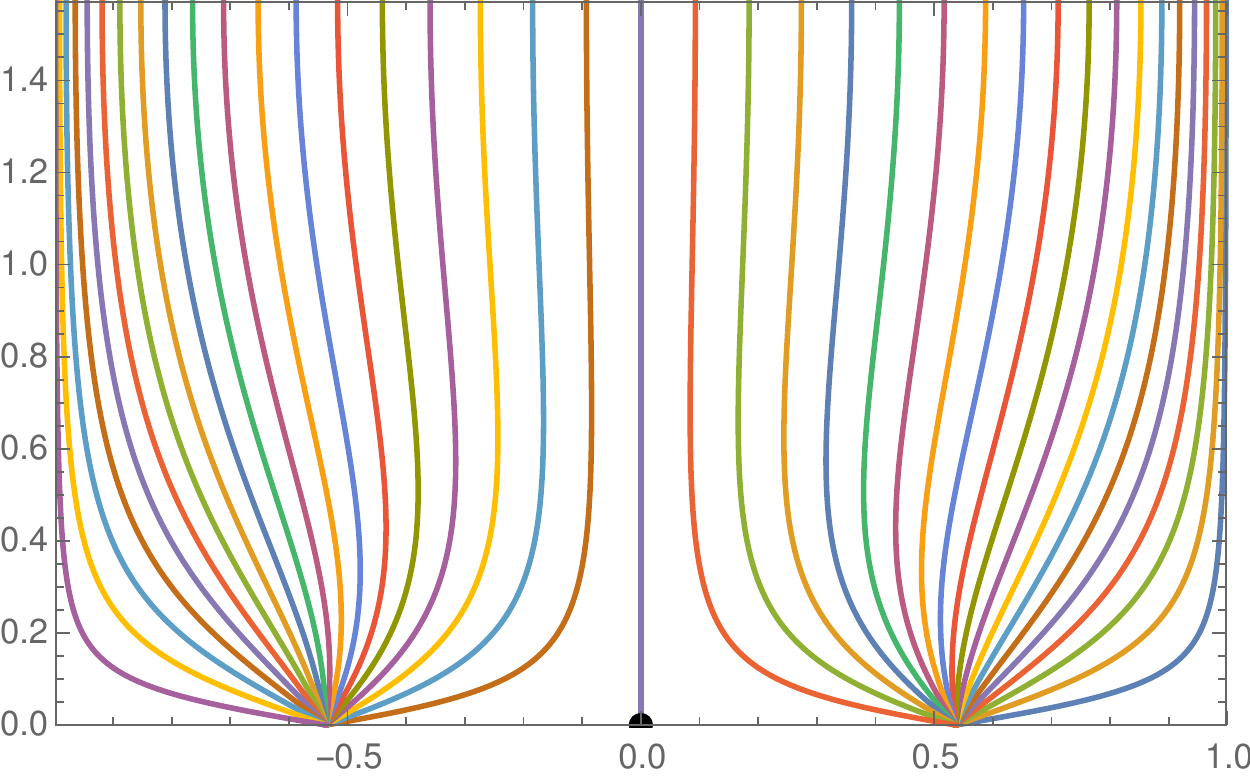}};
		
		\draw [thick] (-0.15-0.97,-1.2) -- +(0,-0.15) node [anchor=north, yshift=0.5mm] {\scriptsize NS5};
		\draw [thick] (0.28+0.97,-1.2) -- +(0,-0.15) node [anchor=north, yshift=0.5mm] {\scriptsize NS5};
		
		\draw [thick] (-0.15-0.97,1.52) -- +(0,-0.15)  node [anchor=south, yshift=0.5mm] {\scriptsize D5};
		\draw [thick] (0.28+0.97,1.52) -- +(0,-0.15)  node [anchor=south, yshift=0.5mm] {\scriptsize D5};
		
	\end{tikzpicture}
\caption{D5$^\prime$ Wilson loop embeddings for the D5$^2$/NS5$^2$ theory on $\Sigma$, with $\tanh(\Re(z))$ on the horizontal axis and $\Im(z)$ on the vertical axis, from left to right for $\delta\in\lbrace 0.1,0.28,0.6\rbrace$.\label{fig:D52NS52-WLD5}}
\end{figure}

For the D5$^\prime$ branes describing antisymmetric Wilson loops we will content ourselves with a numeric discussion in this section.
The Wilson loop D5$^\prime$ branes are, as before, embedded along curves with constant $h_2^D$,
\begin{align}
	h_2^D=\pi\alpha'N c~,
\end{align}
with a constant $c$ normalized such that the D3-brane charge from (\ref{eq:ND3}) is given by $2N c$.
Sample embeddings are shown in fig.~\ref{fig:D52NS52-WLD5}.
Except for one curve, the embeddings start at one of the two NS5 poles at $z=\pm \delta$, depending on the value of $c$. 
The central curve, along $\Re(z)=0$, starts and ends at regular boundary points.
On the lower boundary component it starts at the point where the fundamental string with action (\ref{eq:NS52D52-WF}) is located.
For given $c$ one can choose the end point along the curve freely, and determine the corresponding electric field from the BPS condition (\ref{eq:D5-BPS}). 
The F1 charge, fixing the representation of the Wilson loop, is then given by the integral in (\ref{eq:NF1-int}), and the expectation value by the integral in (\ref{eq:I-int}). 

For a sample of theories and Wilson loops we compared the holographic results to the expectation values obtained from the general field theory expressions (\ref{eq:b_cont}), (\ref{eq:anti_sym_2}) with the eigenvalue density (\ref{eq:varrho_unb_unb}), and found the results to agree.

\paragraph{Information transfer between black holes:}
The D5$^2$/NS5$^2$ solutions (\ref{eq:h1h2-3d-grav}) were used in \cite[sec.~VI]{Uhlemann:2021nhu} to study information transfer between two black holes. 
We review the basic logic briefly, to discuss the results obtained here in that context.
In \cite{Uhlemann:2021nhu} the AdS$_4$ factor in the geometry (\ref{eq:ds2-IIB}) was replaced with a black hole. 
Two subsystems were then defined by splitting the strip $\Sigma$ into the halves with $\Re(z)>0$ and $\Re(z)<0$. 
The entanglement entropy arising from this split was used to quantify the information exchanged between the two subsystems.
This is a string theory version of the analysis in \cite{Geng:2020fxl}, where a braneworld model as shown in fig.~\ref{fig:braneworld-wedge} was studied, with the two subsystems corresponding to the two ETW branes.
The analysis in \cite{Uhlemann:2021nhu} identified a critical value $\delta_c$, with  $\delta_c\approx 0.29$, which separates setups with $\delta<\delta_c$ where the entanglement entropy is time-independent from setups with $\delta_c>\delta_c$ where the entropy follows a non-trivial Page curve.

The plots in fig.~\ref{fig:D52NS52-WLD5} show that the curve $\Re(z)=0$, which divides $\Sigma$ into the two subsystems $\Re(z)>0$ and $\Re(z)<0$, corresponds to the central gauge node. This supports the interpretation of splitting $\Sigma$ put forth in \cite{Uhlemann:2021nhu}, which is that splitting $\Sigma$ along $\Re(z)=0$ corresponds to decomposing the quiver diagram (\ref{eq:D52NS52-quiver}) by cutting it at the central node.
The plots further visualize the qualitative difference between setups with small $\delta$ and those with large $\delta$:  For small $\delta$, splitting the quiver leads to two systems connected by a relatively large number of `bridge' degrees of freedom at the central node.
For large $\delta$, on other hand, one obtains two sectors, with a large number of degrees of freedom in each sector, which are linked by a relatively small number of bridge degrees of freedom.
Correspondingly, in the plots in fig.~\ref{fig:D52NS52-WLD5} the curve $\Re(z)=0$ representing the central node gets more and more isolated for larger $\delta$. The geometry approaches the form of two separate solutions connected by a small bridge.
In the wedge holography picture in fig.~\ref{fig:braneworld-wedge} the regime with large/small $\delta$ corresponds to small/large brane angles.

For the holographic entanglement entropy computations $\Sigma$ has to be split along an extremal curve for corresponding minimal surfaces to exist (as a result of the general constraints found in \cite{Graham:2014iya}). This constrains admissible splits, and the split along the curve $\Re(z)=0$ is one admissible choice. This curve is also distinguished from the perspective of loop operators: For vortex loop operators represented by NS5$^\prime$ branes, the corresponding probe NS5$^\prime$ embeddings can be obtained by a vertical reflection of the plots in fig.~\ref{fig:D52NS52-WLD5}. The only Wilson loop D5$^\prime$ whose mirror-dual NS5$^\prime$ extends along the same curve on $\Sigma$ is the one associated with the central node, with the curve being $\Re(z)=0$. It would be interesting to understand whether there is any connection.

Finally, we note that certain distinguished values for $\delta$ appear naturally from a geometric perspective. For example, the combination $h_1h_2$ is non-negative on $\Sigma$, with one maximum for small $\delta$ and two maxima for large $\delta$. The `critical value' separating the two cases is $\delta=\csch^{-1}\sqrt{2}$, or $\Delta=\frac{2}{3}$.
It would be interesting if the value $\delta_c$ separating the two shapes of the entropy curves found in \cite{Uhlemann:2021nhu} could be understood from such a perspective.

\section{Connection to braneworld models}\label{sec:braneworld}

\begin{figure}
	\centering
	\subfigure[][]{\label{fig:braneworld-BCFT}
	\begin{tikzpicture}[scale=0.8]
		\draw (-2.5,0) -- (2.5,0);	
		
		\draw[thick] (-0.8,0) -- (-2.5,-2/3*2.5);
		
		\draw [white,fill=gray,opacity=0.3] (-0.8,0) -- (-2.5,0) -- (-2.5,-2/3*2.5)--(-0.8,0);
		
		\node at (-0.6-0.8,-0.22) {\footnotesize $\theta$};
		\draw (-0.9-0.8,0) arc (180:225:25pt);
		\node at (3.4,0) {\footnotesize $\partial$AdS$_5$};	
		\node at (1,0.35) {\footnotesize $\RR_+\times \RR^{3}$};
		
		\node at (1,-1.2) {\small AdS$_5$};
		\node[rotate=45] at (-1.5,-1.1) {\footnotesize ETW brane};
	\end{tikzpicture}
	}\hskip 15mm
	\subfigure[][]{\label{fig:braneworld-wedge}
	\begin{tikzpicture}[scale=0.8]
		\draw (-2.5,0) -- (0.9,0);	
		
		\draw[thick] (-0.8,0) -- (-2.5,-2/3*2.5);
		
		\draw [white,fill=gray,opacity=0.3] (-0.8,0) -- (-2.5,0) -- (-2.5,-2/3*2.5)--(-0.8,0);
		
		\draw[thick] (-0.8,0) -- (0.9,-2/3*2.5);
		
		\draw [white,fill=gray,opacity=0.3] (-0.8,0) -- (0.9,0) -- (0.9,-2/3*2.5)--(-0.8,0);

		\node at (-0.6-0.8,-0.22) {\footnotesize $\theta_1$};
		\draw (-0.9-0.8,0) arc (180:225:25pt);
		
		\node at (0.6-0.8,-0.22) {\footnotesize $\theta_2$};
		\draw (0.9-0.8,0) arc (0:-45:25pt);
		
		\node at (1.8,0) {\footnotesize $\partial$AdS$_5$};	
		\node at (-0.8,0.35) {\footnotesize $ \RR^{3}$};
		
		\node at (-0.8,-1.3) {\small AdS$_5$};
		\node[rotate=45,white] at (-1.5,-1.1) {\footnotesize ETW brane};
		
	\end{tikzpicture}
	}
	\caption{Left: braneworld realization of a holographic dual for a 4d BCFT. AdS$_5$ is cut off by an end-of-the-world brane, leaving a half space of the conformal boundary. Right: wedge holography dual for a 3d CFT involving two end-of-the-world branes.}
\end{figure}
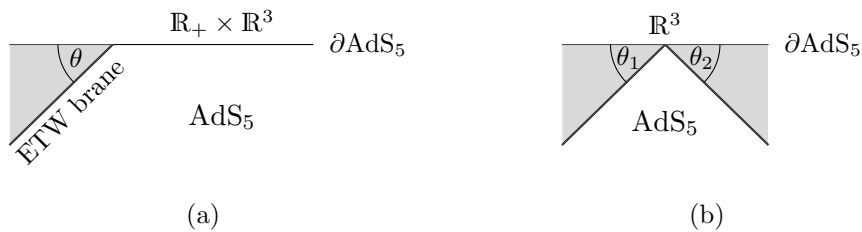

We now use the Wilson loop discussion to make concrete connections between 10d supergravity duals for 4d BCFTs, Janus CFTs, and 3d CFTs on the one hand, and bottom-up braneworld models dual to such field theories on the other.
We start with braneworld models for 4d BCFTs, fig.~\ref{fig:braneworld-BCFT}, in which an AdS$_5$ bulk is cut off by an end-of-the-world (ETW) brane, so that a half space remains of the conformal boundary. The ETW brane is introduced as effective description for the 3d boundary degrees of freedom to which the 4d ambient CFT is coupled. 
The brane angle $\theta$ is set by the brane tension, and encodes the number of 3d defect degrees of freedom relative to the number of 4d ambient degrees of freedom. The related construction of wedge holography, which realizes duals for 3d CFTs with two separate sectors represented by two ETW branes, is shown in fig.~\ref{fig:braneworld-wedge}.

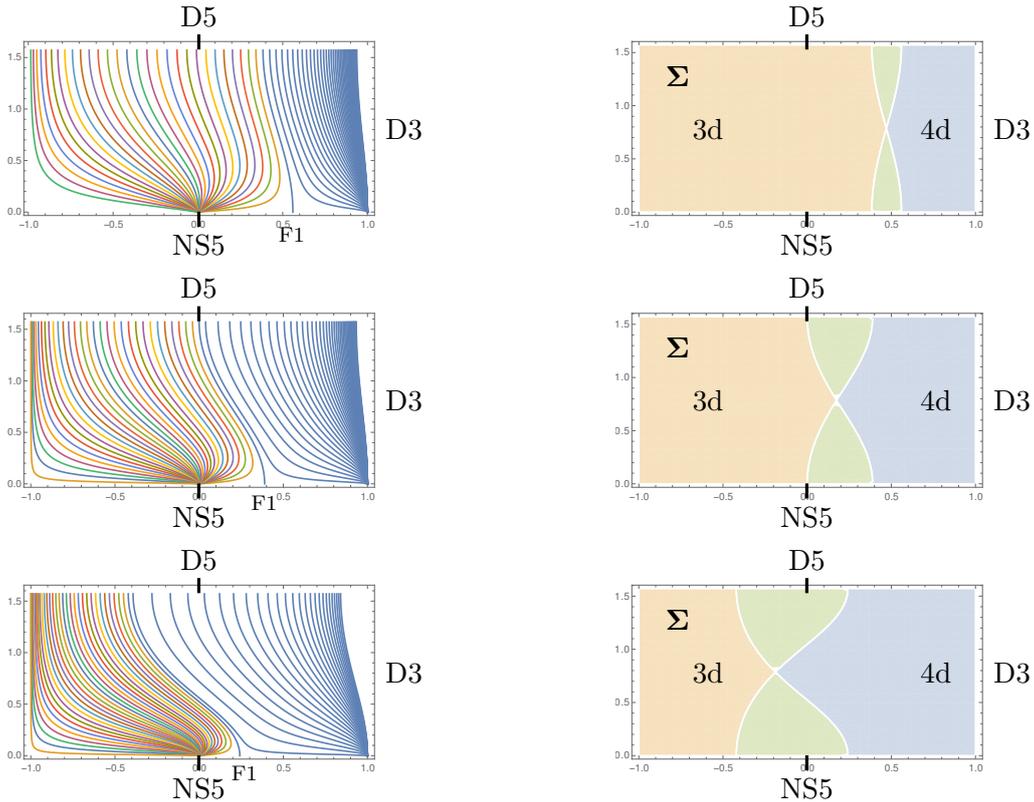
\begin{figure}
	\centering
	\begin{tikzpicture}
		\node at (-4,0){\includegraphics[width=0.3\linewidth]{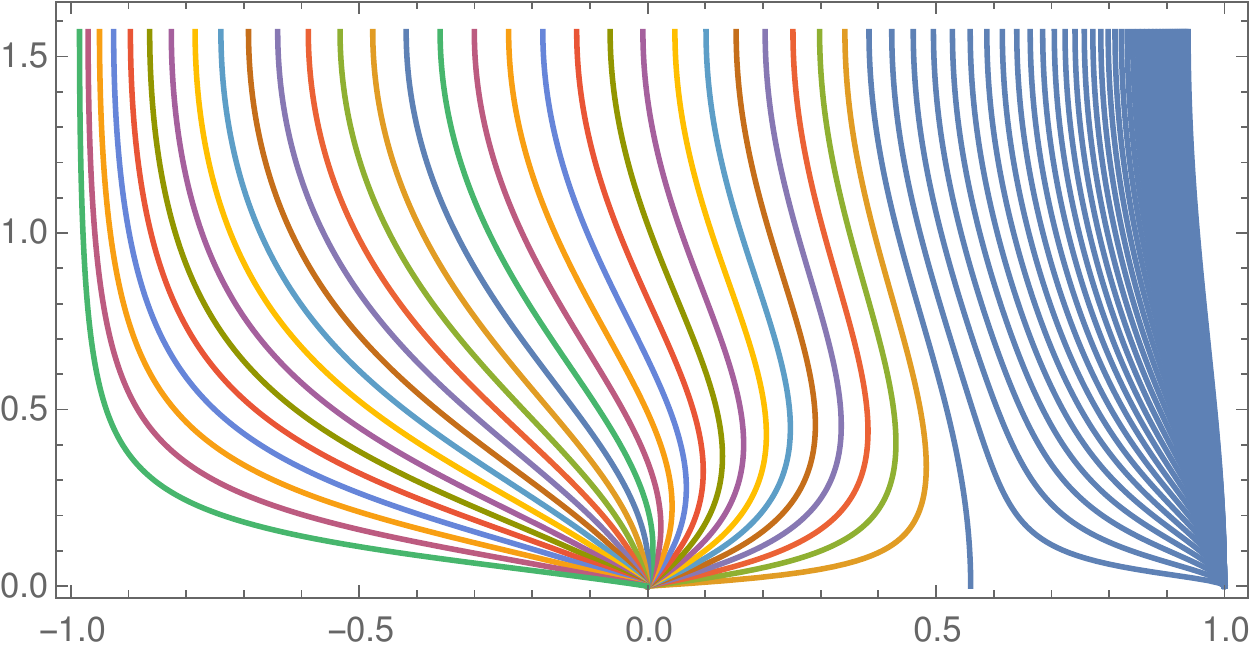}};
		\node at (4,0){\includegraphics[width=0.3\linewidth]{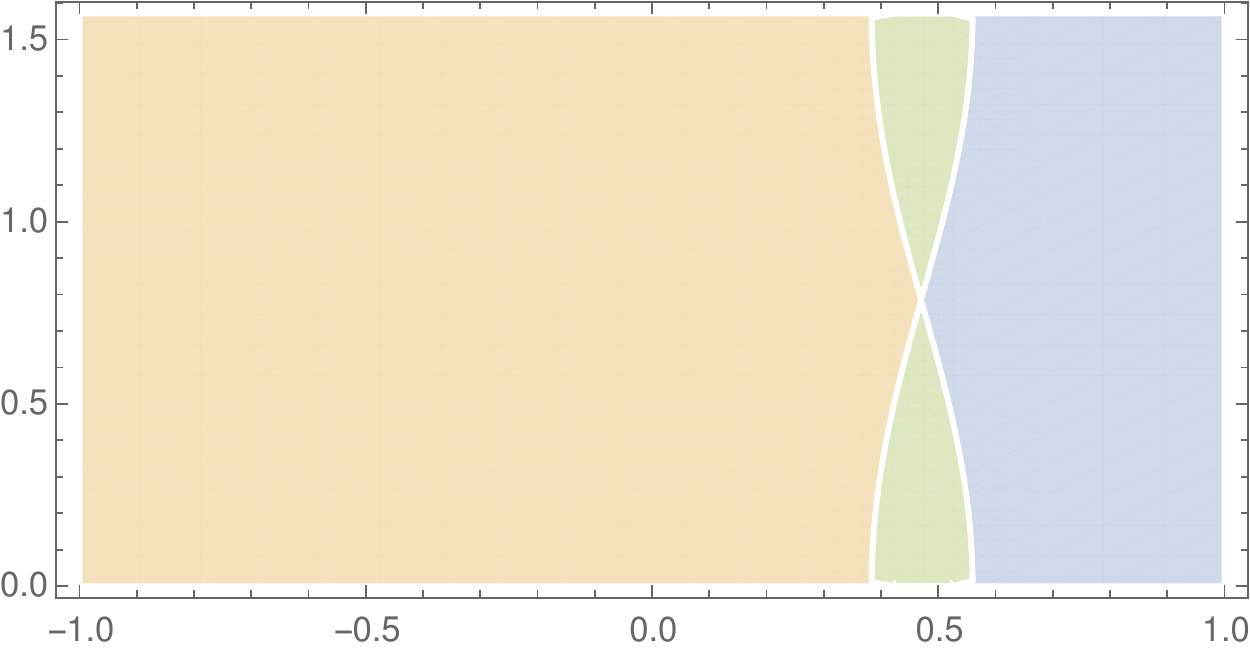}};
		\node at (5.8,0.1) {\small 4d};
		\node at (2.8,0.1) {\small 3d};
		\node at (2.4,0.8) {\small {\boldmath{$\Sigma$}}};
		
		\draw[very thick] (-4+0.1,-1.1+0.1) -- (-4+0.1,-1.1-0.1) node [anchor=north,yshift=0.5mm] {\small NS5};
		\draw[very thick] (-4+0.1,1.25-0.1) -- (-4+0.1,1.25+0.1) node [anchor=south,yshift=-0.5mm] {\small D5};
		\node at (-1.2,0.1) {\small D3};
		
		\draw[very thick] (4+0.1,-1.1+0.1) -- (4+0.1,-1.1-0.1) node [anchor=north,yshift=0.5mm] {\small NS5};
		\draw[very thick] (4+0.1,1.25-0.1) -- (4+0.1,1.25+0.1) node [anchor=south,yshift=-0.5mm] {\small D5};
		\node at (6.8,0.1) {\small D3};
		
		\node [anchor=north,yshift=1.4mm] at  (-4+1.32,-1.1-0.1) {\scriptsize F1};
	\end{tikzpicture}
	\\
	\begin{tikzpicture}
		\node at (-4,0){\includegraphics[width=0.3\linewidth]{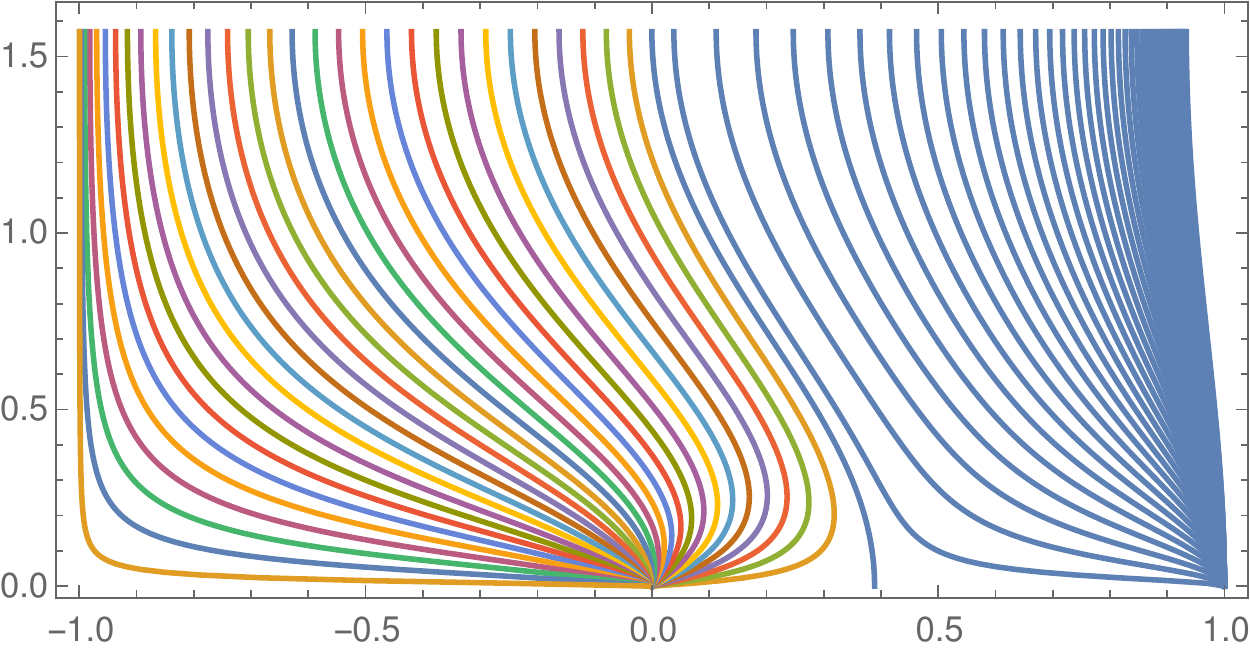}};
		\node at (4,0){\includegraphics[width=0.3\linewidth]{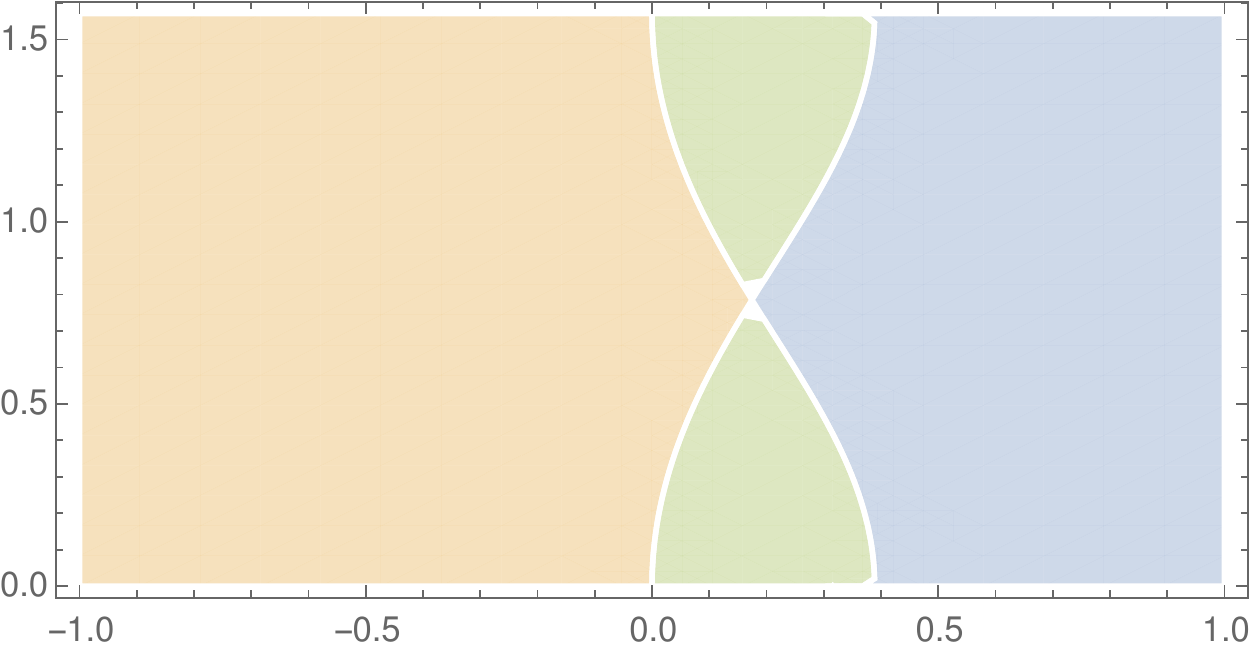}};
		\node at (5.8,0.1) {\small 4d};
		\node at (2.8,0.1) {\small 3d};
		\node at (2.4,0.8) {\small {\boldmath{$\Sigma$}}};
		
		\draw[very thick] (-4+0.1,-1.1+0.1) -- (-4+0.1,-1.1-0.1) node [anchor=north,yshift=0.5mm] {\small NS5};
		\draw[very thick] (-4+0.1,1.25-0.1) -- (-4+0.1,1.25+0.1) node [anchor=south,yshift=-0.5mm] {\small D5};
		\node at (-1.2,0.1) {\small D3};
		
		\draw[very thick] (4+0.1,-1.1+0.1) -- (4+0.1,-1.1-0.1) node [anchor=north,yshift=0.5mm] {\small NS5};
		\draw[very thick] (4+0.1,1.25-0.1) -- (4+0.1,1.25+0.1) node [anchor=south,yshift=-0.5mm] {\small D5};
		\node at (6.8,0.1) {\small D3};
		
		\node [anchor=north,yshift=1.9mm] at  (-4+0.96,-1.1-0.1) {\scriptsize F1};
	\end{tikzpicture}	
	\\
	\begin{tikzpicture}
		\node at (-4,0){\includegraphics[width=0.3\linewidth]{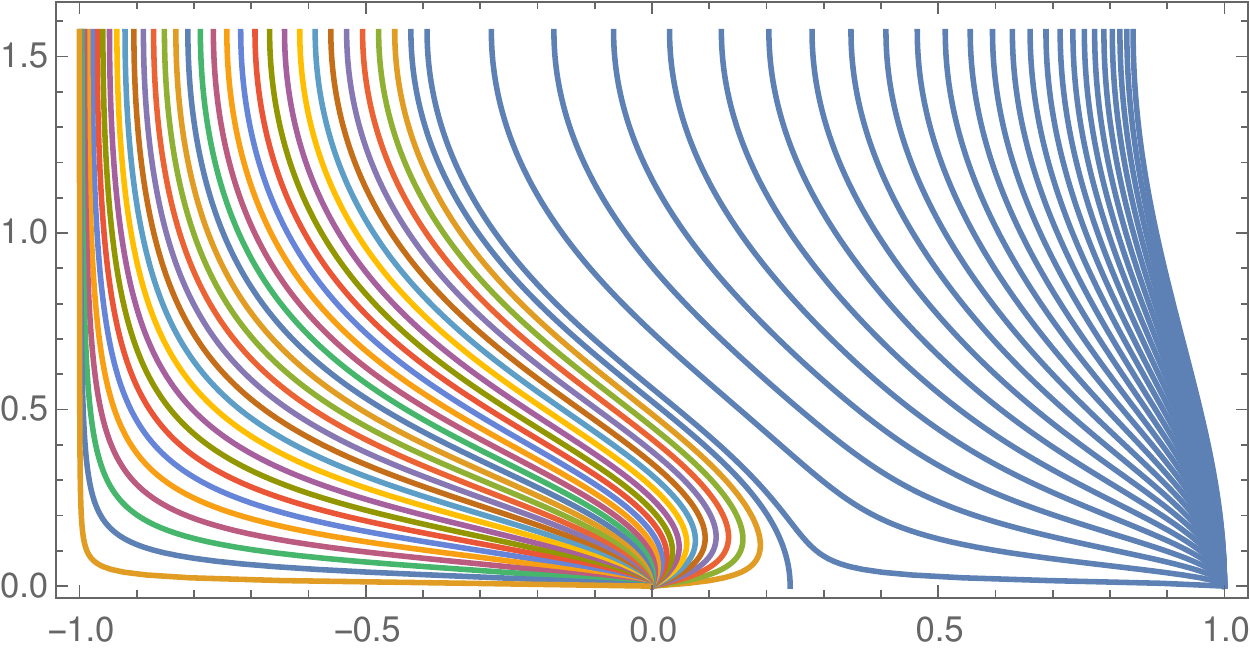}};
		\node at (4,0){\includegraphics[width=0.3\linewidth]{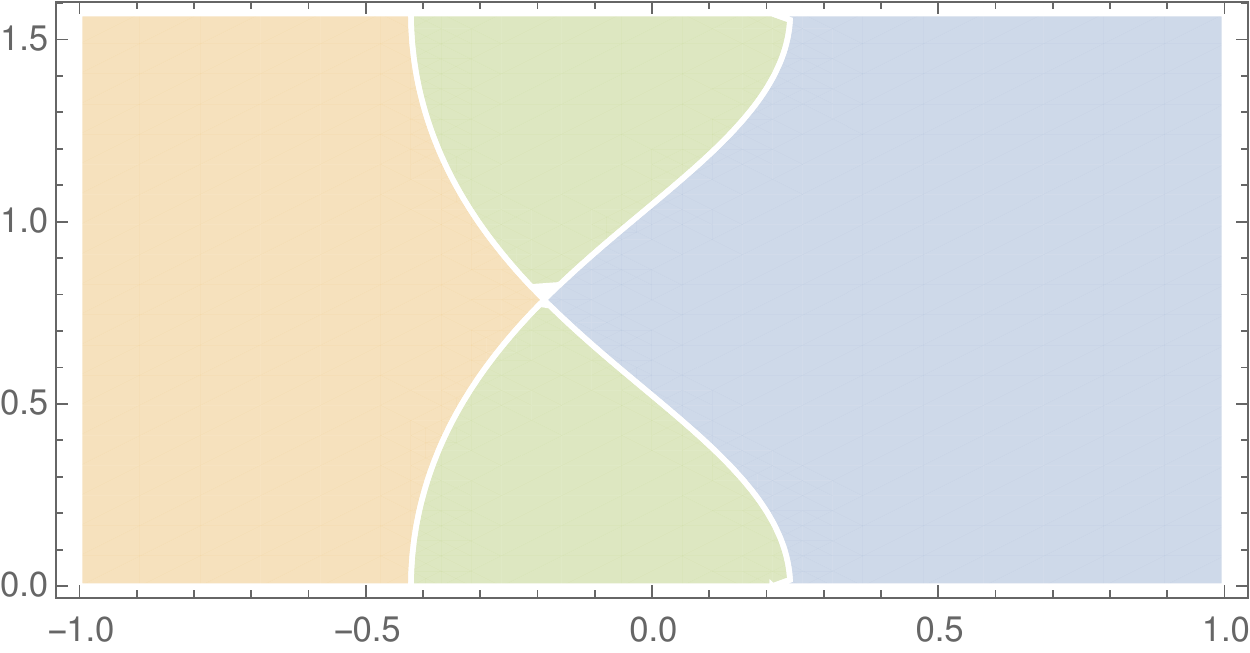}};
		\node at (5.8,0.1) {\small 4d};
		\node at (2.8,0.1) {\small 3d};
		\node at (2.4,0.8) {\small {\boldmath{$\Sigma$}}};
		
		\draw[very thick] (-4+0.1,-1.1+0.1) -- (-4+0.1,-1.1-0.1) node [anchor=north,yshift=0.5mm] {\small NS5};
		\draw[very thick] (-4+0.1,1.25-0.1) -- (-4+0.1,1.25+0.1) node [anchor=south,yshift=-0.5mm] {\small D5};
		\node at (-1.2,0.1) {\small D3};
		
		\draw[very thick] (4+0.1,-1.1+0.1) -- (4+0.1,-1.1-0.1) node [anchor=north,yshift=0.5mm] {\small NS5};
		\draw[very thick] (4+0.1,1.25-0.1) -- (4+0.1,1.25+0.1) node [anchor=south,yshift=-0.5mm] {\small D5};
		\node at (6.8,0.1) {\small D3};
		
		\node [anchor=north,yshift=2mm] at  (-4+0.7,-1.1-0.1) {\scriptsize F1};
	\end{tikzpicture}	
	\caption{Left: D5$^\prime$-brane embeddings in the D5/NS5 $\cN=4$ SYM BCFT (\ref{eq:h1h2-D5NS5-BCFT-z}) with $\Sigma$ a strip, from top to bottom for $N_5/K\in \lbrace 4,2,1\rbrace$. The horizontal axis is $\tanh(\Re(z))$, the vertical axis $\Im(z)$. NS5$^\prime$-brane embeddings are obtained by vertical reflection. Overlaying both leads to the figures on the right: the 3d region hosts D5$^\prime$ and NS$^\prime$ loop operators and is the 10d version of the ETW brane; the 4d region has no 3d loop operators and is the 10d version of the bulk in fig.~\ref{fig:braneworld-BCFT}. The remaining transition regions host one type of 3d loop operator but not both.\label{fig:braneworld}}
\end{figure}

The 10d BCFT solutions discussed in sec.~\ref{sec:BCFT-hol} can be cast in a language which reflects the qualitative features of braneworld models:
Starting point is an asymptotically locally AdS$_5\times$S$^5$ region, with a half space as conformal boundary. This region corresponds to $\Re(z)\rightarrow\infty$ on the strip in fig.~\ref{fig:braneworld} or to $u\rightarrow -1$ in the $u$ coordinate shown in fig.~\ref{fig:WL-D5-NS5-K}.
This region may be seen as dual for the 4d ambient CFT degrees of freedom.
In the 10d solutions the AdS$_5\times$S$^5$ region is not cut off by an effective ETW brane, but instead closes off smoothly by internal cycles collapsing.
This is realized by the fact that on each boundary component of $\Sigma$ one of the $S^2$'s in the $AdS_4\times S^2\times S^2\times\Sigma$ geometry collapses. Upon approaching the boundary of the half space on which the BCFT is defined, the 10d solution interpolates smoothly between the AdS$_5\times$S$^5$ region and an AdS$_4$ solution, which is the holographic dual for the 3d boundary degrees of freedom.
This latter part corresponds to the region around the 5-brane sources in figs.~\ref{fig:WL-D5-NS5-K}, \ref{fig:braneworld}; it is the 10d version of the ETW brane region in the braneworld models.

The discussion of loop operators allows us to make a more quantitative connection between the top-down string theory duals for BCFTs and bottom-up braneworld models. Based on the values of the harmonic functions $h_{1/2}^D$, we can identify the parts of $\Sigma$ in the 10d solutions which correspond to the bulk region in the braneworld models and those which correspond to the  ETW brane (from the perspective of loop operators).

The discussion of Wilson loop D5$^\prime$-brane embeddings in sec.~\ref{sec:BCFT-hol} identified two regions on $\Sigma$: The first region, around the D3-brane sources at $u=-1$ in fig.~\ref{fig:WL-D5-NS5-K}, hosts D5$^\prime$ embeddings which describe surface operators associated with the 4d $\cN=4$ SYM node. This region is separated, by the curve starting at the point marked F1, from the second region on $\Sigma$, which hosts the D5$^\prime$ embeddings describing Wilson loops associated with 3d gauge nodes.
The analogous embeddings in the $z$ coordinate on the strip are shown in the left column of fig.~\ref{fig:braneworld}.
This discussion suggests that -- from the perspective of Wilson loops -- the second region is the 3d part of the holographic dual, corresponding to the ETW brane in the braneworld model in fig.~\ref{fig:braneworld-BCFT}, while the first region is the 4d part of the holographic dual corresponding to the AdS$_5$ bulk in fig.~\ref{fig:braneworld-BCFT}.

Of course Wilson loops are not the only observables that can be used to probe the geometry.
We focus on vortex loops, which are related to Wilson loops by mirror symmetry, as one further natural class of observables.
Vortex loops are represented by NS5$^\prime$ branes, as discussed in sec.~\ref{sec:brane}, and the BPS equations for the corresponding probe NS5$^\prime$ branes in the supergravity solutions can be obtained from those for D5$^\prime$ branes by exchanging $h_1$ and $h_2$. The NS5$^\prime$ embeddings for the D5/NS5 BCFT can be obtained from the D5$^\prime$ embeddings shown in the left column of fig.~\ref{fig:braneworld} by a vertical reflection.\footnote{%
The solution (\ref{eq:h1h2-D5NS5-BCFT-z}) is invariant under S-duality, exchanging $h_1$ and $h_2$ combined with $z\rightarrow \bar z+\frac{i\pi}{2}$.}
Similarly to the discussion for the D5$^\prime$-branes, there are NS5$^\prime$ embeddings representing vortex loops associated with 3d gauge nodes and embeddings representing surface operators associated with 4d $\cN=4$ SYM.
One can then, similarly to the discussion for the D5$^\prime$-branes, identify 3d regions and 4d regions on $\Sigma$ from the perspective of vortex loops.

Overlaying the regions on $\Sigma$ as identified from the perspective of Wilson loops and of vortex loops leads to the figures in the right column of fig.~\ref{fig:braneworld}. 
We obtain a region (I) which only hosts 3d loop operator embeddings (D5$^\prime$ and NS5$^\prime$) and no 4d surface operator embeddings, 
a region (II) which hosts only 4d surface operator embeddings and no 3d loop operators, and a region (III) which hosts both types of embeddings, one type of D5$^\prime$ and the other type of NS5$^\prime$ branes.
From the perspective of loop operators, region (I) is associated with 3d degrees of freedom and the 10d version of the ETW brane region. Region (II) is associated with 4d degrees of freedom and the 10d version of the bulk, while region (III) is a transition region.
The regions are marked accordingly in fig.~\ref{fig:braneworld}, which shows how the 3d region grows at the expense of the 4d region as $N_5/K$ is increased, which increases the number of 3d degrees of freedom relative to the 4d degrees of freedom.

The regions (I), (II) and (III) can be identified based on the values of $h_1^D$ and $h_2^D$ alone, with no need for the actual D5$^\prime$ or NS5$^\prime$ embeddings:
$h_2^D$ and $h_1^D$ have discontinuities at the NS5 and D5 poles, respectively. Depending on the direction from which an NS5 pole is approached, one obtains different limiting values for $h_2^D$, and likewise for D5 poles and $h_1^D$.
The region on $\Sigma$ where $h_2^D$ is in the range of values that can be attained at the NS5 pole is the region which hosts Wilson loop D5$^\prime$-branes. 
This corresponds to $0<h_2^D<\frac{\pi}{2}\alpha^\prime N_5$ for the D5/NS5 $\cN=4$ SYM BCFT with the choice of $\cA_{1/2}$ in (\ref{eq:A12-BCFT}).
Likewise, the region of $\Sigma$ where $h_1^D$ is in the range of values that can be attained at the D5 pole hosts vortex loop NS5$^\prime$-branes. 
For $\cA_{1/2}$ in (\ref{eq:A12-BCFT}), this corresponds to $|h_1^D|\leq \frac{\pi}{4}\alpha^\prime N_5$.
The region where $h_1^D$ and $h_2^D$ are both in the respective intervals is region (I); the region where neither are in the respective intervals is region (II); the region of $\Sigma$ where one and only one is in the respective interval is the transition region (III).

\begin{figure}
	\centering
	\begin{tikzpicture}
		\node at (-4,0){\includegraphics[width=0.3\linewidth]{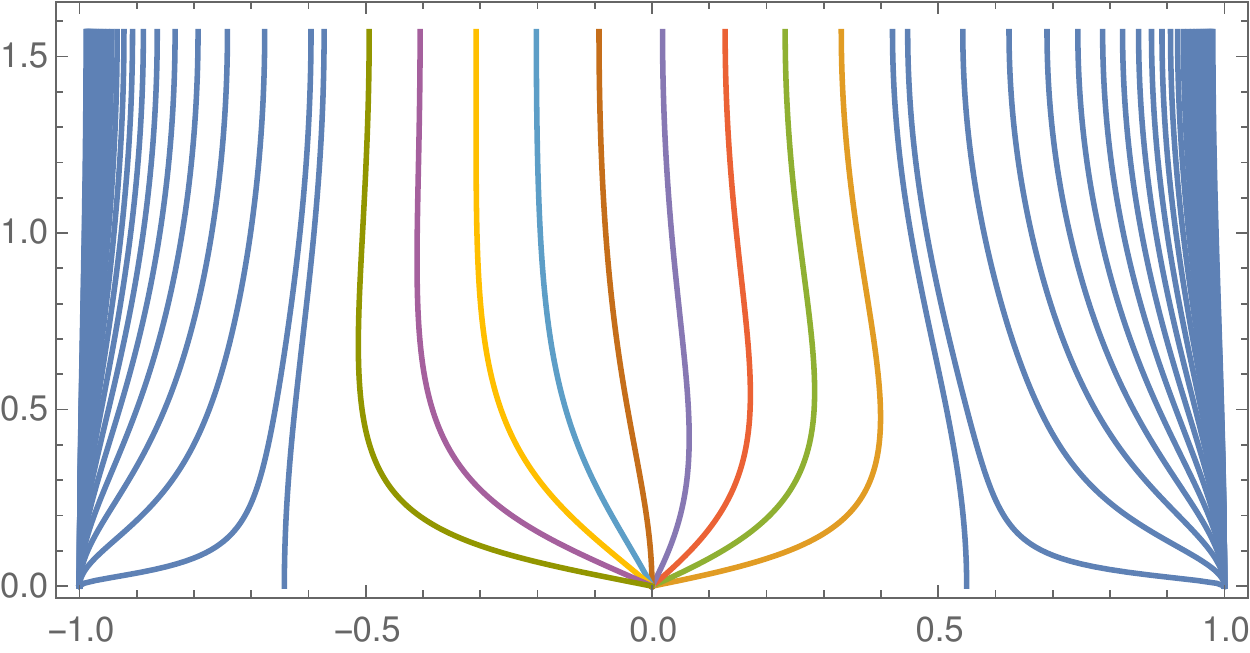}};
		\node at (4,0){\includegraphics[width=0.3\linewidth]{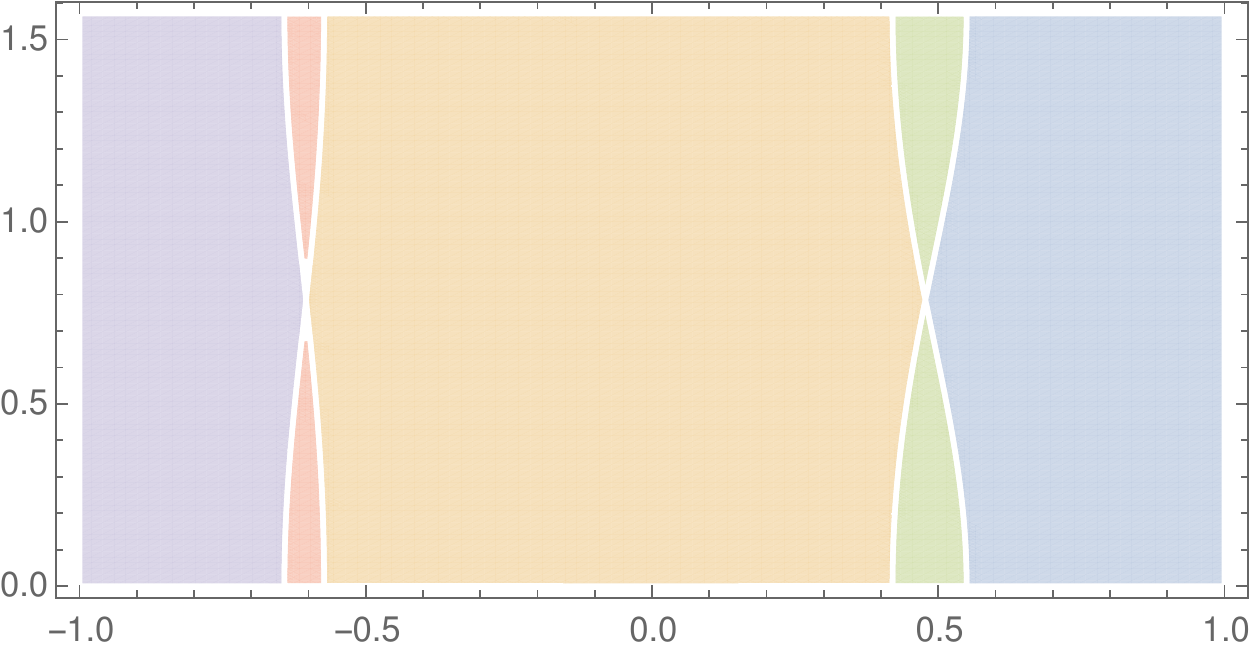}};
		\node at (5.8,0.1) {\small 4d};
		\node at (4,0.1) {\small 3d};
		\node at (2.2,0.1) {\small 4d};
		\node at (2.2,0.8) {\small {\boldmath{$\Sigma$}}};
		
		\draw[very thick] (-4+0.1,-1.1+0.1) -- (-4+0.1,-1.1-0.1) node [anchor=north,yshift=0.5mm] {\small NS5};
		\draw[very thick] (-4+0.1,1.25-0.1) -- (-4+0.1,1.25+0.1) node [anchor=south,yshift=-0.5mm] {\small D5};
		\node at (-1.2,0.1) {\small D3};
		
		\draw[very thick] (4+0.1,-1.1+0.1) -- (4+0.1,-1.1-0.1) node [anchor=north,yshift=0.5mm] {\small NS5};
		\draw[very thick] (4+0.1,1.25-0.1) -- (4+0.1,1.25+0.1) node [anchor=south,yshift=-0.5mm] {\small D5};
		\node at (6.8,0.1) {\small D3};
		
		\node at (-6.8,0.1) {\small D3};
		\node at (1.2,0.1) {\small D3};
	\end{tikzpicture}	
	\caption{D5$^\prime$ embeddings and identification of 3d and 4d regions for a Janus solution, with harmonic functions (\ref{eq:h1h2-gen})
		with $N_5$ D5-branes at $z=\frac{i\pi}{2}$, $N_5$ NS5-branes at $z=0$, and the numbers of semi-infinite D3-branes to the left and right fixed by $L=\frac{1}{4}N_5$ and $K=\frac{1}{3}N_5$.\label{fig:braneworld-Janus}}
\end{figure}
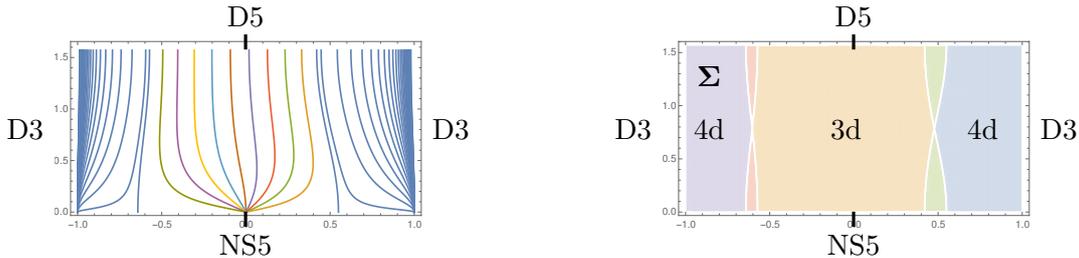

A result of using the combination of probe D5$^\prime$ and NS5$^\prime$ branes to identify 3d and 4d regions is that D5 and NS5 brane sources are both required in the background solution to get proper 3d regions. If only one type of 5-brane source is present the above identification only yields 4d and mixed regions. The requirement for both types of 5-brane sources to be present in order to get a 3d region is in line with the fact that D5 and NS5 sources are both needed to get a solution dual to a proper 3d SCFT (without AdS$_5\times$S$^5$ regions). 

In summary, the discussion above gives a quantitative identification of regions on $\Sigma$ with the ingredients in braneworld models. We note that the proposed identification of 3d and 4d regions is based on the discussion of loop operators. 
Other observables may lead to a different picture. However, we do not see a reason to expect drastic differences. 

The discussion extends to duals of Janus interface CFTs, which have an additional 4d region. The D5$^\prime$ embeddings and identification of 3d and 4d regions for an example solution are shown in fig.~\ref{fig:braneworld-Janus}. The example solution is dual to a field theory composed of two 4d $\cN=4$ SYM nodes on half spaces with gauge groups of different ranks, joined at an interface which hosts (the IR fixed point of) a 3d quiver gauge theory. Bottom-up holographic duals for Janus CFTs may be constructed by joining two slices of AdS$_5$ with different radii along an effective ``interface brane", similar to the discussions in \cite{Erdmenger:2014xya,Simidzija:2020ukv,Bachas:2021fqo}. The result is a geometry with two 4d regions and one 3d region, as in fig.~\ref{fig:braneworld-Janus}.

The wedge holography picture, proposed in \cite{Akal:2020wfl} in the context of braneworld models and illustrated in fig.~\ref{fig:braneworld-wedge}, is realized in 10d by the D5$^2$/NS5$^2$ solutions of sec.~\ref{sec:D52NS52}. 
The two ETW brane regions correspond to $\Re(z)>0$ and $\Re(z)<0$ on the strip $\Sigma$ in 10d, and the perspective gained from Wilson loops on these solutions was discussed in sec.~\ref{sec:D52NS52}.

\section{Discussion}\label{sec:disc}

In this work we studied Wilson and vortex loop operators in 3d SCFTs, 4d BCFTs based on $\cN=4$ SYM coupled to 3d SCFTs, and  Janus interface CFTs using holography and supersymmetric localization. The study of extended operators is of great interest in its own right. Here we used Wilson loops and vortex loops to identify curves on $\Sigma$ in the general $AdS_4\times S^2\times S^2\times \Sigma$ solutions constructed in \cite{DHoker:2007zhm,DHoker:2007hhe,Aharony:2011yc,Assel:2011xz} with individual gauge nodes in the 3d quiver gauge theories that are part of the dual field theories.
The idenitification is particularly interesting for BCFTs and Janus CFTs, since it gives a concrete notion of parts of the geometry which are associated with 3d degrees of freedom and parts which are associated with 4d degrees of freedom.
This identification allows for a quantitative connection to the language used in bottom-up braneworld models. 

For the duals of 3d SCFTs we matched the holographic results for the Wilson loop expectation values to field theory computations using supersymmetric localization and demonstrated perfect agreement.
It would be interesting to extend the field theory computations to BCFTs. The boundary free energies for field theories engineered by D3-branes ending on only D5 or only NS5 branes were obtained in \cite{Raamsdonk:2020tin}, based on gluing 3d and 4d results. 
As discussed in sec.~\ref{sec:braneworld}, the solutions which have the clearest connection to braneworld models involve both D5 and NS5 branes and it would be interesting to study them.
We expect that the formalism of \cite{Coccia:2020wtk}, which underlies the localization results for Wilson loops in 3d SCFTs of sec.~\ref{sec:loc}, can also be extended to BCFTs.

On the supergravity side we identified the holographic representation of Wilson loops in terms of D5$^\prime$ branes. The BPS conditions and expressions for the on-shell action derived here apply also to the holographic duals 3d circular quiver SCFTs  \cite{Assel:2012cj} and to multi-Janus solutions \cite{DHoker:2007hhe}. It would be interesting to study these theories in more detail as well.

The identification of curves on $\Sigma$ with 3d gauge nodes is also of interest in the context of the studies of information transfer between two black hole systems in \cite{Uhlemann:2021nhu}. A quantity of interest is the entanglement entropy associated with decomposing $\Sigma$.
For the theories of sec.~\ref{sec:D52NS52} there is a preferred split of $\Sigma$ which one would expect to be associated with cutting the quiver diagram in the dual field theory at the central node. This is supported by the discussion in sec.~\ref{sec:D52NS52}. It would be interesting to study more general decompositions.
The entropy associated with a decomposition of the quiver can be understood as RG flow from a geometric entanglement entropy, by starting with $\cN=4$ SYM on an interval, with boundary conditions such that the 3d gauge node at which the quiver is split arises from the 4d $\cN=4$ SYM degrees of freedom upon flowing to the IR. Splitting the interval then leads to a geometric entropy in the UV which turns into an internal entropy in the IR. A similar RG flow perspective was recently explored for a different class of theories in  \cite{Uhlemann:2021itz}.

\begin{acknowledgments}
LC thanks Andrea Legramandi for useful discussions. LC is supported by the INFN and by the MIUR-PRIN contract 2017CC72MK003.	
CFU is supported, in part, by the US Department of Energy under Grant No.~DE-SC0007859 and by the Leinweber Center for Theoretical Physics.
Part of this work was completed at the Aspen Center for Physics, which is supported by National Science Foundation grant PHY-1607611.
\end{acknowledgments}

\appendix

\section{BPS conditions}\label{app:BPS}
In this appendix we explain how the fundamental strings and the D5$^\prime$-branes, realizing Wilson loops in fundamental and antisymmetric representations, can be placed on the supergravity background, preserving the correct amount of supersymmetry. 

\subsection{Notation}

\paragraph{Clifford algebra}First of all, we need to review the conventions of \cite{DHoker:2007zhm,DHoker:2007hhe}, introducing gamma matrices adapted to the supergravity background $AdS_4 \times S^2 \times S^2 \times \Sigma$:
\begin{align}
	\Gamma^m & =\gamma^m \otimes \mathds{1}_2 \otimes \mathds{1}_2 \otimes \mathds{1}_2 \ , \qquad & m=0,1,2,3  \ , \\
	\Gamma^{i_1} & =\gamma_{(1)} \otimes \gamma^{i_1} \otimes \mathds{1}_2 \otimes \mathds{1}_2 \ , & i_1=4,5 \ , \\
	\Gamma^{i_2} & =\gamma_{(1)} \otimes \sigma^3 \otimes \gamma^{i_2} \otimes \mathds{1}_2 \ , \quad  \qquad \ & i_2=6,7 \ , \\
	\Gamma^a & =\gamma_{(1)} \otimes \sigma^3 \otimes \sigma^3 \otimes \gamma^a \ , \quad \qquad & a=8,9 \ ,
\end{align}
with
\begin{equation}
	\begin{split}
		i\gamma^0=\sigma^2 \otimes &\mathds{1}_2 \ , \qquad \gamma^1=\sigma^1 \otimes \mathds{1}_2 \ , \qquad \gamma^2=\sigma^3 \otimes \sigma^2 \ , \qquad \gamma^3=\sigma^3 \otimes \sigma^1 \\
		& \gamma^4 =\gamma^6=\gamma^8=\sigma^1 \ , \qquad
		\qquad \qquad \gamma^5=\gamma^7=\gamma^9=\sigma^2 
	\end{split}
\end{equation}
and the associated chirality matrices
\begin{equation}
	\gamma_{(1)}= \sigma^3 \otimes \sigma^3 \ , \qquad \gamma_{(2)}=  \sigma^3 \ , \qquad
	\gamma_{(3)}= \sigma^3 \ ,\qquad
	\gamma_{(4)}= \sigma^3 \ .
\end{equation}
We also need the complex conjugation matrix $\mathcal{B}$, satisfying 
\begin{equation}
	\cB \cB^*= \mathds{1} \ , \qquad \qquad  \left(\Gamma^M \right)^*=\mathcal{B}\Gamma^M \mathcal{B}^{-1} 
\end{equation}
and given by
\begin{equation}
	\mathcal{B}=i\gamma_{(1)}\gamma^2 \otimes \gamma^5 \otimes \gamma^6 \otimes \gamma^9=iB_{(1)}\otimes B_{(2)}\otimes \left(\gamma_{(3)}B_{(3)}\right)\otimes B_{(4)} 
\end{equation}
with
\begin{equation}
	B_{(1)}= i \gamma_{(1)} \gamma^2 \ , \qquad B_{(2)}= \gamma^5  \ , \qquad B_{(3)}= \gamma^7 \ , \qquad B_{(4)}= \gamma^9 \ .
\end{equation}

\paragraph*{Killing spinors and $\tau$-formalism} The next ingredient we need are the Killing spinors.
The explicit forms of the $S_{1/2}^2$ and $AdS_4$ Killing spinors, with constant spinors $\epsilon_{S^2_{1/2},0}^{\eta_2}$, $\epsilon^{\eta_1}_{AdS_4,0}$,
can be chosen as
\begin{align}\label{eq:killing_spin}
	ds^2_{AdS_4}&=dr^2+e^{2r}dx^\mu dx_\mu~,
	&
	\epsilon_{AdS_4}^{\eta_1}&=e^{\frac{\eta_1}{2}r\gamma_r}\left(1+\frac{1}{2}x^\mu\gamma_\mu\left(\eta_1-\gamma_r\right)\right)\epsilon^{\eta_1}_{AdS_4,0}~,
	\nonumber\\
	ds^2_{S^2_{1}}&=d\theta^2+\sin^2\!\theta\,d\phi^2~,
	&
	\epsilon_{S^2_1}^{\eta_2}&=\exp\left(\frac{i\eta_2}{2}\theta\sigma_2\right)\exp\left(-\frac{i}{2}\phi\sigma_3\right)\epsilon_{S^2_1,0}^{\eta_2}~,
	\nonumber\\
	ds^2_{S^2_{2}}&=d\theta^2+\sin^2\!\theta\,d\phi^2~,
	&
	\epsilon_{S_2^2}^{\eta_3}&=\exp\left(\frac{i\eta_3}{2}\theta\sigma_2\right)\exp\left(-\frac{i}{2}\phi\sigma_3\right)\epsilon_{S^2_2,0}^{\eta_3}~,
\end{align}
with $\eta_1, \eta_2, \eta_3 = \pm 1 $. Then, following \cite{DHoker:2007zhm,DHoker:2007hhe}, we decompose the 32-component ten-dimensional Killing spinor $\epsilon$ as
\begin{equation}\label{eq:expansion_spinor}
	\epsilon=\sum_{\eta_1,\eta_2,\eta_3}\chi^{\eta_1,\eta_2,\eta_3}\otimes \zeta_{\eta_1,\eta_2,\eta_3}~, \qquad 
	\chi^{\eta_1,\eta_2,\eta_3}=\epsilon^{\eta_1}_{AdS_4}\otimes \epsilon^{\eta_2}_{S_1^2}\otimes \epsilon^{\eta_3}_{S_2^2}~ \ ,
\end{equation} 
where $\zeta_{\eta_1,\eta_2,\eta_3}$ are 2-component $\Sigma$ dependent spinors. We also choose a basis in which
\begin{equation}\label{eq:expansion_cB}
	\mathcal B^{-1}\epsilon^\star =\sum_{\eta_1,\eta_2,\eta_3}\chi^{\eta_1,\eta_2,\eta_3}\otimes \star\zeta_{\eta_1,\eta_2,\eta_3}~,\qquad \qquad \star\zeta=-i\eta_1\eta_3\zeta_{\eta_1,-\eta_2,-\eta_3}~.
\end{equation}
All the details on the constrains on $\zeta_{\eta_1, \eta_2, \eta_3}$ can be found in \cite{DHoker:2007zhm}.
Finally, we will use the the $\tau$-formalism, defining
\begin{align}\label{eq:tau_formalism}
	(\tau^{(ijk)}\zeta)_{\eta_1\eta_2\eta_3}&\equiv\sum_{\eta_1^\prime\eta_2^\prime\eta_3^\prime}(\tau^i)_{\eta_1\eta_1^\prime}(\tau^j)_{\eta_2\eta_2^\prime}(\tau^k)_{\eta_3\eta_3^\prime}\zeta_{\eta_1^\prime\eta_2^\prime\eta_3^\prime} 
\end{align}
with  $i,j,k= 0, \dots , 3$. As in \cite{DHoker:2007zhm}, we will use two different basis. In the first one, $\tau^0$ is chosen to be the identity and $\tau^{i}$ to be the standard Pauli matrices. In the second one instead, which we will call \emph{rotated basis}, we take
\begin{equation}\label{eq:rotated_basis}
	\tau^0=\mathds{1}_2 \ , 
	\qquad\tau^1=
	\begin{pmatrix}
		1 & 0 \\
		0 & -1 
	\end{pmatrix}  \ ,
	\qquad 
	\tau^2=
	\begin{pmatrix}
		0 & -i \\
		i & 0 
	\end{pmatrix} \ ,
	\qquad
	\tau^3=
	\begin{pmatrix}
		0 & -1 \\
		-1 & 0 
	\end{pmatrix} \ .
\end{equation}
The explicit components of $\zeta$ are given in (6.17) of \cite{DHoker:2007zhm}, in this rotated basis. With $e^{-i\theta}=i$,
\begin{align}\label{eq:zeta_components}
	\zeta_{+++}&=i\nu \zeta_{+--}=\begin{pmatrix}0 \\ \alpha e^{i\theta/2}\end{pmatrix}~,
	&
	\zeta_{---}&=-i\nu \zeta_{-++}=\begin{pmatrix}\bar\beta e^{i\theta/2}\\ 0\end{pmatrix}~,
	\nonumber\\
	\zeta_{+-+}&=i\nu \zeta_{++-}=\begin{pmatrix}\bar\alpha e^{i\theta/2} \\ 0\end{pmatrix}~,
	&
	\zeta_{-+-}&=-i\nu \zeta_{--+}=\begin{pmatrix} 0 \\ \beta e^{i\theta/2}\end{pmatrix}~.
\end{align}

\subsection{Fundamental strings}
To realize the conformal defect symmetry, a F1 must wrap an AdS$_2$ portion of the full AdS$_4$. We choose to place it at $x^1=x^2=0$, so that the AdS$_4$-Killing spinor becomes
\begin{align}\label{eq:Killing_F1_AdS}
	\epsilon_{AdS_4}^{\eta_1}&=e^{\frac{\eta_1}{2}r\gamma_r}\left(1+\frac{1}{2}x^0\gamma_0\left(\eta_1-\gamma_r\right)\right)\epsilon^{\eta_1}_{AdS_4,0}~ \ .
\end{align}
Moreover, from the branes configuration, we see that an F1 should preserve the $SU(2)_H \cong SO(3)_H$ group in the directions (456) wrapped by the D5, while it should break the $SU(2)_C \cong SO(3)_C$ group related with the NS5 branes. These two different $SO(3)$ groups are associated with the spheres $S_1$ and $S_2$, respectively, in the
background  
\begin{equation}\label{eq:geom}
	ds^2=f_4^2 ds^2_{AdS_4}+f_1^2 ds_{S_1^2}+f_2^2 ds_{S_2^2}+ds^2_\Sigma \ ,
\end{equation}
where $\Sigma$ is an infinite strip. Recalling that $S_1$ collapses on the lower boundary of $\Sigma$, we find that fundamental strings should sit on that boundary to preserve $SO(3)_H$ subgroup. Being localized on the other $S_2$, they preserve $U(1)_C$. More precisely, we choose to place the strings on $S_2$ in the position given by $\theta=0$. In this way the $S_2$-Killing spinor reduces to
\begin{align}\label{eq:Killing_F1_S2}
	\epsilon_{S_2^2}^{\eta_3}&=\exp\left(-\frac{i}{2}\phi \sigma_3\right)\epsilon_{S^2_2,0}^{\eta_3} \ .
\end{align}
The remaining spacetime coordinates are then fixed by requiring to preserve half supersymmetries. This can be done by solving a $\kappa$-symmetry condition. Using \cite{Cederwall:1996ri,Bergshoeff:1996tu} with the convention for complex notation of \cite{Karch:2015vra} and incorporating a phase to accommodate the $SU(1,1)$ conventions as discussed in \cite{Gutperle:2018vdd}, the condition imposed by $\kappa$-symmetry is
\begin{align}\label{eq:theta-k}
	\Gamma_\kappa\epsilon&=e^{i\theta_\kappa}\Gamma_{(0)}\cB^{-1}\epsilon^\star=\epsilon~,
	&
	e^{2i\theta_\kappa}&=\frac{1+i\bar \tau}{1-i\tau}~,
\end{align}
with $\tau$ related to the axion $\chi$ and the dilaton $\phi$ via $\tau=\chi+i e^{-2 \phi}$. For the solutions considered here the axion vanishes (see sec.~6.4 of \cite{DHoker:2007zhm}) and $e^{2i\theta_\kappa}=1$.
With our assumptions, we have
\begin{align}\label{eq:F1-kappa}
	\Gamma_{(0)}&=\Gamma^0\Gamma^1=\gamma_r\gamma_0\otimes \mathds{1}_2\otimes \mathds{1}_2\otimes \mathds{1}_2~.
\end{align}
In order to solve \eqref{eq:theta-k} we require that the constant part of the Killing spinors satisfy
\begin{equation}\label{eq:proj_cond}
	\Gamma^{2367}\left(\epsilon^{\eta_1}_{AdS_4,0} \otimes \epsilon_{S^2_1,0}^{\eta_2} \otimes \epsilon_{S^2_2,0}^{\eta_3} \right)=-\lambda \left(\epsilon^{\eta_1}_{AdS_4,0} \otimes \epsilon_{S^2_1,0}^{\eta_2}  \otimes \epsilon_{S^2_2,0}^{\eta_3} \right) \ ,
\end{equation}
with $\lambda^2=1$. On the entire Killing spinor, this becomes
\begin{equation}
	\Gamma^{2367}\epsilon=-\lambda \epsilon
\end{equation}
which can also be rewritten as
\begin{equation}\label{eq:projector}
	\Gamma^{0167}\epsilon=i \lambda \gamma_{(1)}\epsilon \qquad \Rightarrow \qquad  \Gamma^{01}\epsilon =  \lambda \gamma_{(1)}\gamma_{(3)}\epsilon 
\end{equation}
(slight abuse of notation: by $\gamma_{(1)}\gamma_{(3)}$ we mean $\gamma_{(1)} \otimes \mathds{1}_2 \otimes \gamma_{(3)} \otimes \mathds{1}_2$).
Note that, to derive this expression, we used that  $\Gamma^{2367}$ commutes with the spatial dependent part of the Killing spinors, as can be seen from the explicit expressions \eqref{eq:Killing_F1_AdS} and \eqref{eq:Killing_F1_S2}. Hence, writing the $\kappa$-symmetry condition as
\begin{equation}
	e^{i \theta_\kappa}\cB^{-1}\epsilon^*=\Gamma^{01} \epsilon \ ,
\end{equation}
we have
\begin{equation}
	e^{i \theta_\kappa}\sum_{\eta_1,\eta_2,\eta_3} \chi^{\eta_1,\eta_2,\eta_3}\otimes \star \zeta_{\eta_1,\eta_2,\eta_3}= \lambda \gamma_{(1)}\gamma_{(3)}\sum_{\eta_1,\eta_2,\eta_3} \chi^{\eta_1,\eta_2,\eta_3}\otimes \zeta_{\eta_1,\eta_2,\eta_3} \ .
\end{equation}
Recalling now that \cite{DHoker:2007zhm}
\begin{equation}
	\begin{split}
		\gamma_{(1)}\chi^{\eta_1, \eta_2, \eta_3}=\chi^{-\eta_1,\eta_2,\eta_3} \ , \\
		\gamma_{(2)}\chi^{\eta_1, \eta_2, \eta_3}=\chi^{\eta_1,-\eta_2,\eta_3} \ , \\
		\gamma_{(3)}\chi^{\eta_1, \eta_2, \eta_3}=\chi^{\eta_1,\eta_2,-\eta_3} \ , 
	\end{split}
\end{equation}
the $\kappa$-symmetry condition becomes
\begin{equation}
	\sum_{\eta_1,\eta_2,\eta_3} \chi^{\eta_1,\eta_2,\eta_3}\otimes \left( -i e^{i\theta_\kappa}  \eta_1 \eta_3 \zeta_{\eta_1,-\eta_2,-\eta_3}-\lambda \zeta_{-\eta_1,\eta_2,-\eta_3} \right)=0 \ .
\end{equation}
Since the $\chi^{\eta_1,\eta_2,\eta_3}$ are linearly independent, we must require the square brackets to vanish:
\begin{equation}
	-i \lambda e^{i \theta_\kappa}\eta_1 \eta_3 \zeta_{\eta_1,-\eta_2,-\eta_3}=\zeta_{-\eta_1,\eta_2,-\eta_3} \qquad \Rightarrow \qquad   \lambda e^{i\theta_\kappa}\tau^{(312)}\zeta=\tau^{(101)}\zeta \  .
\end{equation}
Using the rotated basis \eqref{eq:rotated_basis} we obtain
\begin{equation}
	i \lambda  e^{i \theta_\kappa} \eta_2 \eta_3 \zeta_{-\eta_1, \eta_2,-\eta_3}= \eta_1 \eta_3 \zeta_{\eta_1, \eta_2 , \eta_3}
\end{equation}
namely
\begin{equation}\label{eq:kappa_cond}
	i \lambda e^{i \theta_\kappa} \eta_1 \eta_2
	\zeta_{-\eta_1, \eta_2,-\eta_3}=  \zeta_{\eta_1, \eta_2 , \eta_3} \ .
\end{equation}
Substituting the explicit components of $\zeta$ \eqref{eq:zeta_components}, one finds that \eqref{eq:kappa_cond} is solved for each choice of $\eta_{1},\eta_2,\eta_3$ by requiring
\begin{equation}\label{eq:alpha_beta_rel}
	\alpha= i \lambda \beta e^{i \theta_{\kappa}} \ .
\end{equation}
In order to interpret this condition, let us recall that the functions in the metric \eqref{eq:geom} can be written as (see equation (6.26) in \cite{DHoker:2007zhm})
\begin{equation}\label{eq:f_def}
	\begin{split}
		f_4 &= \alpha \bar{\alpha}+\beta\bar{\beta} \ , \\ 
		f_1 & = - \nu (\alpha \bar{\beta}+\beta\bar{\alpha}) \ , \\
		f_2 & = i (\beta \bar{\alpha}-\alpha \bar{\beta})  \ .
	\end{split}
\end{equation}
Using the condition \eqref{eq:alpha_beta_rel} we obtain $\alpha \bar{\beta}=-\beta \bar{\alpha}$, which implies $f_1=0$ and $h_1=0$ (see discussion in section 3.4 of \citep{DHoker:2007hhe}). Also, from (9.31) of \cite{DHoker:2007zhm},
\begin{align}\label{eq:h_alpha_beta}
	\nu \rho \bar\alpha^2&=-i e^\phi \partial h_1 - e^{-\phi}\partial h_2~,
	&
	\nu\rho \bar\beta^2&=+i e^\phi \partial h_1-e^{-\phi}\partial h_2
\end{align}
we have 
\begin{equation} \label{eq:position_F1}
	h_1=0  \ , \qquad \partial h_2=0 \ .
\end{equation}
This conclusion is consistent with our expectations. Indeed, we expect the fundamental string to be placed on the lower boundary, on which
\begin{equation}
	h_1=0 \ , \qquad \partial_\bot h_2=0 \ .
\end{equation}
These conditions are actually implied in \eqref{eq:position_F1}. In addition, equation \eqref{eq:position_F1} also requires the derivative tangent to the lower boundary to vanish, identifying the point(s) of the boundary on which fundamental strings can be placed preserving the correct symmetries.

\subsection{\texorpdfstring{D5$^{\prime}$}{D5}-branes}
Adding a D5$^\prime$-branes we should be able to preserve the same symmetries preserved by the fundamental string. 
From the supergravity perspective, this means that the D5$^\prime$ should wrap $AdS_2$, the full $S_1^2$ to preserve $SU(2)_H$ and a one dimensional circumference on $S_2^2$ to preserve $U(1)_C$. This leaves one last direction on the Riemann surface $\Sigma$. 
Introducing coordinates on $S^2_2$ and $\Sigma$
\begin{equation}
	ds^2_{S_2^2}=d\theta^2+\sin^2\theta d \phi^2 \ , \qquad ds^2_{\Sigma}=4 \rho^2\abs{dz}^2 \ ,
\end{equation}
the induced metric (in Einstein frame) on the D5' brane reads
\begin{equation}
	g= f_4^2 ds^2_{AdS_2}+f_1^2 ds^2_{S_1}+f_2^2 \sin^2 \theta (\xi) d\phi^2+\left(f_2^2\theta'^2+4 \rho^2 (x'^2+y'^2)\right)d \xi^2 \ ,
\end{equation}
where $z=x+iy$ and the derivative is with respect to the worldvolume coordinate $\xi$. As for the fundamental string, the position of the D5$^\prime$-branes on the spacetime can be derived requiring to preserve the correct amount of supersymmetry. The general worldvolume flux we can turn on for the D5$^\prime$-brane, preserving the desired symmetries, is
\begin{equation}\label{eq:gen_fluxes}
	F=F_{\text{el}} \text{vol}_{AdS_2}+F_1 \text{vol}_{S_1^2}+F_2(\xi) d \xi \wedge d \phi  \ ,
\end{equation}
with $\text{vol}_{AdS_2}$ and $\text{vol}_{S_1^2}$ worldvolume forms with unitary radius. Resolving the $\kappa$-symmetry condition for this general ansatz, one finds that
\begin{equation}\label{eq:F2_vanishes}
	F_2(\xi)=0 \ .
\end{equation}
So, for the ease of reading, we will assume \eqref{eq:F2_vanishes} from the beginning and we will check the consistency of our solution at the end.

Calling $E^a$ the vielbein in 10 dimensions and denoting $e^a=E_\mu^a\left(\partial_i X^\mu \right)dx^i$ their pullback to the D5' worldvolume, we have
\begin{align}
	&e^a=E^a \ , \ \ a=\underline{0},\underline{1},\underline{4},\underline{5} \\
	& e^{\underline{2}}=e^{\underline{3}}=0 \\
	& e^{\underline{6}}=f_2 \sin \theta d\phi \qquad  e^{\underline{7}}=f_2 \theta' d \xi  \qquad  e^{\underline{8}}= 2 \rho x' d \xi \qquad e^{\underline{9}}=2 \rho y' d \xi \ .
\end{align}
Note that, in case of explicit values, we use an underline to denote Lorentz indices. Formulas that will be used from now on are in the string frame with $\tilde{g}=e^\phi g$. The form of $\Gamma_\kappa$ in our case is
\begin{equation}\label{eq:kappa-D5}
	\begin{split} 
		\Gamma_\kappa\epsilon=\frac{-i}{\sqrt{\det(1+X)}}\Bigg(
		e^{i \theta_\kappa}\Gamma_{(0)}\cB^{-1}\epsilon^\star -\frac{1}{2}\gamma^{ij}& X_{ij}\Gamma_{(0)}\epsilon+\frac{e^{i \theta_\kappa}}{8}\gamma^{ijkl}X_{ij}X_{kl}\Gamma_{(0)}\cB^{-1}\epsilon^\star \Bigg) \ ,
	\end{split}
\end{equation}
where $\gamma_i=e_i^a \Gamma_a$, $X^{i}_{\ j} = \tilde{g}^{ik}\cF_{kj}$ and
\begin{equation}
	\begin{split}
		\Gamma_{(0)}&=\frac{1}{6!\sqrt{-\det \tilde{g}}}\varepsilon^{i_1 \dots i_6}\gamma_{i_1 \dots i_6}=\frac{1}{\sqrt{f_2^2 \theta'^2+4\rho^2(x'^2+y'^2)}}\Gamma^{01456}\Gamma_\xi  \ ,
	\end{split}
\end{equation}
where we introduced
\begin{equation}
	\Gamma_\xi \ \equiv f_2 \theta'\Gamma^7+2 \rho x' \Gamma^8 +2 \rho y'  \Gamma^9  \ .
\end{equation}
Let us now write explicitly the terms in \eqref{eq:kappa-D5}:
\begin{align}
	&\frac{1}{2}\gamma^{ij}X_{ij}=\Gamma^{01}X_{\underline{01}}+\Gamma^{45}X_{\underline{45}} \ ,   \\
	&\frac{1}{8}\gamma^{ijkl}X_{ij}X_{kl}=\Gamma^{0145}X_{\underline{01}}X_{\underline{45}} 
\end{align}
so that we can rewrite the condition $\Gamma_\kappa \epsilon=\epsilon$ as 
\begin{equation}\label{eq:gamma_k_final}
	\begin{split}
		& e^{i \theta_\kappa}\Gamma^{01456}\Gamma_{\xi}\cB^{-1}\epsilon^\star-\left(\Gamma^{456}\Gamma_{\xi}X_{\underline{01}}- \Gamma^{016}\Gamma_{\xi}X_{\underline{45}}\right)\epsilon-e^{i \theta_\kappa}\Gamma^6\Gamma_\xi X_{\underline{01}}X_{\underline{45}} \cB^{-1}\epsilon^\star= h \epsilon \ ,
	\end{split}
\end{equation}
where we defined
\begin{equation}
	h \equiv i\sqrt{\det(1+X)}\sqrt{f_2^2 \theta'^2+4 \rho^2(x'^2+y'^2)} \ .
\end{equation}

We would like to solve \eqref{eq:gamma_k_final} imposing only the projection \eqref{eq:proj_cond}, already used for the fundamental string. In this case however the situation is a bit different. Indeed now, from the condition on the constant part of the killing spinor
\begin{equation}
	\Gamma^{2367}\left(\epsilon^{\eta_1}_{AdS_4,0} \otimes \epsilon_{S^2_1,0}^{\eta_2} \otimes \epsilon_{S^2_2,0}^{\eta_3} \right)=-\lambda \left(\epsilon^{\eta_1}_{AdS_4,0} \otimes \epsilon_{S^2_1,0}^{\eta_2}  \otimes \epsilon_{S^2_2,0}^{\eta_3} \right)  \ ,
\end{equation}
we obtain
\begin{equation}\label{eq:proj_D5}
	\Gamma^{2367}\epsilon=-\lambda e^{-i \eta_3 \theta \sigma_2} \epsilon \ ,
\end{equation}
where $\sigma_2=\mathds{1}_4 \otimes \mathds{1}_2 \otimes \sigma_2 \otimes \mathds{1}_2$ is the matrix present in the expression of the $S_2$-Killing spinor \eqref{eq:killing_spin}, which anticommutes with $\Gamma^{67}=\mathds{1}_4 \otimes \mathds{1}_2 \otimes (-i \sigma_3)\otimes \mathds{1}_2$. Starting from the expression \eqref{eq:gamma_k_final} and using $\Gamma^{01}=-i \Gamma^{23}\gamma_{(1)}$, we rewrite the $\kappa$-symmetry condition as
\begin{equation}
	\begin{split}
		&e^{i \theta_\kappa} \gamma_{(1)}\gamma_{(2)} \Gamma^{236}\Gamma_{\xi}\cB^{-1}\epsilon^\star-\left(i \gamma_{(2)}\Gamma^{6}\Gamma_{\xi}X_{\underline{01}}+i \gamma_{(1)}\Gamma^{236}\Gamma_{\xi}X_{\underline{45}}\right)\epsilon \\
		& \qquad \qquad \qquad -e^{i \theta_\kappa} \Gamma^6\Gamma_\xi X_{\underline{01}}X_{\underline{45}} \cB^{-1}\epsilon^\star-h\epsilon=0 \ .
	\end{split}
\end{equation}
We now collect the terms involving $\cB^{-1}\epsilon^\star$ and those involving $\epsilon$, introducing the expansions \eqref{eq:expansion_cB} and \eqref{eq:expansion_spinor}:
\begin{equation}\label{eq:kappa_interm_1}
	\begin{split}
		&e^{i \theta_\kappa}\bigg( \gamma_{(1)}\gamma_{(2)}\Gamma^{236}\Gamma_{\xi}- \Gamma^6\Gamma_\xi X_{\underline{01}}X_{\underline{45}} \bigg)\sum_{\eta_1,\eta_2,\eta_3}\chi^{\eta_1,\eta_2,\eta_3}\otimes (-i\eta_1\eta_3)\zeta_{\eta_1,-\eta_2,-\eta_3}+ \\
		&-\bigg(i \gamma_{(2)}\Gamma^{6}\Gamma_{\xi}X_{\underline{01}}+i \gamma_{(1)}\Gamma^{236}\Gamma_{\xi}X_{\underline{45}}+h \bigg)\sum_{\eta_1,\eta_2,\eta_3}\chi^{\eta_1,\eta_2,\eta_3}\otimes \zeta_{\eta_1,\eta_2,\eta_3}=0 \ .
	\end{split}
\end{equation}
From \eqref{eq:proj_D5}, we obtain
\begin{equation}
	\Gamma^{23}\chi^{\eta_1,\eta_2,\eta_3}= i\lambda e^{i\eta_3\theta \sigma_2}\chi^{\eta_1,\eta_2,-\eta_3} 
\end{equation}
so that, by expanding $\Gamma_\xi$, we can rewrite \eqref{eq:kappa_interm_1} as
\begin{equation}\label{eq:kappa_interm_2}
	\begin{split}
		&i \lambda e^{i \theta_\kappa} \Gamma^{6}\left(f_2 \theta'\Gamma^7+2 \rho x' \Gamma^8 +2 \rho y'  \Gamma^9\right) \sum_{\eta_1,\eta_2,\eta_3}e^{-i \eta_3 \theta_2 \sigma_2}\chi^{\eta_1,\eta_2,\eta_3}\otimes (-i\eta_1\eta_3)\zeta_{-\eta_1,\eta_2,\eta_3}+ \\
		&-  e^{i \theta_\kappa}X_{\underline{01}}X_{\underline{45}} \Gamma^{6}\left(f_2 \theta'\Gamma^7+2 \rho x' \Gamma^8 +2 \rho y'  \Gamma^9\right) \sum_{\eta_1,\eta_2,\eta_3}\chi^{\eta_1,\eta_2,\eta_3}\otimes (-i\eta_1\eta_3)\zeta_{\eta_1,-\eta_2,-\eta_3}+ \\
		& -i  \Gamma^6 \left(f_2 \theta'\Gamma^7+2 \rho x' \Gamma^8 +2 \rho y'  \Gamma^9\right)  X_{\underline{01}} \sum_{\eta_1,\eta_2,\eta_3}\chi^{\eta_1,\eta_2,\eta_3}\otimes \zeta_{\eta_1,-\eta_2,\eta_3}+\\
		&+\lambda  \Gamma^6 \left(f_2 \theta'\Gamma^7+2 \rho x' \Gamma^8 +2 \rho y'  \Gamma^9\right)  X_{\underline{45}}\sum_{\eta_1,\eta_2,\eta_3}e^{-i \eta_3 \theta \sigma_2}\chi^{\eta_1,\eta_2,\eta_3}\otimes \zeta_{-\eta_1,\eta_2,-\eta_3}+\\
		&-h\sum_{\eta_1,\eta_2,\eta_3}\chi^{\eta_1,\eta_2,\eta_3}\otimes \zeta_{\eta_1,\eta_2,\eta_3} = 0 \ .
	\end{split}
\end{equation}
Using now that,
\begin{equation}
	e^{-i \eta_3 \theta \sigma_2}=\cos\left(\eta_3 \theta \right)+ \sin\left(\eta_3 \theta \right)\Gamma^6\gamma_{(1)}\gamma_{(2)}\gamma_{(3)}=\cos(\theta)+\eta_3\sin(\theta)\Gamma^6\gamma_{(1)}\gamma_{(2)}\gamma_{(3)} \ ,
\end{equation}
we obtain
\begin{equation}\label{eq:kappa_inter}
	\begin{split}
		& \lambda e^{i \theta_\kappa} \Gamma^{6} \left(f_2 \theta'\Gamma^7+2 \rho x' \Gamma^8 +2 \rho y'  \Gamma^9\right) \cos \theta \sum_{\eta_1,\eta_2,\eta_3} \chi^{\eta_1,\eta_2,\eta_3}\otimes \eta_1\eta_3\zeta_{-\eta_1,\eta_2,\eta_3} \\
		& + \lambda e^{i \theta_\kappa} \left(f_2 \theta'\Gamma^7+2 \rho x' \Gamma^8 +2 \rho y' \Gamma^9\right)  \sin \theta \sum_{\eta_1,\eta_2,\eta_3} \chi^{\eta_1,\eta_2,\eta_3}\otimes  \eta_1\zeta_{\eta_1,-\eta_2,-\eta_3}+ \\
		&+i e^{i \theta_\kappa}X_{\underline{01}}X_{\underline{45}} \Gamma^{6} \left(f_2 \theta'\Gamma^7+2 \rho x' \Gamma^8 +2 \rho y'  \Gamma^9\right) \sum_{\eta_1,\eta_2,\eta_3} \chi^{\eta_1,\eta_2,\eta_3}\otimes \eta_1\eta_3 \zeta_{\eta_1,-\eta_2,-\eta_3} + \\
		& -i \Gamma^6 \left(f_2 \theta'\Gamma^7+2 \rho x' \Gamma^8 +2 \rho y'  \Gamma^9\right)  X_{\underline{01}} \sum_{\eta_1,\eta_2,\eta_3} \chi^{\eta_1,\eta_2,\eta_3}\otimes \zeta_{\eta_1,-\eta_2,\eta_3}+\\
		&+\lambda \Gamma^6 \left(f_2 \theta'\Gamma^7+2 \rho x' \Gamma^8 +2 \rho y'  \Gamma^9\right)  X_{\underline{45}} \cos \theta \sum_{\eta_1,\eta_2,\eta_3} \chi^{\eta_1,\eta_2,\eta_3}\otimes  \zeta_{-\eta_1,\eta_2,-\eta_3}+\\
		&+\lambda   X_{\underline{45}} \left(f_2 \theta'\Gamma^7+2 \rho x' \Gamma^8 +2 \rho y' \Gamma^9\right) \sin \theta \sum_{\eta_1,\eta_2,\eta_3} \chi^{\eta_1,\eta_2,\eta_3}\otimes  \eta_3 \zeta_{\eta_1,-\eta_2,\eta_3}+\\
		&-h \sum_{\eta_1,\eta_2,\eta_3} \chi^{\eta_1,\eta_2,\eta_3}\otimes \zeta_{\eta_1,\eta_2,\eta_3} = 0 \ .
	\end{split}
\end{equation}
The previous equation can be simplified with some observations. First of all, from the explicit expressions of the gamma matrices, we see that
\begin{equation}
	(x' \Gamma^8+y'\Gamma^9) \sum_{\eta_1,\eta_2,\eta_3} \chi^{\eta_1,\eta_2,\eta_3}\otimes \zeta_{\eta_1,\eta_2,\eta_3}= \sum_{\eta_1,\eta_2,\eta_3} \chi^{\eta_1,\eta_2,\eta_3} \otimes (x' \sigma^1+y' \sigma^2)\zeta_{-\eta_1,-\eta_2,-\eta_3}
\end{equation}
and
\begin{equation}
	\Gamma^{67} \sum_{\eta_1,\eta_2,\eta_3} \chi^{\eta_1,\eta_2,\eta_3} \otimes\zeta_{\eta_1,\eta_2,\eta_3}=i \sum_{\eta_1,\eta_2,\eta_3} \chi^{\eta_1,\eta_2,\eta_3} \otimes\zeta_{\eta_1,\eta_2,-\eta_3} \ ,
\end{equation}
where we used that $\Gamma^{67}=i\gamma_{(3)}$. What we miss is the knowledge of the action of $\Gamma^6$ and $\Gamma^7=i\Gamma^6 \gamma_{(3)}$ on the spinors. Since we do not want to introduce a further projection condition, we require all the terms involving $\Gamma^6$ and $\Gamma^7$ to vanish together:
\begin{equation}
	\begin{split}
		\sum_{\eta_1,\eta_2,\eta_3} \chi^{\eta_1,\eta_2,\eta_3}&\otimes \Bigg[ 2 \rho \lambda \cos \theta (x'\sigma_1 +y'\sigma_2) \left(e^{i \theta_\kappa}\eta_1\eta_3\zeta_{\eta_1,-\eta_2,-\eta_3}+X_{\underline{45}}\zeta_{\eta_1,-\eta_2,\eta_3}\right)
		\\
		& -2i\rho (x'\sigma_1 +y'\sigma_2)  X_{\underline{01}}\zeta_{-\eta_1,\eta_2,-\eta_3} +2 i \rho (x'\sigma_1 +y'\sigma_2) e^{i \theta_\kappa} X_{\underline{01}}X_{\underline{45}} \eta_1\eta_3 \zeta_{-\eta_1,\eta_2,\eta_3}+ \\
		&+i\lambda f_2 \theta' \sin \theta \left(e^{i \theta_\kappa}\eta_1\zeta_{\eta_1,-\eta_2,\eta_3}- X_{\underline{45}}\eta_3 \zeta_{\eta_1,-\eta_2,-\eta_3}\right)\Bigg]=0 \ .
	\end{split}
\end{equation}
Since the $\chi^{\eta_1,\eta_2,\eta_3}$ are linearly independent, we must require the square brackets to vanish for each choice of $ \eta_1,\eta_2,\eta_3$. Using the $\tau$-formalism in the Pauli matrices basis, we then have
\begin{equation}
	\begin{split}
		&2 \rho \lambda \cos \theta (x'\sigma_1+y'\sigma_2) \left(ie^{i \theta_\kappa}\tau^{(312)}\zeta+X_{\underline{45}}\tau^{(010)}\zeta\right)
		-2i \rho  (x'\sigma_1+y'\sigma_2) X_{\underline{01}}\tau^{(101)}\zeta+ \\
		& - 2 \rho  (x'\sigma_1+y'\sigma_2) e^{i \theta_\kappa} X_{\underline{01}}X_{\underline{45}} \tau^{(203)} \zeta +i\lambda f_2 \theta' \sin \theta \left(e^{i \theta_\kappa} \tau^{(310)}\zeta-i X_{\underline{45}}\tau^{(012)} \zeta\right)=0 \ .
	\end{split}
\end{equation}
In the rotated basis \eqref{eq:rotated_basis} this gives
\begin{equation}\label{eq:D5_kappa_tau_1}
	\begin{split}
		&2 \rho \lambda \cos \theta(x' \sigma_1+y'\sigma_2) \left(-e^{i \theta_\kappa}\eta_2 \eta_3\zeta_{-\eta_1,\eta_2,-\eta_3}+X_{\underline{45}}\eta_2 \zeta_{\eta_1,\eta_2,\eta_3}\right) \\
		&-2 i\rho  ( x' \sigma_1+ y' \sigma_2)\left( X_{\underline{01}}\eta_1\eta_3\zeta_{\eta_1,\eta_2,\eta_3}+ e^{i \theta_\kappa} X_{\underline{01}}X_{\underline{45}}  \eta_1 \zeta_{-\eta_1,\eta_2,-\eta_3}\right) \\
		& +i\lambda f_2
		\theta'\sin \theta \left(-e^{i\theta_\kappa}\eta_2  \zeta_{-\eta_1,\eta_2,\eta_3}-X_{\underline{45}}\eta_2\eta_3\zeta_{\eta_1,\eta_2,-\eta_3} \right)=0 \ .
	\end{split} 
\end{equation}
Substituting the explicit expressions \eqref{eq:zeta_components} of the spinors $\zeta$, it turns out that only two equations are linear independent, and the others can be obtained by complex conjugation. The resulting system coming from \eqref{eq:D5_kappa_tau_1} is
\begin{equation}\label{eq:D5_BPS_1}
	\begin{cases}
		& f_2 \lambda \theta' \sin \theta \left( - X_{\underline{45}} \alpha^*+  e^{i\theta_\kappa} \beta^*\right)+ \\
		&\qquad \qquad +2 \rho \nu (x'-i y') \left(-i X_{\underline{01}}(\alpha+X_{\underline{45}}\beta e^{i\theta_\kappa})+(X_{\underline{45}}\alpha-\beta e^{i\theta_\kappa})\lambda\cos \theta \right)= 0 \\
		& f_2 \lambda \theta' \sin \theta \left( e^{i\theta_\kappa} \alpha+ X_{\underline{45}} \beta \right)+ \\
		& \qquad \qquad + 2 \rho \nu (x'+iy')\left(-i X_{\underline{01}}(\beta^*-X_{\underline{45}}\alpha^* e^{i\theta_\kappa})-(\alpha^*e^{i\theta_\kappa}+X_{\underline{45}}\beta^*)\lambda\cos\theta \right)=0 \ .
	\end{cases}
\end{equation}

The previous equations arise setting to zero the terms in
multiplied by $\Gamma^6$ and $\Gamma^7$ \eqref{eq:kappa_inter}. Requiring also the other terms in \eqref{eq:kappa_inter} to vanish and following steps analogous to those just shown, one obtains, in the rotated basis, the equation
\begin{equation}
	\begin{split}
		& -2 \rho \lambda \sin \theta \left( x' \sigma_1 + y'  \sigma_2 \right)  \left( X_{\underline{45}} \eta_1 \eta_3 \zeta_{\eta_1,\eta_2,-\eta_3} +  e^{i \theta_\kappa}  \eta_1\zeta_{-\eta_1,\eta_2,\eta_3} \right)+ \\
		& +i \lambda  \theta'f_2 \cos \theta \left( X_{\underline{45}} \eta_1 \zeta_{\eta_1,\eta_2,\eta_3}-e^{i \theta_\kappa} \eta_1\eta_3\zeta_{-\eta_1,\eta_2,-\eta_3} \right)+\\ 
		& + f_2 \theta' \left( e^{i \theta_\kappa}X_{\underline{01}}X_{\underline{45}}  \eta_2 \zeta_{-\eta_1,\eta_2,-\eta_3}+  X_{\underline{01}} \eta_2 \eta_3 \zeta_{\eta_1,\eta_2,\eta_3} \right)-h \zeta_{\eta_1,\eta_2,\eta_3} = 0 
	\end{split}
\end{equation}
which, substituting the expressions of $\zeta$, gives
\begin{equation}\label{eq:D5_BPS_2}
	\begin{cases}
		&-h \alpha+2 i  \lambda \rho \nu (x'+i y') \left(X_{\underline{45}} \alpha^*-\beta^* e^{i \theta_\kappa} \right) \sin  \theta+ \\
		&\qquad + f_2 \theta'\left(X_{\underline{01}}(\alpha+X_{\underline{45}}\beta e^{i \theta_\kappa})+i(X_{\underline{45}}\alpha-\beta e^{i \theta_\kappa})\lambda\cos \theta\right)=0 \\
		&-h \beta^*- 2 i  \nu \lambda \rho (x'-i y') (X_{\underline{45}}\beta+\alpha e^{i \theta_\kappa})\sin \theta+ \\
		& \qquad \qquad +f_2 \theta' \left(X_{\underline{01}}(\beta^*-X_{\underline{45}}\alpha^* e^{i \theta_\kappa})-i(\alpha^* e^{i \theta_\kappa}+X_{\underline{45}} \beta^*) \lambda \cos \theta \right)=0 \ .
	\end{cases}
\end{equation}

\paragraph{Solving the BPS conditions} From the previous discussion, we have obtained four equations, given in \eqref{eq:D5_BPS_1} and \eqref{eq:D5_BPS_2}, which have to be solved simultaneously.  A convenient way to understand the structure of these equations is to define
\begin{equation}\label{eq:expr_AB}
	A \equiv \alpha+X_{\underline{45}} \beta e^{i\theta_\kappa} \ , \qquad B \equiv \alpha X_{\underline{45}} - \beta e^{i\theta_\kappa}
\end{equation}
so that the full set of equations becomes (taking the conjugate of first and fourth)
\begin{equation}\label{eq:set1}
	\begin{cases}
		& -f_2 \lambda \theta' \sin \theta B+2 \rho \nu (x'+i y') \left(i X_{\underline{01}} A^* +\lambda B^* \cos \theta \right)= 0 \\
		& f_2 \lambda \theta' \sin \theta e^{i\theta_\kappa} A + 2 \rho \nu (x'+iy')\left(i X_{\underline{01}}e^{i\theta_\kappa} B^*- e^{i\theta_\kappa}\lambda A^* \cos\theta \right)=0 \\
		
		&-h \alpha+2 i \lambda \rho \nu  B^* \sin \theta (x'+i y') + f_2 \theta'\left(X_{\underline{01}} A+i B \lambda\cos \theta\right)=0 \\
		&h \beta+2 i \nu \lambda \rho e^{i\theta_\kappa} A^* \sin \theta (x'+i y')+f_2 \theta' ( -X_{\underline{01}}e^{i\theta_\kappa}B+i e^{i\theta_\kappa} A \lambda \cos \theta) =0 \ ,
	\end{cases}
\end{equation}
where we used that $e^{i\theta_\kappa}=e^{-i\theta_\kappa}$. We now define
\begin{equation}
	\vartheta \equiv \cos \theta \quad \Rightarrow \quad \vartheta'=-\theta' \sin \theta \ , \ \ (1-\vartheta^2)=\sin^2 \theta \ .
\end{equation} 
We now multiply the last two equations in \eqref{eq:set1} by $\sin \theta$ and take a combination of them such that the terms with $h$ disappear. The resulting set is
\begin{equation}\label{eq:set2}
	\begin{cases}
		& f_2 \lambda \vartheta' B+2 \rho \nu z' \left( i X_{\underline{01}}A^* +  \lambda B^* \vartheta \right)= 0 \\
		& -f_2 \lambda \vartheta' A + 2 \rho \nu z' \left(i X_{\underline{01}} B^*- \lambda  A^* \vartheta \right)=0 \\
		&2 i \nu \lambda \rho  B^* \beta (1-\vartheta^2)z' - f_2 \vartheta' \beta \left(X_{\underline{01}} A+i B \lambda \vartheta\right)+ \\
		&\qquad \qquad +2 i \nu \lambda \rho e^{i\theta_\kappa} A^* \alpha (1-\vartheta^2)z'+f_2 \vartheta' \alpha e^{i\theta_\kappa} (X_{\underline{01}} B-i A \lambda \vartheta)=0 \\
		&\tilde{h} \beta+2 i \nu \lambda \rho e^{i\theta_\kappa} A^* (1-\vartheta^2)z'+f_2 \vartheta' e^{i\theta_\kappa} (X_{\underline{01}} B-i A \lambda \vartheta)=0 \ ,
	\end{cases}
\end{equation}
where $\tilde{h}=h \sin \theta$ and we reintroduced the complex variable $z=x+i y$.
We focus on the first two equations and consider them as a system in $\vartheta'$ and $z'$. In order to have a non trivial solution we require the determinant to vanish
\begin{equation}
	B\left(i X_{\underline{01}} B^*- \lambda A^* \vartheta \right)+A\left( i X_{\underline{01}}A^* +  \lambda B^* \vartheta \right)=0 
\end{equation}
namely
\begin{equation} \label{eq:first_cond_set}
	\vartheta=\frac{i X_{\underline{01}}}{\lambda}\left(\frac{\abs{A}^2+\abs{B}^2}{A^*B-AB^*}\right)= i \lambda X_{\underline{01}}\left(\frac{\abs{\alpha}^2+\abs{\beta}^2}{\alpha \beta^*-\alpha^*\beta}\right)  \ .
\end{equation}
Under this condition, the first two equations are linear dependent and we can get rid of one of them.  So, after imposing \eqref{eq:first_cond_set}, we have 
\begin{equation}\label{eq:subset}
	\begin{cases}
		&  -f_2 \lambda \vartheta' A + 2 \rho \nu z' \left(i X_{\underline{01}}B^*- \lambda A^* \vartheta \right)=0 \\
		&2 i \nu \lambda \rho  B^* \beta (1-\vartheta^2)z' - f_2 \vartheta' \beta \left(X_{\underline{01}} A+i B \lambda \vartheta\right)+ \\
		&\quad \quad +2 i \nu \lambda \rho e^{i\theta_\kappa} A^* \alpha (1-\vartheta^2)z'+f_2 \vartheta' \alpha e^{i\theta_\kappa} (X_{\underline{01}} B-i A \lambda \vartheta)=0  \\
		&\tilde{h} \beta+2 i \nu \lambda \rho e^{i\theta_\kappa} A^* (1-\vartheta^2)z'+f_2 \vartheta' e^{i\theta_\kappa} (X_{\underline{01}} B-i A \lambda \vartheta)=0 \ .
	\end{cases}
\end{equation}
Again, in order to have a non trivial solution in $z'$ and $\vartheta'$ of the first two equations, we should require the vanishing of the determinant. This can be done by choosing $e^{i \theta_k}=1$ and
\begin{equation}\label{eq:second_cond_set}
	X_{\underline{45}}= \frac{\abs{\beta}^2-\abs{\alpha}^2}{\alpha^* \beta+\alpha \beta^*} \ .
\end{equation} 
Now, if we plug this condition into \eqref{eq:subset}, together with \eqref{eq:first_cond_set}, we are left with two equations
\begin{equation}\label{eq:BPS_final_2}
	\begin{cases}
		&  -f_2 \lambda \vartheta' A + 2 \rho \nu z' \left(i X_{\underline{01}}B^*- \lambda A^* \vartheta \right)=0 \\
		&\tilde{h} \beta=-2 i \nu \lambda \rho A^* (1-\vartheta^2)z'-f_2 \vartheta'  (X_{\underline{01}} B-i A \lambda \vartheta) \ .
	\end{cases}
\end{equation}
Substituting the conditions \eqref{eq:first_cond_set} and \eqref{eq:second_cond_set} into the first equation we find
\begin{equation}\label{eq:third_cond_set}
	z'= i \frac{\vartheta'}{2 \vartheta  \nu \rho}\left(\frac{\alpha^* \beta^*(\alpha^2-\beta^2)-\alpha \beta({\alpha^*}^2-{\beta^*}^2)}{{\alpha^*}^2+{\beta^*}^2}\right) \ .
\end{equation}
Plugging this into the second equation of \eqref{eq:BPS_final_2} and squaring both sides leads to an identity. All in all, in order to solve the full set of equations \eqref{eq:set1}, the fluxes and the position of the D5$^\prime$-branes are chosen as in \eqref{eq:first_cond_set}, \eqref{eq:second_cond_set} and \eqref{eq:third_cond_set}, namely
\begin{equation}\label{eq:solution_flux}
	\begin{split}
		X_{\underline{01}}=\lambda \cos \theta\frac{f_2}{f_4} \ , \qquad X_{\underline{45}}= \nu \left(\frac{\abs{\alpha}^2-\abs{\beta}^2}{f_1}\right) 
	\end{split}
\end{equation}
and 
\begin{equation}\label{eq:solution_w}
	4 z'e^{-\phi} \partial_w h_2 =-f_2 f_4 \theta' \tan \theta \ ,
\end{equation}
where we used \eqref{eq:f_def} and \eqref{eq:h_alpha_beta}.

\paragraph{Interpreting the solution}
In order to interpret the previous result, we start considering the expression of $z'$, which identifies how the D5$^\prime$ is placed on the Riemann surface $\Sigma$. The real part of \eqref{eq:solution_w} is
\begin{equation}
	4e^{-\phi}(z'\partial_z h_2 + \bar z'\partial_{\bar z}h_2) = -2f_2f_4 \theta' \tan \theta
\end{equation} 
which, using \eqref{eq:solution_flux}, can be rewritten
\begin{equation}
	2e^{-\phi} (h_2)' = -f_2 f_4 (\ln \abs{f_2f_4})' \ .
\end{equation}
Squaring both sides and using $f_2^2 f_4^2 = 4 e^{-2\phi} h_2^2$, this equation turns out to be an identity. The imaginary part of \eqref{eq:solution_w}, instead, is given by
\begin{equation}
	z'\partial_z h_2 - \bar z'\partial_{\bar z}h_2 =0 
\end{equation}
and it actually defines a curve on $\Sigma$. In order to better characterize it, recall that since $h_2$ is a harmonic function we can write it as the imaginary part of a holomorphic function,
\begin{align}
	h_2&=\cA_2+\bar\cA_2~, &	h_2^D &= i (\mathcal A_2-\bar{ \mathcal A}_2)~.
\end{align}
Using $z=x(y)+i y$ the BPS equation for the embedding becomes
\begin{align}
	\frac{d}{dy} \left(\cA_2(x(y)+i y)-\bar \cA_2(x(y)-iy)\right)&=0 \ ,
\end{align}
which is solved when the dual of $h_2$ is constant along the embedding,
\begin{align}\label{eq:curve_BPS}
	h_2^D&={\rm const}~.
\end{align}

Let us now move to the conditions on $X_{\underline{45}}$ and $X_{\underline{01}}$. In terms of the fluxes in \eqref{eq:gen_fluxes}, the solution in  \eqref{eq:solution_flux} becomes
\begin{equation} \label{eq:fluxes}
	F_1=b_1 + \nu e^{\phi}f_1(|\alpha|^2-|\beta|^2)~, \qquad  F_{\rm el}=\lambda f_4 f_2 e^\phi \cos\theta~.
\end{equation}
In order for this solution to be consistent with the Bianchi identity, these expressions have to be constant along the curve defined by \eqref{eq:curve_BPS}. One can numerically check that, for all the theories considered in this paper, this condition is actually satisfied.

Note that, since the value of fluxes is constant along \eqref{eq:curve_BPS}, we can choose the points along the curves where to evaluate the expressions in \eqref{eq:fluxes}. We can apply this argument to simplify the expression of $F_1$. When we are on $y=0$ (lower boundary of the strip), we find that $f_1=0$ and the second term in the expression of $F_1$ in \eqref{eq:fluxes} vanishes. It is then enough to find the value of $b_1$ on the lower boundary, along the curve \eqref{eq:curve_BPS}. But,
\begin{equation}
	b _1  =  2 h_2^D
\end{equation}
on the lower boundary, as one can check from the explicit expression in \eqref{eq:b1b2}. So we have
\begin{equation}
	F_1=2 h_2^D 
\end{equation}
along \eqref{eq:curve_BPS}, which is consistent with the Bianchi identity.

\subsubsection{On-shell action}\label{app:D5-action-charge}

The Legendre-transformed on-shell action is given by
\begin{align}
	S_{\rm D5}-F_{\rm el}\frac{\delta S_{\rm D5}}{\delta F_{\rm el}}
	&=\frac{8}{\pi^2{\alpha'}^3}I~,
	&
	I&=\int d\xi\, h_1 h_2 (\partial_z h_2) z'~.
\end{align}
Since $h_2^D$ is constant along the embedding, we have  $z'\partial_z \cA_2 = \bar z'\partial_{\bar z}\bar\cA_2$.
Therefore, 
\begin{align}
	\frac{dh_2}{d\xi}&=z'\partial_z\cA_2+\bar z'\partial_{\bar z}\bar\cA_2=2z'\partial_zh_2~.
\end{align}
We find
\begin{align}
	I&=\frac{1}{2}\int d\xi\, h_1 h_2 \frac{dh_2}{d\xi}
	=\frac{1}{4}\left[h_1h_2^2\right]_{\xi_0}^{\xi_1}-\frac{1}{4}\int d\xi\, h_2^2 \frac{dh_1}{d\xi}~,
\end{align}
where integration by parts has been used for the second equality. Evaluating this more explicitly, using  $h_2=2\cA_2+ih_2^D$ , leads to
\begin{align}
	I&=\frac{1}{4}\left[h_1h_2^2\right]_{\xi_0}^{\xi_1}+\frac{i}{4}\int d\xi\, h_2^2\,\partial_z \cA_1 z'
	-\frac{i}{4}\int d\xi\, h_2^2\,\partial_{\bar z}\bar\cA_1 \bar z'
	\nonumber\\
	&=\frac{1}{4}\left[h_1h_2^2\right]_{\xi_0}^{\xi_1}-\frac{1}{2}\Im\left[\int  \big(2\cA_2+ih_2^D\big)^2\partial \cA_1\right]~.
\end{align}
Since $h_2^D$ is constant along the embedding, the remaining integrand is holomorphic.
To integrate it we introduce
\begin{align}
	\partial \cW &= \cA_2^2 \partial \cA_1~.
\end{align}
This leads to (\ref{eq:I-def}).

\bibliographystyle{JHEP.bst}
\bibliography{3d-Wilson}

\providecommand{\href}[2]{#2}\begingroup\raggedright\begin{thebibliography}{10}

\bibitem{Karch:2000gx}
A.~Karch and L.~Randall, \emph{{Open and closed string interpretation of SUSY
  CFT's on branes with boundaries}},
  \href{https://doi.org/10.1088/1126-6708/2001/06/063}{\emph{JHEP} {\bfseries
  06} (2001) 063} [\href{https://arxiv.org/abs/hep-th/0105132}{{\ttfamily
  hep-th/0105132}}].

\bibitem{Karch:2000ct}
A.~Karch and L.~Randall, \emph{{Locally localized gravity}},
  \href{https://doi.org/10.1088/1126-6708/2001/05/008}{\emph{JHEP} {\bfseries
  05} (2001) 008} [\href{https://arxiv.org/abs/hep-th/0011156}{{\ttfamily
  hep-th/0011156}}].

\bibitem{Takayanagi:2011zk}
T.~Takayanagi, \emph{{Holographic Dual of BCFT}},
  \href{https://doi.org/10.1103/PhysRevLett.107.101602}{\emph{Phys. Rev. Lett.}
  {\bfseries 107} (2011) 101602}
  [\href{https://arxiv.org/abs/1105.5165}{{\ttfamily 1105.5165}}].

\bibitem{Fujita:2011fp}
M.~Fujita, T.~Takayanagi and E.~Tonni, \emph{{Aspects of AdS/BCFT}},
  \href{https://doi.org/10.1007/JHEP11(2011)043}{\emph{JHEP} {\bfseries 11}
  (2011) 043} [\href{https://arxiv.org/abs/1108.5152}{{\ttfamily 1108.5152}}].

\bibitem{Almheiri:2019hni}
A.~Almheiri, R.~Mahajan, J.~Maldacena and Y.~Zhao, \emph{{The Page curve of
  Hawking radiation from semiclassical geometry}},
  \href{https://doi.org/10.1007/JHEP03(2020)149}{\emph{JHEP} {\bfseries 03}
  (2020) 149} [\href{https://arxiv.org/abs/1908.10996}{{\ttfamily
  1908.10996}}].

\bibitem{Almheiri:2019yqk}
A.~Almheiri, R.~Mahajan and J.~Maldacena, \emph{{Islands outside the horizon}},
   \href{https://arxiv.org/abs/1910.11077}{{\ttfamily 1910.11077}}.

\bibitem{Rozali:2019day}
M.~Rozali, J.~Sully, M.~Van~Raamsdonk, C.~Waddell and D.~Wakeham,
  \emph{{Information radiation in BCFT models of black holes}},
  \href{https://doi.org/10.1007/JHEP05(2020)004}{\emph{JHEP} {\bfseries 05}
  (2020) 004} [\href{https://arxiv.org/abs/1910.12836}{{\ttfamily
  1910.12836}}].

\bibitem{Almheiri:2019psy}
A.~Almheiri, R.~Mahajan and J.E.~Santos, \emph{{Entanglement islands in higher
  dimensions}},
  \href{https://doi.org/10.21468/SciPostPhys.9.1.001}{\emph{SciPost Phys.}
  {\bfseries 9} (2020) 001} [\href{https://arxiv.org/abs/1911.09666}{{\ttfamily
  1911.09666}}].

\bibitem{Geng:2020qvw}
H.~Geng and A.~Karch, \emph{{Massive islands}},
  \href{https://doi.org/10.1007/JHEP09(2020)121}{\emph{JHEP} {\bfseries 09}
  (2020) 121} [\href{https://arxiv.org/abs/2006.02438}{{\ttfamily
  2006.02438}}].

\bibitem{Chen:2020uac}
H.Z.~Chen, R.C.~Myers, D.~Neuenfeld, I.A.~Reyes and J.~Sandor, \emph{{Quantum
  Extremal Islands Made Easy, Part I: Entanglement on the Brane}},
  \href{https://doi.org/10.1007/JHEP10(2020)166}{\emph{JHEP} {\bfseries 10}
  (2020) 166} [\href{https://arxiv.org/abs/2006.04851}{{\ttfamily
  2006.04851}}].

\bibitem{Uhlemann:2021nhu}
C.F.~Uhlemann, \emph{{Islands and Page curves in 4d from Type IIB}},
  \href{https://doi.org/10.1007/JHEP08(2021)104}{\emph{JHEP} {\bfseries 08}
  (2021) 104} [\href{https://arxiv.org/abs/2105.00008}{{\ttfamily
  2105.00008}}].

\bibitem{Cooper:2018cmb}
S.~Cooper, M.~Rozali, B.~Swingle, M.~Van~Raamsdonk, C.~Waddell and D.~Wakeham,
  \emph{{Black hole microstate cosmology}},
  \href{https://doi.org/10.1007/JHEP07(2019)065}{\emph{JHEP} {\bfseries 07}
  (2019) 065} [\href{https://arxiv.org/abs/1810.10601}{{\ttfamily
  1810.10601}}].

\bibitem{Antonini:2019qkt}
S.~Antonini and B.~Swingle, \emph{{Cosmology at the end of the world}},
  \href{https://doi.org/10.1038/s41567-020-0909-6}{\emph{Nature Phys.}
  {\bfseries 16} (2020) 881}
  [\href{https://arxiv.org/abs/1907.06667}{{\ttfamily 1907.06667}}].

\bibitem{Akal:2020wfl}
I.~Akal, Y.~Kusuki, T.~Takayanagi and Z.~Wei, \emph{{Codimension two holography
  for wedges}}, \href{https://doi.org/10.1103/PhysRevD.102.126007}{\emph{Phys.
  Rev. D} {\bfseries 102} (2020) 126007}
  [\href{https://arxiv.org/abs/2007.06800}{{\ttfamily 2007.06800}}].

\bibitem{Gaiotto:2008sa}
D.~Gaiotto and E.~Witten, \emph{{Supersymmetric Boundary Conditions in N=4
  Super Yang-Mills Theory}},
  \href{https://doi.org/10.1007/s10955-009-9687-3}{\emph{J. Statist. Phys.}
  {\bfseries 135} (2009) 789}
  [\href{https://arxiv.org/abs/0804.2902}{{\ttfamily 0804.2902}}].

\bibitem{Gaiotto:2008ak}
D.~Gaiotto and E.~Witten, \emph{{S-Duality of Boundary Conditions In N=4 Super
  Yang-Mills Theory}},
  \href{https://doi.org/10.4310/ATMP.2009.v13.n3.a5}{\emph{Adv. Theor. Math.
  Phys.} {\bfseries 13} (2009) 721}
  [\href{https://arxiv.org/abs/0807.3720}{{\ttfamily 0807.3720}}].

\bibitem{DHoker:2007zhm}
E.~D'Hoker, J.~Estes and M.~Gutperle, \emph{{Exact half-BPS Type IIB interface
  solutions. I. Local solution and supersymmetric Janus}},
  \href{https://doi.org/10.1088/1126-6708/2007/06/021}{\emph{JHEP} {\bfseries
  06} (2007) 021} [\href{https://arxiv.org/abs/0705.0022}{{\ttfamily
  0705.0022}}].

\bibitem{DHoker:2007hhe}
E.~D'Hoker, J.~Estes and M.~Gutperle, \emph{{Exact half-BPS Type IIB interface
  solutions. II. Flux solutions and multi-Janus}},
  \href{https://doi.org/10.1088/1126-6708/2007/06/022}{\emph{JHEP} {\bfseries
  06} (2007) 022} [\href{https://arxiv.org/abs/0705.0024}{{\ttfamily
  0705.0024}}].

\bibitem{Aharony:2011yc}
O.~Aharony, L.~Berdichevsky, M.~Berkooz and I.~Shamir, \emph{{Near-horizon
  solutions for D3-branes ending on 5-branes}},
  \href{https://doi.org/10.1103/PhysRevD.84.126003}{\emph{Phys. Rev. D}
  {\bfseries 84} (2011) 126003}
  [\href{https://arxiv.org/abs/1106.1870}{{\ttfamily 1106.1870}}].

\bibitem{Assel:2011xz}
B.~Assel, C.~Bachas, J.~Estes and J.~Gomis, \emph{{Holographic Duals of D=3 N=4
  Superconformal Field Theories}},
  \href{https://doi.org/10.1007/JHEP08(2011)087}{\emph{JHEP} {\bfseries 08}
  (2011) 087} [\href{https://arxiv.org/abs/1106.4253}{{\ttfamily 1106.4253}}].

\bibitem{Bachas:2017rch}
C.~Bachas and I.~Lavdas, \emph{{Quantum Gates to other Universes}},
  \href{https://doi.org/10.1002/prop.201700096}{\emph{Fortsch. Phys.}
  {\bfseries 66} (2018) 1700096}
  [\href{https://arxiv.org/abs/1711.11372}{{\ttfamily 1711.11372}}].

\bibitem{Bachas:2018zmb}
C.~Bachas and I.~Lavdas, \emph{{Massive Anti-de Sitter Gravity from String
  Theory}}, \href{https://doi.org/10.1007/JHEP11(2018)003}{\emph{JHEP}
  {\bfseries 11} (2018) 003}
  [\href{https://arxiv.org/abs/1807.00591}{{\ttfamily 1807.00591}}].

\bibitem{Coccia:2020cku}
L.~Coccia, \emph{{Topologically twisted index of $T[SU(N)]$ at large $N$}},
  \href{https://doi.org/10.1007/JHEP05(2021)264}{\emph{JHEP} {\bfseries 05}
  (2021) 264} [\href{https://arxiv.org/abs/2006.06578}{{\ttfamily
  2006.06578}}].

\bibitem{Raamsdonk:2020tin}
M.V.~Raamsdonk and C.~Waddell, \emph{{Holographic and localization calculations
  of boundary F for $ \mathcal{N} $ = 4 SUSY Yang-Mills theory}},
  \href{https://doi.org/10.1007/JHEP02(2021)222}{\emph{JHEP} {\bfseries 02}
  (2021) 222} [\href{https://arxiv.org/abs/2010.14520}{{\ttfamily
  2010.14520}}].

\bibitem{Coccia:2020wtk}
L.~Coccia and C.F.~Uhlemann, \emph{{On the planar limit of 3d
  $T_\rho^\sigma[SU(N)]$}},  \href{https://arxiv.org/abs/2011.10050}{{\ttfamily
  2011.10050}}.

\bibitem{Reeves:2021sab}
W.~Reeves, M.~Rozali, P.~Simidzija, J.~Sully, C.~Waddell and D.~Wakeham,
  \emph{{Looking for (and not finding) a bulk brane}},
  \href{https://arxiv.org/abs/2108.10345}{{\ttfamily 2108.10345}}.

\bibitem{VanRaamsdonk:2021duo}
M.~Van~Raamsdonk and C.~Waddell, \emph{{Finding AdS$^{5}$\texttimes{} S$^{5}$
  in 2+1 dimensional SCFT physics}},
  \href{https://doi.org/10.1007/JHEP11(2021)145}{\emph{JHEP} {\bfseries 11}
  (2021) 145} [\href{https://arxiv.org/abs/2109.04479}{{\ttfamily
  2109.04479}}].

\bibitem{Akhond:2021ffz}
M.~Akhond, A.~Legramandi and C.~Nunez, \emph{{Electrostatic description of 3d $
  \mathcal{N} $ = 4 linear quivers}},
  \href{https://doi.org/10.1007/JHEP11(2021)205}{\emph{JHEP} {\bfseries 11}
  (2021) 205} [\href{https://arxiv.org/abs/2109.06193}{{\ttfamily
  2109.06193}}].

\bibitem{DeLuca:2021ojx}
G.B.~De~Luca, N.~De~Ponti, A.~Mondino and A.~Tomasiello, \emph{{Cheeger bounds
  on spin-two fields}},
  \href{https://doi.org/10.1007/JHEP12(2021)217}{\emph{JHEP} {\bfseries 12}
  (2021) 217} [\href{https://arxiv.org/abs/2109.11560}{{\ttfamily
  2109.11560}}].

\bibitem{Uhlemann:2019ypp}
C.F.~Uhlemann, \emph{{Exact results for 5d SCFTs of long quiver type}},
  \href{https://doi.org/10.1007/JHEP11(2019)072}{\emph{JHEP} {\bfseries 11}
  (2019) 072} [\href{https://arxiv.org/abs/1909.01369}{{\ttfamily
  1909.01369}}].

\bibitem{Uhlemann:2020bek}
C.F.~Uhlemann, \emph{{Wilson loops in 5d long quiver gauge theories}},
  \href{https://doi.org/10.1007/JHEP09(2020)145}{\emph{JHEP} {\bfseries 09}
  (2020) 145} [\href{https://arxiv.org/abs/2006.01142}{{\ttfamily
  2006.01142}}].

\bibitem{Hanany:1996ie}
A.~Hanany and E.~Witten, \emph{{Type IIB superstrings, BPS monopoles, and
  three-dimensional gauge dynamics}},
  \href{https://doi.org/10.1016/S0550-3213(97)00157-0}{\emph{Nucl. Phys. B}
  {\bfseries 492} (1997) 152}
  [\href{https://arxiv.org/abs/hep-th/9611230}{{\ttfamily hep-th/9611230}}].

\bibitem{Nishioka:2011dq}
T.~Nishioka, Y.~Tachikawa and M.~Yamazaki, \emph{{3d Partition Function as
  Overlap of Wavefunctions}},
  \href{https://doi.org/10.1007/JHEP08(2011)003}{\emph{JHEP} {\bfseries 08}
  (2011) 003} [\href{https://arxiv.org/abs/1105.4390}{{\ttfamily 1105.4390}}].

\bibitem{Cremonesi:2014uva}
S.~Cremonesi, A.~Hanany, N.~Mekareeya and A.~Zaffaroni,
  \emph{{T$_{\rho}^{\sigma}$ (G) theories and their Hilbert series}},
  \href{https://doi.org/10.1007/JHEP01(2015)150}{\emph{JHEP} {\bfseries 01}
  (2015) 150} [\href{https://arxiv.org/abs/1410.1548}{{\ttfamily 1410.1548}}].

\bibitem{Assel:2015oxa}
B.~Assel and J.~Gomis, \emph{{Mirror Symmetry And Loop Operators}},
  \href{https://doi.org/10.1007/JHEP11(2015)055}{\emph{JHEP} {\bfseries 11}
  (2015) 055} [\href{https://arxiv.org/abs/1506.01718}{{\ttfamily
  1506.01718}}].

\bibitem{Dey:2021jbf}
A.~Dey, \emph{{Line Defects in Three Dimensional Mirror Symmetry beyond Linear
  Quivers}},  \href{https://arxiv.org/abs/2103.01243}{{\ttfamily 2103.01243}}.

\bibitem{Dey:2021gbi}
A.~Dey, \emph{{Line Defects in Three Dimensional Mirror Symmetry beyond ADE
  quivers}},  \href{https://arxiv.org/abs/2112.04969}{{\ttfamily 2112.04969}}.

\bibitem{Kapustin:2009kz}
A.~Kapustin, B.~Willett and I.~Yaakov, \emph{{Exact Results for Wilson Loops in
  Superconformal Chern-Simons Theories with Matter}},
  \href{https://doi.org/10.1007/JHEP03(2010)089}{\emph{JHEP} {\bfseries 03}
  (2010) 089} [\href{https://arxiv.org/abs/0909.4559}{{\ttfamily 0909.4559}}].

\bibitem{Assel:2012cp}
B.~Assel, J.~Estes and M.~Yamazaki, \emph{{Large N Free Energy of 3d N=4 SCFTs
  and $AdS_4/CFT_3$}},
  \href{https://doi.org/10.1007/JHEP09(2012)074}{\emph{JHEP} {\bfseries 09}
  (2012) 074} [\href{https://arxiv.org/abs/1206.2920}{{\ttfamily 1206.2920}}].

\bibitem{Drukker:2019bev}
N.~Drukker et~al., \emph{{Roadmap on Wilson loops in 3d
  Chern\textendash{}Simons-matter theories}},
  \href{https://doi.org/10.1088/1751-8121/ab5d50}{\emph{J. Phys. A} {\bfseries
  53} (2020) 173001} [\href{https://arxiv.org/abs/1910.00588}{{\ttfamily
  1910.00588}}].

\bibitem{Estes:2012nx}
J.~Estes, A.~O'Bannon, E.~Tsatis and T.~Wrase, \emph{{Holographic Wilson Loops,
  Dielectric Interfaces, and Topological Insulators}},
  \href{https://doi.org/10.1103/PhysRevD.87.106005}{\emph{Phys. Rev. D}
  {\bfseries 87} (2013) 106005}
  [\href{https://arxiv.org/abs/1210.0534}{{\ttfamily 1210.0534}}].

\bibitem{Gutperle:2020gez}
M.~Gutperle and C.F.~Uhlemann, \emph{{Janus on the Brane}},
  \href{https://doi.org/10.1007/JHEP07(2020)243}{\emph{JHEP} {\bfseries 07}
  (2020) 243} [\href{https://arxiv.org/abs/2003.12080}{{\ttfamily
  2003.12080}}].

\bibitem{Rey:1998ik}
S.-J.~Rey and J.-T.~Yee, \emph{{Macroscopic strings as heavy quarks in large N
  gauge theory and anti-de Sitter supergravity}},
  \href{https://doi.org/10.1007/s100520100799}{\emph{Eur. Phys. J. C}
  {\bfseries 22} (2001) 379}
  [\href{https://arxiv.org/abs/hep-th/9803001}{{\ttfamily hep-th/9803001}}].

\bibitem{Maldacena:1998im}
J.M.~Maldacena, \emph{{Wilson loops in large N field theories}},
  \href{https://doi.org/10.1103/PhysRevLett.80.4859}{\emph{Phys. Rev. Lett.}
  {\bfseries 80} (1998) 4859}
  [\href{https://arxiv.org/abs/hep-th/9803002}{{\ttfamily hep-th/9803002}}].

\bibitem{Yamaguchi:2006tq}
S.~Yamaguchi, \emph{{Wilson loops of anti-symmetric representation and
  D5-branes}}, \href{https://doi.org/10.1088/1126-6708/2006/05/037}{\emph{JHEP}
  {\bfseries 05} (2006) 037}
  [\href{https://arxiv.org/abs/hep-th/0603208}{{\ttfamily hep-th/0603208}}].

\bibitem{Drukker:2008wr}
N.~Drukker, J.~Gomis and S.~Matsuura, \emph{{Probing N=4 SYM With Surface
  Operators}}, \href{https://doi.org/10.1088/1126-6708/2008/10/048}{\emph{JHEP}
  {\bfseries 10} (2008) 048} [\href{https://arxiv.org/abs/0805.4199}{{\ttfamily
  0805.4199}}].

\bibitem{Geng:2021mic}
H.~Geng, A.~Karch, C.~Perez-Pardavila, S.~Raju, L.~Randall, M.~Riojas et~al.,
  \emph{{Entanglement Phase Structure of a Holographic BCFT in a Black Hole
  Background}},  \href{https://arxiv.org/abs/2112.09132}{{\ttfamily
  2112.09132}}.

\bibitem{Balasubramanian:2017hgy}
V.~Balasubramanian, A.~Lawrence, A.~Rolph and S.~Ross, \emph{{Entanglement
  shadows in LLM geometries}},
  \href{https://doi.org/10.1007/JHEP11(2017)159}{\emph{JHEP} {\bfseries 11}
  (2017) 159} [\href{https://arxiv.org/abs/1704.03448}{{\ttfamily
  1704.03448}}].

\bibitem{Geng:2020fxl}
H.~Geng, A.~Karch, C.~Perez-Pardavila, S.~Raju, L.~Randall, M.~Riojas et~al.,
  \emph{{Information Transfer with a Gravitating Bath}},
  \href{https://doi.org/10.21468/SciPostPhys.10.5.103}{\emph{SciPost Phys.}
  {\bfseries 10} (2021) 103}
  [\href{https://arxiv.org/abs/2012.04671}{{\ttfamily 2012.04671}}].

\bibitem{Graham:2014iya}
C.R.~Graham and A.~Karch, \emph{{Minimal area submanifolds in AdS x compact}},
  \href{https://doi.org/10.1007/JHEP04(2014)168}{\emph{JHEP} {\bfseries 04}
  (2014) 168} [\href{https://arxiv.org/abs/1401.7692}{{\ttfamily 1401.7692}}].

\bibitem{Erdmenger:2014xya}
J.~Erdmenger, M.~Flory and M.-N.~Newrzella, \emph{{Bending branes for DCFT in
  two dimensions}}, \href{https://doi.org/10.1007/JHEP01(2015)058}{\emph{JHEP}
  {\bfseries 01} (2015) 058} [\href{https://arxiv.org/abs/1410.7811}{{\ttfamily
  1410.7811}}].

\bibitem{Simidzija:2020ukv}
P.~Simidzija and M.~Van~Raamsdonk, \emph{{Holo-ween}},
  \href{https://doi.org/10.1007/JHEP12(2020)028}{\emph{JHEP} {\bfseries 12}
  (2020) 028} [\href{https://arxiv.org/abs/2006.13943}{{\ttfamily
  2006.13943}}].

\bibitem{Bachas:2021fqo}
C.~Bachas and V.~Papadopoulos, \emph{{Phases of Holographic Interfaces}},
  \href{https://doi.org/10.1007/JHEP04(2021)262}{\emph{JHEP} {\bfseries 04}
  (2021) 262} [\href{https://arxiv.org/abs/2101.12529}{{\ttfamily
  2101.12529}}].

\bibitem{Assel:2012cj}
B.~Assel, C.~Bachas, J.~Estes and J.~Gomis, \emph{{IIB Duals of D=3 N=4
  Circular Quivers}},
  \href{https://doi.org/10.1007/JHEP12(2012)044}{\emph{JHEP} {\bfseries 12}
  (2012) 044} [\href{https://arxiv.org/abs/1210.2590}{{\ttfamily 1210.2590}}].

\bibitem{Uhlemann:2021itz}
C.F.~Uhlemann, \emph{{Information transfer with a twist}},
  \href{https://arxiv.org/abs/2111.11443}{{\ttfamily 2111.11443}}.

\bibitem{Cederwall:1996ri}
M.~Cederwall, A.~von Gussich, B.E.W.~Nilsson, P.~Sundell and A.~Westerberg,
  \emph{{The Dirichlet super p-branes in ten-dimensional type IIA and IIB
  supergravity}},
  \href{https://doi.org/10.1016/S0550-3213(97)00075-8}{\emph{Nucl. Phys. B}
  {\bfseries 490} (1997) 179}
  [\href{https://arxiv.org/abs/hep-th/9611159}{{\ttfamily hep-th/9611159}}].

\bibitem{Bergshoeff:1996tu}
E.~Bergshoeff and P.K.~Townsend, \emph{{Super D-branes}},
  \href{https://doi.org/10.1016/S0550-3213(97)00072-2}{\emph{Nucl. Phys. B}
  {\bfseries 490} (1997) 145}
  [\href{https://arxiv.org/abs/hep-th/9611173}{{\ttfamily hep-th/9611173}}].

\bibitem{Karch:2015vra}
A.~Karch, B.~Robinson and C.F.~Uhlemann, \emph{{Supersymmetric D3/D7 for
  holographic flavors on curved space}},
  \href{https://doi.org/10.1007/JHEP11(2015)112}{\emph{JHEP} {\bfseries 11}
  (2015) 112} [\href{https://arxiv.org/abs/1508.06996}{{\ttfamily
  1508.06996}}].

\bibitem{Gutperle:2018vdd}
M.~Gutperle, A.~Trivella and C.F.~Uhlemann, \emph{{Type IIB 7-branes in warped
  AdS$_{6}$: partition functions, brane webs and probe limit}},
  \href{https://doi.org/10.1007/JHEP04(2018)135}{\emph{JHEP} {\bfseries 04}
  (2018) 135} [\href{https://arxiv.org/abs/1802.07274}{{\ttfamily
  1802.07274}}].

\end{thebibliography}\endgroup
\end{document}